\documentclass[prd,aps,showpacs,nofootinbib]{revtex4}
\usepackage{amssymb}
\usepackage{graphicx}
\usepackage{amsmath}

\newcommand{\be}{\begin{equation}}
\newcommand{\ee}{\end{equation}}
\newcommand{\bq}{\begin{eqnarray}}
\newcommand{\eq}{\end{eqnarray}}

\begin{document}
\title{On the electrodynamics of moving particles in a quasi flat spacetime with Lorentz violation and its cosmological implications} 
\author{*Cl\'audio Nassif Cruz}

\affiliation{\small{{\bf*CPFT}:{\bf Centro de Pesquisas em F\'isica Te\'orica}, Rua Rio de Janeiro 1186, Lourdes, CEP:30.160-041,
 Belo Horizonte-MG, Brazil.\\
  {\bf cnassif@cbpf.br}; {\bf cncruz777@yahoo.com.br}}}

\begin{abstract}
This research aims to develop a new approach toward a consistent coupling of electromagnetic and gravitational fields, by
using an electron that couples with a weak gravitational potential by means of its electromagnetic field. To accomplish this, we must first
build a new model which provides the electromagnetic nature of both the mass and the energy of the electron, and which is implemented with the
idea of ​​$\gamma$-photon decay into an electron-positron pair. After this, we place the electron (or positron) in the presence of a
weak gravitational potential given in the intergalactic medium, so that its electromagnetic field undergoes a very small perturbation,
thus leading to a slight increase in the field's electromagnetic energy density. This perturbation takes place by means of a tiny coupling
constant $\xi$ because gravity is a very weak interaction compared with the electromagnetic one. Thus we realize that $\xi$ is a new 
dimensionless universal constant, which reminds us of the fine structure constant $\alpha$; however, $\xi$ is much smaller than $\alpha$ 
because $\xi$ takes into account gravity, i.e., $\xi\propto\sqrt{G}$. We find $\xi=V/c\cong 1.5302\times 10^{-22}$,
where $c$ is the speed of light and $V\propto\sqrt{G}(\cong 4.5876\times 10^{-14}$m/s) is a universal minimum speed that represents
the lowest limit of speed for any particle. Such a minimum speed, unattainable by particles, represents a preferred reference frame 
associated with a background field that breaks the Lorentz symmetry. The metric of the flat space-time shall include the presence of a 
uniform vacuum energy density, which leads to a negative pressure at cosmological scales (cosmological anti-gravity). The tiny values of the 
cosmological constant and the vacuum energy density will be successfully obtained in agreement with the observational data. 
\end{abstract}

\pacs{{\bf 11.30.Qc}
     {\bf DOI:10.1142/S0218271816500966}\\
     http://www.worldscientific.com/worldscinet/ijmpd?journalTabs=read}
\maketitle

\section{Introduction}

After Einstein who has tried to find in vain the unified field theory, where General Relativity (GR), Electromagnetism (EM) and 
Quantum Mechanics (QM) are put into together, many theoretical physicists have been attempting for decades to find a satisfactory 
marriage of quantum field theory and gravity. While the full theory
of quantum gravity is far from being achieved, a great deal of work has been done and some efforts to couple the electron minimally 
with gravity have also been made by using the Dirac equation\cite{1}. In sum, Electrodynamics in curved space-time has been well explored,
however this subject is still an open question in view of the need of searching for alternative possibilities of coupling electromagnetic 
fields with gravity, so that the Lorentz symmetry could be broken by emerging new invariant quantities in space-time. 

In this work, let us search for a new line of research of coupling of fields by using new basic concepts, where, for instance, the 
influence of gravity on electromagnetic fields is already capable of breaking the Lorentz symmetry only at much lower energies due
to the existence of an invariant minimum speed for subatomic particles, leading to a Doubly Special Relativity (DSR) with two invariant
speeds, namely the well-known speed of light $c$ and a new invariant speed, i.e., a minimum speed $V$, providing a new symmetry in 
space-time. So, we admit that a modified extension of ``On the Electrodynamics of Moving Bodies"\cite{2} due to the presence of a weak
gravitational potential (a quasi flat space-time) leads to such a breaking of the Lorentz symmetry when a week gravity is dully coupled to
electrodynamics, thus deforming the Minkowsky space-time by including an invariant minimum speed in the subatomic world. This leads to a 
fundamental energy state connected to a universal background field that has deep cosmological implications such as the cosmological 
anti-gravity, thus leading to the solution of the well-known cosmological constant problem\cite{3}. 

We know that the density of matter of the universe as a whole is too small (a quasi flat space-time) and the whole universe is flat due 
to the presence of the dark energy, i.e.,$\Omega_{M}(\approx 0.3)+\Omega_{\Lambda}(\approx 0.7)=1$\cite{3}. Therefore, as we are only 
interested to focus our attention in electrodynamics of moving particles in the presence of gravity within the modern cosmological context in
order to search for the comprehension of the origin of the low vacuum energy density and its connection with the cosmological anti-gravity,
we will take into account the case of a charged particle coupled to a weak gravitational potential given in the intergalactic medium. 

 The current research attempts to combine quantum effects of electromagnetic fields with effects of weak gravity in the 
intergalactic medium. However, we use an alternative (heuristic) approach 
in order to extend, by means of the idea of complementarity, the classical concept of electromagnetic fields (vectorial fields to represent waves) to a new 
quantum aspect of such fields (scalar fields to represent corpuscules) and, thus, explain the nature of the relativistic mass as being of 
electromagnetic origin, given in a scalar form (corpuscular aspect of matter). With this heuristic approach, it becomes possible to establish 
a new form of coupling between electromagnetic and gravitational fields in order to emphasize the fact that gravity is the weakest
interaction in nature (the hierarchical problem), since such a new coupling is given by a tiny pure number with direct dependence on the 
small constant of gravitation ($G$). So, the present approach will allow us to realize that a minuscule influence of gravity on electromagnetic fields 
leads to a violation of the Lorentz symmetry, which should be thoroughly investigated. Such minuscule apparent violations of the Lorentz invariance
might be observable in nature\cite{4}\cite{5}. The basic idea is that the violations would arise as suppressed effects from a more fundamental 
theory of space-time with quantum-gravity effects. 

 In the present work, a theory with Lorentz symmetry breakdown will be investigated by searching for a new kind of coupling, i.e., a 
 dimensionless coupling constant, working like a kind of ``fine structure" constant which couples the gravitational field with the
 electromagnetic one, in such a way as to extend our notion of flat (empty) space-time by means of the emergence of a background 
 field related to a zero-point energy (a vacuum energy), which is associated to a minimum speed ($V$) as an invariant and universal constant 
 for lower energy scales. This leads to a violation of the Lorentz symmetry because $V$ establishes the existence of a preferential reference
 frame for the background field (vacuum energy). 

  During the last 30 years of Einstein's life, he attempted to bring the principles of QM and EM into his theory of gravity (GR) 
 by means of a unified field theory\cite{6}. Unfortunately his effort was not
 successful in establishing a consistent theory between QM, EM and GR, from where the uncertainty principle should naturally emerge, i.e.,
 a consistent quantum gravity theory. Currently string theories inspired by an old idea of Kaluza\cite{7} and Klein\cite{8} regarding
 extra dimensions in space-time have been prevailing in the scenario of attempts to find a unified theory\cite{9}.

  In the next section, a new model will be built to describe the electromagnetic nature of the electron mass. It is based on Maxwell
 theory used for investigating the electromagnetic nature of a photon when the amplitudes of the fields of a certain electromagnetic wave
 are normalized just for one single photon with energy $\hbar w$\cite{10}. Thus, due to reasonings of reciprocity and symmetry that give
 support to our heuristic approach, we should extend such an alternative model of the photon to be applied to matter (electron),
 implemented with the idea of pair materialization ($e^-|e^+$), after $\gamma$-photon decay. So we define an electromagnetic mass for the 
 pair $e^-|e^+$ in order to compute the energy of the pair, such that it must be equivalent to its own energy $m_e^-c^2$($m_e^+c^2)\cong 0.51$MeV. 

 In Section 3, we will consider an electron (or positron) moving in the presence of a weak gravitational potential, so that the electromagnetic
field created in the space around it undergoes a very small perturbation due to gravity, thus leading to a slight increase of the electromagnetic
energy density. Such an increase occurs by means of a dimensionless coupling constant whose value is infinitesimal, since it has its origin
in gravity, the weakest interaction. This coupling constant is called the {\it fine adjustment constant} $\xi$, which behaves like a kind
of minuscule fine structure, which will be obtained from the hydrogen atom (Section 4). Nevertheless, unlike the usual fine structure constant 
of quantum electrodynamics, the constant $\xi$ plays a coupling role of gravitational and electromagnetic origin rather than simply being an
electric interaction between two electronic charges. 

In Section 4, we will obtain the value of the tiny dimensionless coupling constant $\xi$, where we find that $\xi=V/c=
\sqrt{Gm_em_p/4\pi\epsilon_0}~q_e/\hbar c\cong 1.5302\times 10^{-22}$, $V(\cong 4.5876\times 10^{-14}$m/s) being interpreted as the
universal minimum speed unattainable for any particle at lower energies, and of the same status as the speed of light $c$ (maximum speed)
for higher energies, in the sense that both speeds $c$ and $V$ are invariant. Thus, we have a new kind of DSR,
that was denominated as Symmetrical Special Relativity (SSR)\cite{11}\cite{12}\cite{13}, where the unattainable minimum speed is 
associated with a privileged reference frame of background field (a non-luminiferous aether). 

 In Section 5, we will investigate the transformations of space-time and velocity in the presence of the reference frame of
such a background field connected to the invariant minimum speed $V$. 

 In Section 6, we will generalize the space-time transformations for $(3+1)D$ with the presence of such a background field 
connected to the background velocity vector $\vec V$ that breaks the Lorentz symmetry. 

 In Section 7, we will investigate whether the new transformations of SSR form a group and what are their deep implications. 

 In Section 8, we will study the energy and momentum of a particle with the presence of the minimum speed. The Lagrangian of a particle
in SSR will be obtained for some important cases. 

 In Section 9, we will investigate how the electromagnetic fields ($F_{\mu\nu}$) are transformed with the change of the reference frames
 in space-time of SSR. The simple case $(1+1)D$ and an important extension of this case will be explored. 

 In Section 10, we will investigate the cosmological implications of the homogeneous background field connected to the invariant 
minimum speed $V$ that breaks the Lorentz symmetry, leading to a negative pressure at cosmological scales (cosmological anti-gravity).
The tiny values of the cosmological constant and the vacuum energy density will be successfully obtained, being in agreement with the 
observational results.

The last section will be dedicated to some theoretical and experimental prospects by proposing two experiments to test both of the 
existence of such a universal minimum speed and any crosstalk between gravitation and electromagnetism.

\section{Electromagnetic Nature of the Photon and Electron (or Positron)}

\subsection{Electromagnetic nature of the photon}

In accordance with some laws of Electrodynamics introduced by Feynman\cite{10}, we obtain that the electric field of 
a plane electromagnetic wave, whose amplitude is normalized for just one single photon\cite{10}, is of the form, namely: 

\begin{equation}
\vec{E}(z,t)=\sqrt{8\pi\hbar w}\sin(wt-kz)\vec e,
\end{equation}
where $\vec k.\vec r=kz$, admitting that the wave propagates in the direction of $z$, $\vec e$ being the unitary vector of
polarization. 

From Eq.(1), we see that $e_0(=\sqrt{8\pi\hbar w})$ could be thought of as an electric field amplitude normalized for one
single photon, since we consider the Fock space. So we also have $b_0=e_0$ (Gaussian units) to be the magnetic field amplitude normalized
for one single photon. Thus we write $\vec{E}(z,t)= e_0\sin(wt-kz)\vec e$. 

In the Gaussian System of units, we have $|\vec E|=|\vec B|$. So, the average energy density of the electromagnetic wave normalized for 
one single photon with a unitary volume ($v_{ph}=1$)\cite{10} is 

\begin{equation}
\left<E_{em}\right>=\frac{1}{4\pi}e_m^2\equiv\hbar w,
\end{equation}
where we consider an average quadratic electric field normalized for one single photon, namely $e_m=\sqrt{\left<|\vec E|^2\right>}=
e_0/\sqrt{2}=\sqrt{4\pi\hbar w}$. 

 Here it is important to emphasize that, although the field given in Eq.(1) is normalized for only one photon, it is still a classical
 field of Maxwell in the sense that its value oscillates like a classical wave [Eq.(1)]; the only difference here is that we have 
considered a small amplitude field for just one photon, given in the Fock space. 

   Actually, the amplitude of the microscopic field ($e_0$) cannot be measured directly. Only in the
 classical approximation (macroscopic case), where we have a very large number of photons ($N\rightarrow\infty$), can we somehow measure
 the macroscopic field $\vec E$ of the wave. Therefore, although we could idealize the case of one single photon as if it were a Maxwell
 electromagnetic wave of small amplitude, Eq.(1) is still a classical solution since the field $\vec E$ presents oscillation.

   On the other hand, we already know that the photon wave is a quantum wave, i.e., it is a de-Broglie wave, where its wavelength
 ($\lambda=h/p$) is not interpreted classically as the oscillation frequency (wavelength due to oscillation) of a classical field because,
 if it were so, using the classical solution in Eq.(1), we would have the electromagnetic energy density, namely: 

\begin{equation}
E_{em}=\frac{1}{4\pi}|\vec{E}(z,t)|^2=
\frac{1}{4\pi}e_0^2\sin{^2}(wt-kz).
\end{equation}

If the wave of a photon were really a classical wave, then its energy would not have a fixed value according to Eq.(3).
 Consequently, its energy $\hbar w$ would be only an average value [see Eq.(2)]. Hence, in order to achieve consistency between the 
 result in Eq.(2) and the quantum wave (de-Broglie wave), we must interpret Eq.(2) as being related to the de-Broglie wave of the
 photon with a discrete and fixed value of energy $\hbar w$ instead of an average energy value of an oscilating classical field,
 since we should consider the wave of one single photon as being a non-classical wave, namely a de-Broglie (quantum) wave. 
 Thus we simply rewrite Eq.(2), as follows:

\begin{equation}
E_{em}=E=pc=\frac{hc}{\lambda}=\hbar w\equiv\frac{1}{4\pi}e_{ph}^2,
\end{equation}
from where we conclude

\begin{equation}
\lambda\equiv\frac{4\pi hc}{e_{ph}^2},
\end{equation}
where we replace $e_m$ by $e_{ph}$, $\lambda$ being the de-Broglie wavelength.

 According to Eq.(5), the single photon field $e_{ph}$ 
should not be assumed as a mean value for an oscillating classical (vectorial) field, and so now we shall preserve it in order to be
interpreted as a {\it quantum electric field}, i.e., a {\it microscopic scalar field} of one single photon. Thus, let us also call $e_{ph}$
a {\it scalar electric field} for representing the corpuscular (or quantum-mechanical) aspect of the field normalized for one single photon. 
So, the quantity $e_{ph}$ is taken just as the magnitude of its mean electric field ($e_m$), being simply $e_m=e_{ph}$. 

As the scalar field $e_{ph}$ provides the energy of the photon ($E\propto e_{ph}^2$), with $w\propto e_{ph}^2$ and
$\lambda\propto 1/e_{ph}^2$, we realize that $e_{ph}$ presents a quantum behavior since it provides the dual aspect (wave-particle) 
of the photon, so that its mechanical momentum may be written as $p=\hbar k=2\pi\hbar/{\lambda}$=$\hbar e_{ph}^2/2hc$ [refer to Eq.(5)],
or simply $p=e_{ph}^2/4\pi c$.

\subsection{Electromagnetic nature of the electron (positron) mass}

   Our goal is to extend the idea of photon electromagnetic energy [Eq.(4)] to matter. By doing this, we shall provide
 heuristic arguments that rely directly on the de-Broglie reciprocity postulate, which has extended the idea of wave (photon wave) to 
 matter (electron), which also behaves like wave. Thus, Eq.(5) considers that the photon quantum field $e_{ph}$ given by the alternative
 de-Broglie relation ($p=h/\lambda=e_{ph}^2/4\pi c$), may also be extended to matter (electron) in accordance
 with the very idea of de-Broglie reciprocity. In order to strengthen such an
 argument, besides this, we are going to assume the phenomenon of pair formation, where the $\gamma$-photon decays into two charged massive
 particles, namely the electron ($e^{-}$) and its anti-particle, the positron ($e^{+}$). Such an example will enable us to understand
 better the need of extending, by means of the idea of reciprocity, the concept of scalar field of a photon and its electromagnetic mass
 (relativistic energy) [Eq.(4)] to be also introduced into the matter (massive particles like $e^{-}$ and $e^{+}$). So, we use
 the heuristic assumption about {\it scalar electromagnetic fields} to simply represent the magnitudes of such fields for matter, 
 such magnitudes being associated with the corpuscular aspect of the fields, so that these magnitudes are now given by scalar quantities
 related to the mass-energy [see Eq.(10)]. 

  In short, we intend to extend the idea of de-Broglie reciprocity to be applied specifically to electromagnetic fields, and to expand
our notion of {\it wave-field-matter} so that, whereas the well-known vectorial (classical) fields represent waves, we introduce the
heuristic concept of scalar (quantum) fields that provides the corpuscular aspect for matter. Thus, the electromagnetic field itself 
should also have an aspect of duality, namely the {\it corpuscular-wave field}. 

  Now consider the phenomenon of pair formation in the presence of a nucleus, i.e., $\gamma\rightarrow e^{-}+e^{+}$.
Here, for our purpose, let us just take into account the conservation of energy for $\gamma$-decay, although the momentum must also
be conserved. So we write the following balanced energy equation:

\begin{equation}
E_{\gamma}=\hbar w = m_{\gamma}c^2=m_0^{-}c^2+m_0^{+}c^2 + K^{-}+ K^{+},
\end{equation}
with $m_0^{-}c^2$ (or $m_0^{+}c^2$) being the electron (or positron) mass, where $m_0^{-}c^2+m_0^{+}c^2=2m_0c^2$, since the electron and 
positron have the same mass ($m_0$). $K^{-}$ and $K^{+}$ represent the kinetic energies of the electron and positron respectively.
We have $E_0=m_0^{-}c^2=m_0^{+}c^2\cong 0,51$MeV.

Since the electromagnetic energy of the $\gamma$-photon is $E_{\gamma}=h\nu=m_{\gamma}c^2=\frac{1}{4\pi}e_{\gamma}^2=\frac{1}{4\pi}
e_{\gamma}b_{\gamma}$, or likewise, in IS (International System) of units, we have $E_{\gamma}=\epsilon_0e_{\gamma}^2$, and also knowing that
 $e_{\gamma}=cb_{\gamma}$ (in IS), where $b_{\gamma}$ is the {\it magnetic scalar field} of the $\gamma$-photon, we may also write

\begin{equation}
E_{\gamma}=\hbar w=c\epsilon_0(e_{\gamma})(b_{\gamma}), 
\end{equation}
where we already have considered the volume of the photon being normalized to $1$, i.e., $v_{\gamma}=1$\cite{10}. 

 The photon has no charge, however, when the $\gamma$-photon is materialized into an electron-positron pair, its electromagnetic content
 (the scalar fields $e_{\gamma}$ and $b_{\gamma}$) given in Eq.(7) ceases to be free or purely kinetic (pure relativistic mass) to become 
 massive due to the materialization of the pair. Since such massive particles ($v_{(+,-)}<c$) also behave like waves in accordance with
 the de-Broglie idea, it would be natural to extend Eq.(5) of the photon for representing now the wavelengths of matter 
 (electron or positron) after $\gamma$-decay, namely:

\begin{equation}
\lambda_{(+,-)}\propto\frac{hc}{\epsilon_0[e_s^{(+,-)}]^2}=
\frac{h}{\epsilon_0[e_s^{(+,-)}][b_s^{(+,-)}]},
\end{equation}
where the fields $e_s^{(+,-)}$ and $b_s^{(+,-)}$ play the role of the electromagnetic content (scalar electromagnetic fields) 
that provides the total mass (energy) of the particle (electron or positron), its mass being essentially of electromagnetic origin,
as follows:

\begin{equation}
m\equiv m_{em}\propto e_sb_s,
\end{equation}
where $E=mc^2\equiv m_{em}c^2$.

Using Eq.(7) and Eq.(8) as a basis, we may write Eq.(6) in the following way:

\begin{equation}
E_{\gamma}=c\epsilon_0e_{\gamma}b_{\gamma}=
c\epsilon_0e_s^{-}b_s^{-}v_e^{-}+c\epsilon_0e_s^{+}b_s^{+}v_e^{+}, 
\end{equation}
with $v_e$ being the volume of the electron (or positron), where $c\epsilon_0e_s^{-}b_s^{-}v_e^{-}=(c\epsilon_0e_{s0}^{-}b_{s0}^{-}v_e
 + K^{-})=(m_0^{-}c^2+ K^{-})$ and $c\epsilon_0e_s^{+}b_s^{+}v_e^{+}=(c\epsilon_0e_{s0}^{+}b_{s0}^{+}v_e + K^{+})=(m_0^{+}c^2+ K^{+})$.

 The quantities $e_{s0}^{(+,-)}$ and $b_{s0}^{(+,-)}$ represent the scalar electromagnetic fields that provide the mass ($m_0$) or 
 energy ($E_0$) of the electron or positron.  

   A fundamental point which the present heuristic model challenges is that, in accordance with Eq.(10), we realize that the electron
 is not necessarily an exact point-like particle. Quantum Electrodynamics, based on Special Relativity (SR), deals with the electron as a
 point-like particle. The well-known classical theory of the electron foresees for the electron radius the same order of magnitude
 as the proton radius, i.e., $R_e\sim 10^{-15}m$.

  Some experimental evidences of the scattering of electrons by electrons at very high kinetic energies, indicate that the
 electron can be considered approximately as a point-like particle. Actually, electrons have an extent smaller than collision distance,
 which is about $R_e\sim 10^{-16}$m\cite{14}. Of course, such an extent is negligible in comparison to the dimensions of an atom
 ($\sim 10^{-10}$m) or even the dimensions of a nucleus ($\sim 10^{-14}$m), but the extent is not exactly a point. For this reason, 
 the present model can provide a very small
 non-null volume $v_e$ of the electron. But, if we just consider $v_e=0$ according to Eq.(10), we would have an absurd result, i.e,
 divergent scalar fields ($e_{s0}=b_{s0}\rightarrow\infty$). However, for instance, if we consider $R_e\sim 10^{-16}$m ($v_e\propto
 R_e^3\sim 10^{-48}m^3$) in our model, and knowing that $m_0c^2\cong 0,51$MeV$(\sim 10^{-13}$J), hence, in this case [see Eq.(10)], we should
 obtain $e_{s0}\sim 10^{23}$V/m.  This value is extremely high and therefore we may conclude that the electron is extraordinarily compact,
 having a high mass (energy) density. If we imagine over the ``surface'' of the electron or even inside it, we would detect a
 constant and finite scalar field $e_{s0}\sim 10^{23}$V/m instead of a divergent value given by the classical theory. So, according to the present model that
 attempts to include gravity into electrodynamics with violation of the Lorentz symmetry, the electron should not be exactly a point-like
 particle, having an internal structure, since the quantum scalar field $e_{s0}$ inside the non-classical electron, that is almost 
 point-like with a small radius ($\sim 10^{-16}$m), is finite and constant ($e_{s0}\sim10^{23}$V/m) instead of a function like $1/r^2$ 
 with divergent behavior. Of course, for $r>R_e$, we have the external vectorial (classical) field $\vec E$, decreasing with $1/r^2$ (Fig.1).

 We will see that the advantage of this new model for describing the electron with such internal structure $e_{s0}$ (non-punctual particle)  
 is now the possibility to include gravity by coupling it with the electromagnetic fields of the electron, disconsidering its spin for 
 while. The next section will clarify better this question.

 Here we must stress that, since gravity is an extremely weak force, the new results of the present model will not affect some important
 results of QED as, for instance, the high accuracy result predicted for the giro-magnetic ratio of the electron. In the future, we should 
 verify how the present theory influences quantitatively the predictions of QED. So, in view of the presence of gravity, we believe that
 the modifications on the significant results of QED are minimal; however, the point in which we should call attention is the advantage
 of the present model with respect to its new predictions, which are not taken into account in QED-theory as, for instance, the 
 predictions with respect to the cosmological puzzle about the very low vacuum energy density and the tiny positive value of the
 cosmological constant (Section 10). 

\section{Electron (or positron) coupled to gravity}

For our purpose that aims to introduce a new model of electrodynamics modified by gravity at cosmological length scales (intergalactic medium),
a massive particle (electron) with mass $m_0$ should move in a very weak gravitational potential $\phi$, so that its total energy $E$ 
can be written as  

\begin{equation}
E=mc^2=m_0c^2\sqrt{g_{00}} + K,
\end{equation}
where we have $\sqrt{g_{00}}=\left(1+\frac{\phi}{c^2}\right)$ with $\phi<<c^2$. $K$ is the kinetic energy of the particle. In Section
10, we will justify better such purpose of considering a weak gravity (a quasi flat space-time). 

We can think that $m_0(=m_0^{(+,-)})$ represents the mass of the electron $e^{-}$ (or positron $e^{+}$) emerging from the 
$\gamma$-photon decay in the presence of a very weak gravitational potential $\phi$, i.e., $\gamma\rightarrow e^{-} + e^{+}$ with
$\phi<<c^2$. 

In order to facilitate the understanding of what we are proposing, let us consider $K<<m_0c^2$ ($v<<c$), since we are interested only
in the influence of a weak gravitational potential $\phi$. Therefore, we simply write

\begin{equation}
E=E_0^{(+,-)}\sqrt{g_{00}}=m_0^{(+,-)}c^2\sqrt{g_{00}}.  
\end{equation}

Since we alternatively write $E_0^{(+,-)}=m_0^{(+,-)}c^2=c\epsilon_0e_{s0}^{(+,-)}b_{s0}^{(+,-)}v_e$, we can also write Eq.(12) as
follows:

\begin{equation} 
E =c\epsilon_0 e_s^{(+,-)}b_s^{(+,-)}v_e=
c\epsilon_0 e_{s0}^{(+,-)}b_{s0}^{(+,-)}v_e\sqrt{g_{00}},
\end{equation}
from where, we get

\begin{equation}
e_s^{(+,-)}=e_{s0}^{(+,-)}\sqrt[4]{g_{00}},~ ~
b_s^{(+,-)}=b_{s0}^{(+,-)}\sqrt[4]{g_{00}}.
\end{equation}

From Eq.(13) and Eq.(14), we obtain the increments of the internal quantum fields in the presence of $\phi$ as follow: 

\begin{equation}
\Delta e_s=e_{s}^{(+,-)}-e_{s0}^{(+,-)}=e_{s0}^{(+,-)}(\sqrt[4]{g_{00}} - 1); ~
\Delta b_s=b_{s}^{(+,-)}-b_{s0}^{(+,-)}=b_{s0}^{(+,-)}(\sqrt[4]{g_{00}} - 1),
\end{equation}
where $\Delta e_s=c\Delta b_s$.

  As the energy of the particle (electron or positron) can be obtained from the scalar electromagnetic fields with magnitudes
 $e_{s0}$ and $b_{s0}$ that undergo an increase in the presence of gravity [Eq.(14)], this heuristic model is capable of assisting us to 
imagine that the external fields $\vec{E}$ and $\vec{B}$ created by the moving charge, by storing an energy density 
($\propto |\vec{E}|^2 +|\vec{B}|^2$), should also undergo perturbations (shifts) due to the presence of gravity (Fig.1). However, as
the gravitational and electomagnetic strength differs by a factor of about $10^{−40}$ in a hydrogen atom, we expect that there should be
tiny perturbations on electromagnetic fields $\vec{E}$ and $\vec{B}$ due to the presence of gravity. This assumption does not contradict
General Relativity (GR), since, according to GR, any kind of energy is also a source of gravitational field. So, a very strong 
electromagnetic field is also a source of a very weak gravitational field as shown in a recent paper\cite{15}, since gravity is the weakest
interaction. This recent research is an interesting motivation for our reasoning that states that it would be required a very strong 
gravitational field to generate a
small change in an electromagnetic field by means of a tiny coupling constant. In other words, our reasoning represents an idea of
reciprocity with respect to that assumption investigated recently\cite{15}, i.e., a curved spacetime changes electromagnetic fields
such as electromagnetic fields change spacetime geometry.

  The recent paper quoted in ref.\cite{15} still deserves a special attention, because this research states that there should be a generation
 of artificial gravitational fields with electric currents, which could be in principle detected through the induced change in space-time 
geometry that results in a purely classical deflexion of light by magnetic fields. Such an effect does not invoke any new physics, as it
is a consequence of the equivalence principle of GR. In spite of a very weak effect, it could be detectable. In sum, the amplitude 
of the space-time deformation due to electric currents ($I$) is extremely tiny, of order of a dimensionless magneto-gravitational
constant $C_{I}=8\pi(I/I_{Pl})^2$, where $I_{Pl}=c^2/\sqrt{G\mu_0}=9.8169\times 10^{24}$A is the Planck current. 

 As it was shown in the mentioned paper\cite{15}, where a strong electromagnetic field deforms space-time, our assumption of reciprocity is
 plausible, because there is a non-linearity which is inherent to gravitational fields, leading us to think that the classical (external) 
 fields $\vec E$ and $\vec B$ should undergo tiny shifts like
 $\delta\vec E$ and $\delta\vec B$ in the presence of a gravitational potential $\phi$. As such shifts are too small due to the fact that
gravity is a very weak interaction, and as these shifts have positive magnitudes
 with the same direction of $\vec E$ and $\vec B$, they lead to a slight increase of the electromagnetic energy density around the
 particle. And, since the internal energy of the particle given by the fields $e_{s0}$ and $b_{s0}$ also increases in the presence of 
$\phi$ according to Eq.(13) and Eq.(14), we expect that the magnitudes of the external shifts $\delta\vec E$ and $\delta\vec B$ should 
 be proportional to the increments of the internal (scalar) fields of the particle ($\Delta e_s$ and $\Delta b_s$ in Eq.(15)), namely:

\begin{equation}
\delta E]_{ext.}\propto\Delta e_s[=(e_s-e_{s0})]_{int.},~ ~\delta B]_{ext.}\propto\Delta b_s[=(b_s-b_{s0})]_{int.},
\end{equation}
where $\delta E]_{ext.}$=$\delta E=\delta E(\phi)=(E^{\prime}-E)>0$ and  $\delta B]_{ext.}$=$\delta B=\delta B(\phi)=(B^{\prime}-B)>0$, 
where $\phi$ being a weak gravitational potential (Fig.1). Here we have omitted the signs $(+,-)$ just for the purpose of simplifying the
notation.

 In accordance with Eqs.(16), we may conclude that there is a constant of proportionality that couples the external electromagnetic fields
$\vec E$ and $\vec B$ of the moving particle (electron) with gravity by means of the external small shifts $\delta\vec E$ and
$\delta\vec B$ that are proportional to the internal increments $\Delta e_s$ and $\Delta b_s$. So we write Eqs.(16) as follow:

\begin{equation}
\delta\vec E=\vec\epsilon\xi\Delta e_s,~ ~
\delta\vec B=\vec\epsilon\xi\Delta b_s,
\end{equation}
where $\vec\epsilon$ is the unitary vector given in the same direction of $\vec E$ (or $\vec B$). So the external small shift $\delta\vec E$
(or $\delta\vec B$) has the same direction as $\vec E$ (or $\vec B$) (Fig.1). 

From Eq.(17), it is easy to conclude that the coupling $\xi$ is a dimensionless proportionality constant. We expect that
$\xi<<1$ due to the fact that the gravitational interaction is much weaker than the electromagnetic one, since we are motivated
by the assumption that states that a very strong electromagnetic field is needed to create a very weak gravitational field, i.e., 
{\it``How Current Loops and Solenoids Curve Space-time''}\cite{15}. 

The external shifts $\delta\vec E$ and $\delta\vec B$ depend only on gravitational potential ($g_{00}$) over the electron (Fig.1).

\begin{figure}
\begin{center}
\includegraphics[scale=0.40]{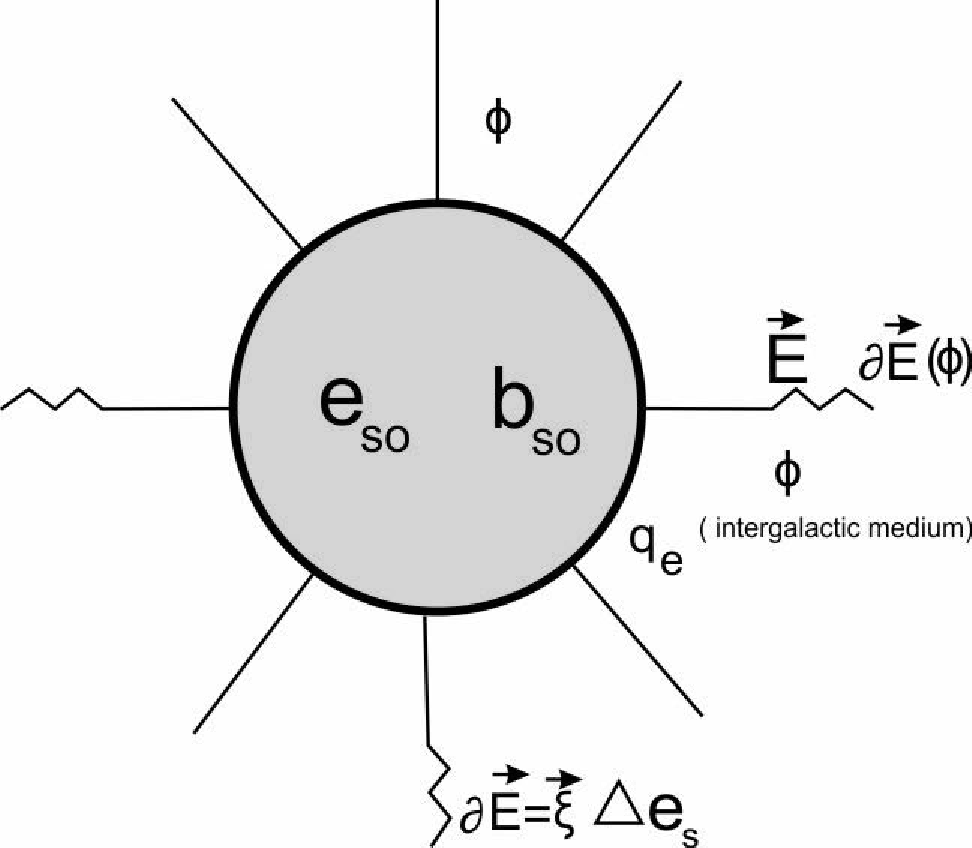}
\end{center}
\caption{This pictorial illustration shows us that the magnitude of the classical (external) field $\vec E$ of the particle undergoes a
small shift like $\delta\vec E(\phi)$ due to the gravitational potential $\phi$. The vector $\delta\vec E(\phi)$ has the same direction 
as $\vec E$ so that the electric energy density around the charge increases in the presence of $\phi$. The magnitude
of the external tiny shift $\delta\vec E(\phi)$ (or $\delta\vec B(\phi)$) is proportional to the increment $\Delta e_s$ (or $\Delta b_s$)
of the internal (scalar) field, i.e., $\delta E=\xi\Delta e_s$ (or $\delta B=\xi\Delta b_s$), $\xi$ being a tiny dimensionless
coupling constant of gravito-electric origin. The fields $e_{s0}$ and $b_{s0}$ are internal (scalar) quantum fields of the electron,
i.e., they represent the magnitudes of internal fields, where $e_{s0}b_{s0}\propto m_0$ (electron mass).}
\end{figure}

 Inserting Eq.(15) into Eq.(17), we obtain

\begin{equation}
\delta\vec E=\delta\vec E(\phi)=\vec\epsilon\xi e_{s0}(\sqrt[4]{g_{00}}-1), ~ ~
\delta\vec B=\delta\vec B(\phi)=\vec\epsilon\xi b_{s0}(\sqrt[4]{g_{00}}-1).
\end{equation}

 Due to the tiny positive shifts with magnitudes $\delta E(\phi)$ and $\delta B(\phi)$, the total electromagnetic energy density in 
the space around the charged particle (electron or positron) is slightly increased, as follows:

\begin{equation}
\rho_{em}^{total}=\frac{1}{2}\epsilon_0[E+\delta E(\phi)]^2+\frac{1}{2\mu_0}
[B +\delta B(\phi)]^2.
\end{equation}  

Inserting the magnitudes $\delta E$ and $\delta B$ from Eqs.(18) into Eq.(19) and performing the calculations, we finally obtain the
total electromagnetic energy density in view of the very weak coupling $\xi$, namely:

\begin{equation}
\rho^{total}_{em}=\frac{1}{2}\left[\epsilon_0 E^2+\frac{1}{\mu_0}B^2\right]+
\xi\left[\epsilon_0 Ee_{s0}+\frac{1}{\mu_0}Bb_{s0}\right](\sqrt[4]{g_{00}}-1)+\frac{1}{2}\xi^2\left[\epsilon_0(e_{s0})^2
+\frac{1}{\mu_0}(b_{s0})^2\right](\sqrt[4]{g_{00}}-1)^2.
\end{equation}

   We may assume that $\rho_{em}^{total}=\rho_{em}^{(0)}+\rho_{em}^{(1)}+\rho_{em}^{(2)}$ to represent Eq.(20), where 
$\rho_{em}^{(0)}$ is the free electromagnetic energy density (of zero order) for the ideal case of a charged particle uncoupled to 
gravity ($\xi=0$), i.e, the ideal case of a free charge. We have $\rho^{(0)}_{em}\propto 1/r^4$ (coulombian term).

  The coupling term $\rho^{(1)}_{em}$ represents an electromagnetic energy density of first order, since it contains a dependence
 on $\delta E$ and $\delta B$, i.e., it is proportional to $\delta E$ and $\delta B$ due to the influence of gravity.
 Therefore, it is a mixed term that behaves essentially like a radiation term for representing a density of radiant energy. So we find
 $\rho^{(1)}_{em}\propto 1/r^2$ as we have
 $E~(B)\propto 1/{r^2}$ and $e_{s0}~(b_{s0})\propto 1/r^0\sim constant$. It is interesting to notice that this radiation term has its origin
 in the non-inertial aspect of gravity that couples weakly with the electromagnetic field of the moving charge, since, according to the
 classical electromagnetic theory, it would already be expected that such a coupling term, acting like a radiation field 
($\propto 1/r$), comes from an accelerated charge. In fact, the term $\rho^{(1)}_{em}$ reflects a quite weak non-inertial effect of the
 charged particle, since it is weakly coupled to gravity. 

  The last coupling term ($\rho^{(2)}_{em}$) is purely interactive due to the presence of gravity only, namely it is a 2nd order interactive
 electromagnetic energy density term, since it is proportional to $(\delta E)^2$ and $(\delta B)^2$. And, as $e_{s0}~(b_{s0})\propto
 1/r^0\sim constant$, we find $\rho^{(2)}_{em}\propto 1/r^0\sim constant $, where we can also write $\rho^{(2)}_{em}
 =\frac{1}{2}\epsilon_0(\delta E)^2+\frac{1}{2\mu_0}(\delta B)^2$, which depends only on the weak gravitational potential ($\phi$)
 [see Eqs.(18)].

  As we have $\rho^{(2)}_{em}\propto 1/r^0$, this term has non-locality behavior. Non-locality behavior means that
 $\rho^{(2)}_{em}$ behaves like a kind of non-local field inherent to space (a background field). This term $\rho^{(2)}_{em}$ is
 purely of gravitational origin. It does not depend on the distance $r$ from the charged particle. Therefore, $\rho^{(2)}_{em}$ is a
 uniform energy density for a given weak gravitational potential $\phi$ on the particle.

   In reality, we generally have $\rho^{(0)}_{em}>>\rho^{(1)}_{em}>>\rho^{(2)}_{em}$. For a very weak gravitational potential, we can make a good practical
 approximation as $\rho_{em}^{total}\approx\rho^{(0)}_{em}$; however, from a fundamental viewpoint, we cannot neglect the coupling terms
 $\rho^{(1)}_{em}$ and $\rho^{(2)}_{em}$, especially the latter ($\rho^{(2)}_{em}$) for large distances, since
 $\rho^{(2)}_{em}(\propto 1/r^0\sim constant$) has a vital importance in this work, allowing us to understand the uniform energy
 density of the background field. 

  The term $\rho^{(2)}_{em}$ does not have $r$-dependence, since $e_{s0}$ or $b_{s0}\sim constant$, and thus it still remains constant
  even when $r\rightarrow\infty$. That is the reason why it represents a uniform density of background energy (vacuum energy). 

  The term $\rho^{(2)}_{em}$ has deep implications in our understanding of spacetime for large scales of length (Section 10). 

   In the next section, based on the Bohr model of hydrogen atom, from where we get the well-known fine structure constant $\alpha$, 
 we will estimate the minuscule value of the coupling constant $\xi$, whose value depends on six fundamental constants like $G$, 
 $\hbar$, $c$; $q_e$, $m_e$ (electron mass) and $m_p$ (proton mass), where the first three fundamental constants ($G$, $\hbar$ and $c$)
 allow us to build the three fundamental quantities like length (Planck length $l_{Pl}=\sqrt{G\hbar/c^3}$), 
 time (Planck time $t_{Pl}=\sqrt{G\hbar/c^5}$) and mass (Planck mass $m_{Pl}=\sqrt{\hbar c/G}$). 

 \section{The coupling $\xi$ and the space-time with an invariant minimum speed $V$}

  Let us begin this section by considering the well-known problem that deals with the electron in the bound state of a coulombian
 potential of a proton (hydrogen atom), i.e., the problem of the fine structure constant $\alpha$ (Bohr model). We start from this issue 
 because it poses an important analogy with the present model of the electron charge coupled to a weak gravitational field (e.g: of a proton) 
 by means of the tiny coupling constant $\xi$ to be estimated. 

   We know that the {\it fine structure constant} ($\alpha\cong 1/137$) plays an important role for obtaining the energy levels that
bind the electron to the nucleus (proton) in the hydrogen atom. The reason for considering the hydrogen atom instead of any other bound 
state of two particles is that the hydrogen atom is the most elementary and stable bound state that exists, since, as for instance, the
positronion ``atom'' ($e^{-}|e^{+}$) could not be used to estimate $\xi$ because it is not stable, although it is more elementary than hydrogen.
Thus, hydrogen satisfies such two conditions at the same time, i.e., it is the most elementary and stable bound state of two particles,
and therefore it should be used to estimate the fine coupling constant $\xi$. In other words, we say that a universe without baryons
(protons) would not have stable atoms like hydrogen atoms and thus it would colapse, so that spacetime would colapse too. This is the 
reason why we cannot estimate $\xi$ by using only the mass of electron (or positron), since we will realize that the strong force
(quarks), that is responsible for the proton mass, leads to stability in the atom, and thus it is a fundamental condition for obtaining the coupling $\xi$. This question will deserve a 
deeper investigation by considering the so-called Dirac's Large Number Hypothesis (LNH)\cite{16}\cite{17}\cite{18}\cite{19}
\cite{20}\cite{21} that states that the ratio between the electrical and gravitational forces in the hydrogen atom can be written as $F_e/F_g=e^2/Gm_pm_e\sim m_pc^2/H_0h\sim 10^{40}$,
$10^{40}$ being the Dirac's number and $H_0$ being the Hubble's constant.

  Is interesting to note that Dirac's LNH makes a connection between micro (quantum) and macro (cosmological) scales by means
 of the own mass of the proton $m_p$ and the constant $H_0$ that appears in the term $m_pc^2/H_0h$, which is the Dirac's scale factor
 $R_u/\lambda_{cp}$, $R_u(=c/H_0)$ being the radius of the universe and $\lambda_{cp}(=h/m_pc)$ being the Compton's length for the proton. 
 This indicates that the proton mass $m_p$ in the cosmological scenario is used for obtaining the ratio $F_e/F_g$, which leads us
 to think that specially the proton mass $m_p$ and not only the electron mass $m_e$ should play a fundamental role for determining the tiny coupling
 constant $\xi$. A deeper physical reason for having only the proton mass $m_p$ in the Dirac's scale factor instead of the electron
 mass $m_e$ will be investigated. So we will realize that Dirac's LNH will be essential for understanding the origin of the 
 gravito-electromagnetic coupling constant $\xi$ and its cosmological implications. 

  Since Dirac's scale factor in his LNH allows us to connect directly the charge $e(=q_e/\sqrt{4\pi\epsilon_0})$ with the masses
 $m_p$ and $m_e$ in the hydrogen atom, i.e., $e^2\sim Gm_p^2m_ec^2/H_0h$, then we will be motivated to follow
two steps for achieving the tiny gravito-electromagnetic coupling constant $\xi$, namely the real hydrogen atom
(coulombian interaction ``$e^2$'') and a hypothetical hydrogen atom, where only the gravitational interaction between proton and 
electron is taken into account (gravitational interaction ``$m_pm_e$''). 

  We notice that the electron spin is not considered here. Thus, in this model, the external magnetic field of the electron has its 
origin only in its translational or orbital motion. 

\subsection{The hydrogen atom}

Let's initially consider the energy that binds an electron to a proton in the fundamental state of the hydrogen atom, as follows:

\begin{equation}
\Delta E=\frac{1}{2}\alpha^2 m_ec^2,
\end{equation}
where $\Delta E$ is assumed as module. We have $\Delta E<<m_ec^2$, where $m_e$ is the electron mass. 

 We have $\alpha=e^2/\hbar c=q_e^2/4\pi\epsilon_0\hbar c\approx1/137$ (fine structure constant). 
Since $m_ec^2\cong 0,51$MeV, from Eq.(21) we get $\Delta E\approx 13,6$eV.

 As we already know that $E_0=m_0c^2=m_ec^2=c\epsilon_0e_{s0}b_{s0}v_e$, we may write Eq.(21) in the following alternative way:

\begin{equation}
\Delta E=\frac{1}{2}\alpha^2c\epsilon_0e_{s0}b_{s0}v_e\equiv\frac{1}{2}c\epsilon_0(\Delta e_{s})(\Delta b_{s})v_e,
\end{equation}
from where we extract

\begin{equation}
\Delta e_{s}\equiv\alpha e_{s0}=\frac{e^2}{\hbar c}e_{s0};~~
\Delta b_{s}\equiv\alpha b_{s0}=\frac{e^2}{\hbar c}b_{s0}.
\end{equation}

 It is interesting to observe that Eqs.(23) maintains a certain analogy with Eq.(17); however, first of all we must emphasize that
 the variations (increments) $\Delta e_s$ and $\Delta b_s$ of the electron internal fields, given in Eq.(23), are of purely coulombian
 origin, since the fine structure constant $\alpha$ depends solely on the electron charge. Thus, we could express the electric force 
 between two electrons in the following way:

\begin{equation}
F_e=\frac{e^2}{r^2}=\frac{q_e^2}{4\pi\epsilon_0 r^2}=\frac{\alpha\hbar c}{r^2}.
\end{equation}

According to Bohr's model of the hydrogen atom, we can write

\begin{equation}
 \alpha=\frac{v_B}{c}=\frac{e^2}{\hbar c}\cong\frac{1}{137}, 
\end{equation}
where $v_B=e^2/\hbar(\cong c/137)$ is the well-known Bohr velocity obtained in the fundamental bound state of the hydrogen atom. 
Thus, we can alternatively write Eq.(24), namely $F_e=\hbar v_B/r^2$. 

\subsection{The gravitational ``hydrogen atom''}

 If we now consider only the gravitational interaction between a proton and an electron in the hydrogen atom, thus, in a similar way 
to Eq.(24), we write

\begin{equation}
F_g=\frac{Gm_pm_e}{r^2}=\frac{\alpha_g\hbar c}{r^2}=\frac{\hbar v_g}{r^2},
 \end{equation}
from where we get

\begin{equation}
\alpha_g=\frac{v_g}{c}=\frac{Gm_pm_e}{\hbar c}, 
 \end{equation}
where $v_g(=\alpha_g c=Gm_pm_e/\hbar)$ is a very low speed obtained in the fundamental bound state of the hypothetical gravitational 
``hyhrogen atom''. 

  We find $\alpha_g<<\alpha$ and $v_g<<v_B$ due to the fact that the gravitational interaction is much weaker than the electric 
 interaction, i.e., $F_g/F_e=\alpha_g/\alpha=v_g/v_B=Gm_pm_e/e^2\sim 10^{-40}$, where $\alpha_g\sim 10^{-43}$. Therefore, we shall 
 denominate $\alpha_g$ as the {\it superfine structure constant} of the gravitational ``hydrogen atom'', since the gravitational 
 interaction creates a bonding energy extremely smaller than the coulombian interaction given to the fundamental state ($\Delta E$)
 in the hydrogen atom. 

 To sum up, we say that, whereas the fine structure constant $\alpha(e^2)$ provides the adjustment for the coulombian bonding energies between proton and electron
 in the hydrogen atom, the tiny constant $\alpha_g(Gm_pm_e)$ gives the adjustment for the gravitational bonding energies between 
 the masses of the proton and electron in the gravitational ``hydrogen atom''. Both bonding energies of electric ($\alpha$) and
 gravitational ($\alpha_g$) origin lead to an increment of energy in the particle by means of variations of the internal
 fields,i.e., $\Delta e_s$ and $\Delta b_s$.

 \subsection{Dirac's LNH and the coupling constant $\xi$}

  Dirac's LNH can be written in the following way: 

\begin{equation}
\frac{\alpha}{\alpha_g}=\frac{v_B}{v_g}\sim\frac{m_pc^2}{H_0h}\sim 10^{40}, 
\end{equation}
from where we find 

\begin{equation}
 \alpha\sim\left(\frac{m_pc^2}{H_0h}\right)\alpha_g=\left(\frac{m_pc^2}{H_0h}\right)\left(\frac{Gm_pm_e}{\hbar c}\right) 
\end{equation}
 
and 

\begin{equation}
\alpha_g\sim\left(\frac{H_0h}{m_pc^2}\right)\alpha=\left(\frac{H_0h}{m_pc^2}\right)\left(\frac{e^2}{\hbar c}\right),  
\end{equation}
where $\alpha=e^2/\hbar c$ and $\alpha_g=Gm_pm_e/\hbar c$. 

 Here it is important to mention that Dirac's LNH predicted the idea of variation of the gravitational constant $G$\cite{16}
\cite{17}\cite{20}\cite{21}, such that $G$ have descended from a very high initial value (early universe) to its present small value. 

 Although so many years have passed after the inception of LNH, it has not lost its significance. Instead, perhaps it has gained a new
momentum after the discovery of accelerating universe. The present cosmological picture emerging out of SN Ia observations\cite{22}
\cite{23} reveals that the present universe is accelerating. Some kind of exotic repulsive force in the form of vacuum energy is supposed to be 
responsible for this acceleration which started about $7$ Gyr.ago. This repelling force is termed as dark energy and is designated
by $\Lambda$ that will be investigated in Section 10 as an effect of the gravito-electromagnetic coupling $\xi$ at cosmological scales. 

 On the way of investigating dark energy, many variants of $\Lambda$ have been proposed including a constant $\Lambda$. But, compatibility
 with other areas of physics demands that $\Lambda$ should be slowly time decreasing\cite{24}. Moreover, the currently observed small 
value of $\Lambda$\cite{25}\cite{26} suggests that it has decreased slowly from a very high value to its present nearly zero value. This
type of time dependency of $\Lambda$ has a similarity with that of the gravitational constant $G$ as proposed by Dirac in 
his LNH\cite{24}. This is the reason why LNH should be used for searching for the value of the coupling constant $\xi$, whose cosmological
effect is a cosmological ``constant'' (parameter) $\Lambda$ depending on the radius of the expanding universe, as we will verify
in Section 10. 

 In fact, a combined framework of LNH and the cosmological parameter is used to address a number of important issues such as an explanation 
of flat galactic rotation curves\cite{27}, unification of LNH with GR\cite{28}, etc. Even the Higgs scalar-tensor theory, in which the
 mass of the particles appeared through gravitational interaction\cite{29}\cite{30}\cite{31}, is also found to be compatible
 with LNH\cite{32}. Also, the problem of unifying all other forces with
gravity demands an understanding of the coincidence of large numbers and hence LNH\cite{33}, which becomes specially a strong motivation 
for  unifying electromagnetism with gravity in a cosmological scenario, as is proposed in the present work by obtaining the coupling constant
$\xi$, which is responsible for the background field in Eq.(20), leading to cosmological implications (Section 10).

 Here it is important to stress that Dirac\cite{24} noted clearly that the variability of $G(\sim t^{-1})$ does not agree with general 
relativity (GR), but the physical significance of the agreement between the large numbers had demonstrated to be more significant than
the the logical harmony of GR. This motivates us to think that a deep connection between gravity and EM could lead to a new 
cosmological implication such as
the dark energy (cosmological constant); however GR does not provide a satisfactory explanation for this question. Furthermore, an
extension of LNH explored by Zeldovich\cite{34} will be presented here in order to show a connection of the cosmological parameter with the
radius of the universe $R_u$ and also the proton mass $m_p$. This will strengthen the argument in favor of obtaining the coupling $\xi$
as a function of the masses $m_e$ and $m_p$, allowing us to connect particle physics with space-time, unification of fields and vacuum
energy (cosmology) in the spirit of Zeldovich.       

 Thanks to Dirac's LNH, it is possible to write the electrical coupling constant (fine structure constant $\alpha$) in function of 
the masses $m_p$ and $m_e$, according to Eq.(29), i.e., $\alpha(e^2)=F[m_p,H_0,\alpha_g(Gm_pm_e)]$. The inverse transformation, where
the tiny gravitational coupling ($\alpha_g$) is written as function of the charge $e$, is also obtained in Eq.(30), i.e., 
$\alpha_g=F^{\prime}[m_p,H_0,\alpha(e^2)]$. Such crossover transformations between the masses $m_p$, $m_e$ and the charge $e$ by means
of Dirac's scale factor and its inverse allow us to understand the charge and mass as two aspects of the same reality where the
cosmological scenario takes place, so that the presence of a background field in the space-time represents such scenario. Thus,
due to such crossover transformations, it is natural to conclude that when we multiply Eq.(29) by Eq.(30), a gravito-electromagnetic
coupling squared is found, i.e.: 

 \begin{equation}
\alpha^{2}_{gel}=\alpha_g\alpha=F^{\prime}(e^2)F(Gm_pm_e)=\frac{Gm_pm_ee^2}{\hbar^2 c^2}, 
\end{equation}
where $e^2=q_e^2/4\pi\epsilon_0$, such that, from Eq.(31), we get 

\begin{equation}
\xi=\alpha_{gel}=\sqrt{\alpha_g\alpha}=\sqrt{\frac{Gm_pm_e}{4\pi\epsilon_0}}\frac{q_e}{\hbar c} 
\end{equation}

From Eq.(32), we obtain $\xi\cong 1.5302\times 10^{-22}$. Let us denominate $\xi$ as the {\it fine adjustment constant}, 
which represents the tiny gravito-electromagnetic coupling constant. Such coupling appears due to a gravitational potential $\phi$ that couples
to an electric field $\vec E$ (electric energy density) created by the charge $q_e$, according to Eq.(20). However, it must be stressed that
we have considered a very weak gravitational potential because the charged particle is placed in the intergalactic medium (cosmological 
vacuum), where the density of matter (energy) is roughly equal to the dark energy density. We have made such appproximation since it is
valid for our purpose of estimating the tiny positive value of the cosmological constant and the low vacuum energy density.  
  
The quantity $\sqrt{Gm_pm_e}$ can be termed as an effective {\it gravitational charge $e_g$} of the electron in the presence of the proton, 
so that we simply define $e_g=\sqrt{Gm_pm_e}$. Thus we alternatively write Eq.(32), namely $\xi=e_ge/\hbar c$. 

Finally, from Eq.(32), we rewrite Eq.(20) as follows:

\begin{equation}
\rho=\frac{1}{2}\left[\epsilon_0 E^2+\frac{1}{\mu_0}B^2\right]+
\frac{e_ge}{\hbar c}\left[\epsilon_0 Ee_{s0}+\frac{1}{\mu_0}Bb_{s0}\right](\sqrt[4]{g_{00}}-1)+
\frac{1}{2}\frac{e_g^2e^2}{\hbar^2 c^2}\left[\epsilon_0 (e_{s0})^2+\frac{1}{\mu_0}(b_{s0})^2\right](\sqrt[4]{g_{00}}-1)^2.
\end{equation}

\subsection{An extension of Dirac's LNH and the coupling constant $\xi$}

Zeldovich\cite{34} extended Dirac's LNH by including the cosmological parameter $\Lambda$. The question is to know how does the situation
change with the coincidences of the large numbers in a theory with a cosmological constant, i.e., in the $\Lambda$ model of the
universe. To do that, let us write down the analogous relation for Dirac's LNH, replacing the radius $R_u(=c/H_0)$ by the quantity 
$\Lambda^{-1/2}$, which has the same dimensionality, so that $R_u\sim\Lambda^{-1/2}$\cite{34}. This yields the ratio

\begin{equation}
 \frac{\Lambda^{-1/2}}{\hbar/m_pc}\sim\frac{m_pcR_u}{\hbar}=\frac{m_pc^2}{H_0\hbar}\cong\frac{\hbar c}{Gm_p^2}, 
\end{equation}
from where we get

\begin{equation}
 \Lambda\cong\frac{G^2m_p^6}{\hbar^4}
\end{equation}

The vacuum energy density is $\rho_{\Lambda}=\Lambda c^2/8\pi G$. Thus, from Eq.(35), we find 

\begin{equation}
 \rho_{\Lambda}\sim\frac{Gm_p^6 c^2}{\hbar^4}=\frac{Gm_p^2}{\hbar c}~m_p\left(\frac{m_pc}{\hbar}\right)^3=\frac{1}{c^2}
\left(\frac{Gm_p^2}{\lambda_p}\right)\left(\frac{1}{\lambda_p}\right)^3,
\end{equation}
where $\lambda_p(=\hbar/m_pc)$ is the reduced Compton's length for the proton and $\lambda_p^3$ represents a volume. 

Eq.(36) was intuitively interpreted by Zeldovich as follows: virtual particles with mass $m_p$, the distance between which is $\lambda_p$,
are produced in the vacuum; their self-energy is identically equal to zero, but the energy of gravitational interaction of two 
particles per volume $\lambda_p^3$ causes the vacuum energy density. This reasoning extends Dirac's LNH by considering that the proton mass 
has a virtual representation in the vacuum, creating a gravitational energy of the vacuum connected to the cosmological constant.
Therefore, we realize that this extension of Dirac's LNH introduced by Zeldovich help us to understand better how the electron charge 
(electromagnetism) is coupled to gravity in a cosmological scenario (Eq.(33)), since the proton mass plays an important role in 
obtaining a background energy (vacuum energy) of gravitational origin, leading us to conclude that a connection between such fields should
require a coupling $\xi$ depending on the proton mass. Such conclusion can also be more sophisticated if we take into account the fact 
that the quantity $\rho_{\Lambda}(\sim 10^{-22}g/cm^3)$ (Eq.(36)) is still $10^{7}$ times larger than the observable value of 
$\rho_{\Lambda}(\sim 10^{-29}g/cm^3)$. Numerical agreement could be obtained by replacing for instance $m_p^6$ with $m_p^4m_e^2$\cite{35}
in Eq.(36), so that we get $\rho_{\Lambda}\sim Gm_p^4m_e^2c^2/\hbar^4(\sim 10^{-29}g/cm^3)$. Thus, now if we write this correct relation for
$\rho_{\Lambda}$ in an analagous form of Eq.(36), we find 

\begin{equation}
 \rho_{\Lambda}\sim\frac{Gm_p^4m_e^2c^2}{\hbar^4}=\frac{Gm_pm_e}{\hbar c}~m_p\left(\frac{m_pc}{\hbar}\right)^2
\left(\frac{m_ec}{\hbar}\right)=\frac{1}{c^2}\left(\frac{Gm_pm_e}{\lambda_p}\right)\left(\frac{1}{\lambda_p}\right)^2
\left(\frac{1}{\lambda_e}\right),
\end{equation}
where $\lambda_p(=\hbar/m_pc)$ and $\lambda_e(=\hbar/m_ec)$ are the reduced Compton's lengths for the proton and electron respectively, 
$\lambda_p^2\lambda_e$ being a volume. 

Eq.(37) is an analogous relation for Eq.(36) and thus it should be interpreted as follows: pairs of virtual particles with masses 
$m_p$ and $m_e$ separeted by the distance $\lambda_p$ are produced in the vacuum, such that the energy of gravitational interaction 
for each pair ``$m_p|m_e$'' per volume $\lambda_p^2\lambda_e$ causes the permissible vacuum energy density $\rho_{\Lambda}$, which is now in agreement with
the observational results $(\sim 10^{-29}g/cm^3)$. This means that the interpretation given to Eq.(37) is more accurate to mimic the 
vacuum, leading us to conclude that both virtual particles with masses $m_p$ and $m_e$ should contribute for the observable vacuum energy 
density. In other words, this means that the hydrogen atom is the most fundamental system of two particles in the sense that its particles 
with masses $m_p$ and $m_e$ have a virtual representation for mimicing the minimal energy density of the vacuum with accuracy, leading to 
the observable cosmological constant. Thus, for this reason, now it becomes still more clear to realize that the coupling $\xi$ depends on 
$m_p$ and $m_e$, since $\xi$ has cosmological implications in obtaining the permissible vacuum energy density and the 
cosmological constant in agreement with the observational data. 

The improvement of the relations of Zeldovich to be in agreement with the recent observational results of $\rho_{\Lambda}$
and $\Lambda$ is necessary for searching for a deeper comprehension of $\xi$. We will show that the new relation given in Eq.(37) 
is consistent with the coupling constant $\xi(m_pm_e)$ in the sense that $\xi$ is related to Dirac's scale factor. 
To do that, we first replace $m_p^6$ by $m_p^4m_e^2$ in Eq.(35) or, from Eq.(37), by making $\Lambda\sim G\rho_{\Lambda}/c^2$, we find 

\begin{equation}
 \frac{\Lambda^{1/2}}{m_p}\sim\frac{Gm_pm_e}{\hbar^2}
\end{equation}

By substituting Eq.(38) into Eq.(31), we write Eq.(31) as follows: 

\begin{equation}
\xi^2=\alpha^{2}_{gel}=\left(\frac{Gm_pm_e}{\hbar^2}\right)\left(\frac{e^2}{c^2}\right)\sim\frac{\Lambda^{1/2}e^2}{m_pc^2}
\end{equation}

According to Zeldovich\cite{34}, we already know that $R_u\sim\Lambda^{-1/2}$. Thus, we can write Eq.(39) as follows:

\begin{equation}
\xi^2\sim\frac{e^2}{m_pc^2R_u}
\end{equation}

As we can make $e^2=\alpha\hbar c$ and $R_u=(c/H_0)\sim\Lambda^{-1/2}$ in Eq.(40), we find

\begin{equation}
\xi\sim\sqrt{\alpha\left(\frac{\hbar H_0}{m_pc^2}\right)}=\sqrt{\alpha\left(\frac{\lambda_p}{R_u}\right)}
\sim\frac{1}{c}\sqrt{\frac{e^2}{m_p}\Lambda^{1/2}}, 
\end{equation}
where $\lambda_p=\hbar/m_pc$ and $R_u=c/H_0$. 

As we have $\Lambda^{1/2}\sim (G\rho_{\Lambda})^{1/2}/c$, we can alternatively write Eq.(41), namely:
 $\xi\sim (1/c)\sqrt{(e^2/m_pc)(G\rho_{\Lambda})^{1/2}}$. 

Eq.(41) shows that the coupling $\xi$ is related to the inverse Dirac's scale factor $(m_pc^2/\hbar H_0)^{-1}$, i.e., actually we have
 $\xi\sim\sqrt{\alpha(\hbar H_0/m_pc^2)}\approx 1/\sqrt{137}\times\sqrt{\hbar H_0/m_pc^2}\approx 10^{-22}$.

Eq.(41) shows that the coupling $\xi$ is directly related to the vacuum energy density $\rho_{\Lambda}$ or $\Lambda$. In fact, this result is
consistent with Eq.(20) in the sense that, if we consider a null vacuum energy density ($\rho_{\Lambda}=0$), we would find $\xi=0$ in
Eq.(41), such that the last term of Eq.(20) would also vanish ($\rho_{em}^{(2)}=0$), since $\rho_{em}^{(2)}$ represents the own background
(vacuum) energy density due to the coupling $\xi$ of the gravitational and electromagnetic fields. Thus Eq.(41) is consistent with the 
fact that $\rho_{em}^{(2)}$ will lead to $\Lambda(\propto R_u^{-2})$ in the space-time of SSR (Section 10). 

Eq.(40) and Eq.(41) indicate that the coupling $\xi$ could vary with the cosmological time according to Dirac's LNH, since LNH and
its extension for $\Lambda$ model lead us to conclude that $G$ and $\Lambda$ should decrease with time and thus $\xi$ could 
decrease with time, however such possible variation of $\xi$ would also depend on how other fundamental constants such as $e$, 
$\hbar$, $c$, $m_e$ and $m_p$ should vary with the cosmological time\cite{17}. This question will be deeper investigated elsewhere. 

\subsection{The universal minimum speed $V$} 

 In the hydrogen atom we get the fine structure constant $\alpha=e^2/\hbar c=v_B/c$, where $v_B=e^2/\hbar\cong c/137$, which is the
 speed of the electron at the fundamental atomic level (Bohr velocity). At this level, the electron does not radiate because it is in a
 kind of balanced state in spite of its electrostatic interaction with the nucleus (centripete force), that is to say, it acts as if it
 were effectively an inertial system. With analogous reasoning applied to the fundamental case of the electron charge coupled to a 
 gravitational field, we have obtained the gravito-electromagnetic coupling constant $\xi$ in Eq.(32). So we may write Eq.(32) as follows:

\begin{equation}
\xi=\frac{V}{c}=\sqrt{\frac{Gm_pm_e}{4\pi\epsilon_0}}\frac{q_e}{\hbar c},
\end{equation}
from where we find the very low speed $V$, namely: 

\begin{equation}
V=\frac{e_ge}{\hbar}=\sqrt{\frac{Gm_pm_e}{4\pi\epsilon_0}}\frac{q_e}{\hbar},
\end{equation}
where $V\cong 4.5876\times 10^{-14}$m/s. This seems to be a fundamental constant of nature with the same status of the invariance of
the speed of light ($c=e^2/\alpha\hbar$). Therefore, we will realize that $V$ plays the role of an invariant minimum speed 
in the space-time, so that a modified SR (SSR) arises (next Sections). 

  Similarly to the Bohr velocity $v_B(=\alpha c=e^2/\hbar)$, the speed $V(=\xi c=e_ge/\hbar)$ is also a universal constant; however the
 crucial difference between $v_B$ and $V$ is that the minimum speed $V$ is related to the most fundamental bound state in the
 universe represented by the vacuum energy density $\rho_{\Lambda}$ that appears in Eq.(41), from where we get 
 $V=\xi c\sim\sqrt{(e^2/m_pc)(G\rho_{\Lambda})^{1/2}}\approx\sqrt{(e^2/m_p)\Lambda^{1/2}}$, leading to the 
 cosmological anti-gravity (Section 10).

 It is interesting to verify that the minimum speed $V$ is directly related to the minimum length (Planck length), namely 
 $V=\frac{\sqrt{Gm_pm_e} e}{\hbar}=\left(e\sqrt{\frac{m_pm_ec^3}{\hbar^3}}\right)l_{Pl}=\frac{l_{Pl}}{\tau}\sim 10^{-14}$m/s, where
 $l_{Pl}=\sqrt{\frac{G\hbar}{c^3}}\sim{10^{-35}}$m and $\tau=\left(e\sqrt{\frac{m_pm_ec^3}{\hbar^3}}\right)^{-1}\sim 10^{-21}$s, such that,
 if we make $l_{Pl}\rightarrow 0$, this implies $V\rightarrow 0$ and thus we recover the classical space-time of SR. This leads us to realize
 that there should be a deep connection between the minimum speed and new quantum gravity effects for much lower energies, since
 we have $V\propto\sqrt{G}$ in Eq.(43). 

 In summary, we have shown that there is a tiny dimensionless coupling constant ($\xi=V/c\sim 10^{-22}$) that couples gravity with
electromagnetic fields, where $V(\sim 10^{-14}$m/s) is the universal minimum speed related to a preferred background frame (Section 5), 
which is associated to a uniform background field with low vacuum energy density connected to the cosmological 
``constant'' $\Lambda$\cite{11}\cite{13}. This connection between $\Lambda$ and $V$ can already be obtained directly from Eq.(41) 
and Eq.(42) ($V=c\xi$), such that we find $\Lambda\sim(m_p/e^2)^{2}V^{4}$. 

 In Section 10, we will obtain $\Lambda$ within the formalism of SSR-theory that emerges from such unification of 
 fields in a quasi-flat space-time (cosmological scales), so that it contains the invariant minimum speed $V$ and the cosmological ``constant'' $\Lambda$ 
 governed by physics of particles like the proton (baryon) and electron (lepton). 

 General theory of unification of fields (a SGR-theory) given in curved space-time with the presence of the coupling $\xi$ will be 
 deeply investigated elsewhere, since our purpose here is to explore only relevant cosmological implications of this special theory of
 unification given in a flat space-time filled by a very low energy density of gravitational origin (SSR-theory). 

Here it is important to mention that the minimum speed $V$ given in the scenario of a flat space-time filled by a very low energy density
(SSR) has led to a fundamental understanding of the uncertainty principle\cite{12} of QM within a context of quantum-gravity for very 
low energies. 

\subsection{Flat space-time with a uniform low energy density of background field and minimum speed $V$}

 As the last term in Eq.(33) ($\rho^{(2)}_{em}$) does not depend on the distance from the particle, $\rho^{(2)}_{em}$ represents a
 uniform background field energy density. Let us rewrite this term, as follows:  

 \begin{equation}
 \rho^{(2)}_{em}=\frac{V^2}{2c^2}\left[\epsilon_0(e_{s0})^2+\frac{1}{\mu_0}(b_{s0})^2\right](\sqrt[4]{g_{00}}-1)^2,
 \end{equation}
 where $V/c=\xi(\sim 10^{-22})$ according to Eq.(42), $V(\sim 10^{-14}$m/s) being the minimum speed. Thus, we conclude that the 
background energy density is extremely low, since, from Eq.(44), we get $\xi^2=V^2/c^2\sim 10^{-44}$.

In a perfectly flat space-time of Special Relativity (SR), we would have $g_{00}=1$ ($\phi=0$) and, thus, we would get $\rho^{(2)}_{em}=0$; 
however, in physical reality, the existence of a background field, even if it is too weak or almost negligible, should be taken into 
account because there is no exact zero gravity anywhere.

The minuscule gravitational effects that arise by means of a background field related to a vacuum energy density $\rho^{(2)}_{em}$,
in a quasi-flat space-time, lead to a very low zero-point energy at cosmological scales, since we should remember that our universe has
a quasi-flat space-time with a very low vacuum energy density\cite{11}\cite{13}\cite{36}\cite{37}. Such a zero-point energy
of gravitational origin, i.e., a vacuum energy that fills the whole universe, leads to the existence of the universal minimum speed $V$. 
This effect is consistent with the fact that the minimum speed $V$ has direct dependence on the gravitational constant in accordance 
with Eq.(43), where $V\propto\sqrt{G}$.

The background field energy density $\rho^{(2)}_{em}$ brings a new aspect of gravitation that needs to be thoroughly investigated. This
new feature brings with it a change in our notion of space-time and, thus, also the concept of inertial frames (galilean frameworks)
with regard to the motion of subatomic particles\cite{11}\cite{12}\cite{13}. This subject will be investigated in the next section. 

Our attention here must be given to the existence of the minimum speed $V$ of gravitational origin, which produces a symmetrization 
of space-time together with the speed of light $c$ of electromagnetic origin, in the sense that, while $c$ is the upper limit of speed
(photon), unattainable by any massive particle, $V$ is the lowest limit of speed, being unattainable and, hence, invariant; however, 
the point is that no particle is found at $V$, because this speed is associated with a privileged reference frame coinciding with a
cosmic background frame, which is inaccessible and presents non-locality\cite{11}\cite{12}\cite{13}. 

It is very curious to notice that the idea of a universal background field was sought in vain by Einstein\cite{38} during the last
four decades of his life. Einstein has coined the term {\it ultra-referential} as the fundamental aspect of space for representing a 
universal background field\cite{39}, although he had abandoned the older concept of ether in 1905. Inspired by this new concept of space, 
let us simply call {\it ultra-referential} $S_V$ the universal background field ($\rho^{(2)}_{em}$) of the preferred reference frame
connected to $V$.

 The dynamics of particles in the presence of the universal background reference frame associated to $V$ is within a context of the ideas 
 of Sciama\cite{40}, Schr\"{o}dinger\cite{41} and Mach\cite{42}, where there should be an ``absolute" inertial reference frame in relation 
 to which we have the speeds and the inertia of all bodies. However, we must emphasize that the approach used here is not classical like
 Mach's idea, since the lowest (unattainable) limit of speed $V$ plays the role of a privileged reference frame of background field
 instead of the ``inertial" frame of fixed stars. Thus, in this space-time, all the speeds ($v$) are given in relation to the 
 ultra-referential $S_V$ as being the absolute background frame, where we have the interval of speeds with $V<v\leq c$ (Fig.2).

 The doubly special relativity (DSR) with an invariant minimum speed $V$ emerging from space-time with a very low vacuum energy density
 ($\rho^{(2)}_{em}$) was denominated as Symmetrical Special Relativity (SSR)\cite{11}\cite{12}\cite{13}, which is a new kind of Deformed
 Special Relativity (DSR) with two invariant scales of speed ($c$ and $V$). Here it is important to mention that DSR-theory was first 
 proposed by Camelia\cite{43}\cite{44}\cite{45}\cite{46}. It contains two invariant scales, namely the speed of light $c$ and the minimum 
 length scale (Planck length $l_{Pl}$). An alternative approach to DSR-theory, inspired by that of Camelia, was proposed later by Smolin and
 Magueijo\cite{47}\cite{48}\cite{49}. There is also another extension of SR, which is known as triply special relativity, being 
 characterized by three invariant scales, namely the speed of light $c$, a mass $k$ and a length $R$\cite{50}. Still another
 generalization of SR is the quantizing of speeds\cite{51}, where the Barrett-Crane spin foam model for quantum gravity with a
 positive cosmological constant was used to aid the authors search for a discrete spectrum of velocities and the physical
 implications of this effect, namely an effective deformed Poincar\'e symmetry. 

\section{Transformations of space-time and velocity in the presence of the ultra-referential $S_V$ connected to the invariant minimum
speed $V$}

 The existence of a lowest non-null limit of speed in the space-time results in the following physical reasoning:

- In non-relativistic QM, the plane wave wave-function ($Ae^{\pm ipx/\hbar}$) which represents a free particle is an 
idealisation that is impossible to conceive under physical reality. In the event of such an idealized plane wave, it would be possible 
to find with certainty the reference frame that cancels its momentum ($p=0$), so that the uncertainty on its position would be 
$\Delta x=\infty$. However, the presence of the unattainable minimum (non-zero) limit of speed $V$ emerges in order to prevent the
ideal case of a plane wave wave-function ($p=constant$ or $\Delta p=0$ with $\Delta x=\infty$). This means that there is no perfect
inertial motion such as a plane wave. Such impossibility is due to the presence of non-null gravity given by means of very weak
gravitational potentials in the intergalactic medium. As the minimum speed $V(\sim\sqrt{G})$ (preferred background frame $S_V$) is unattainable for 
the particles with very low energies (large wavelengths), their momentum can never vanish when one tries to be closer to such a
preferred frame ($V$).

 On the other hand, according to Special Relativity (SR), the momentum cannot be infinite since the maximum speed $c$ is also unattainable
 for a massive particle, except the photon ($v=c$) as it is a massless particle.

 This reasoning allows us to think that the electromagnetic radiation (photon:$``c-c"=c$) as well as the massive particles  
($``v-v">V$ for $v<c$) are in equal-footing in the sense that it is not possible to find a reference frame at rest
 ($v_{relative}=0$) for both through any speed transformation in a space-time with a maximum and a minimum limit of speed. 

 The classical notion we have about the inertial (galilean) reference frames, where the system at rest exists, is eliminated in SSR, where
 $v>V(S_V)$ (Fig.2). However, if we consider classical systems composed of macroscopic bodies, the minimum speed $V$ is neglected ($V=0$)
 and so we can reach a vanishing velocity ($v=0$), i.e., in the classical approximation ($V\rightarrow 0$), the
 ultra-referential (background frame) $S_V$ is eliminated and simply replaced by the galilean reference frame $S$ connected 
 to a classical system at rest.

 Since we cannot consider a reference system made up of a set of infinite points at rest in quantum space-time 
 with an invariant minimum speed, then we should define a new status of referentials in SSR, namely a non-galilean 
 reference system, which is given essentially as a set of all the particles having the same state of movement (speed $v$) with respect 
 to the ultra-referential $S_V$ (preferred reference frame of the background field), so that $v>V$, $V$ being unapproachable and connected
 to $S_V$. So, a set of particles with the same speed $v$ with respect to the ultra-referential $S_V$ provides a given non-galilean
 framework. Hence, SSR should contain three postulates, namely:

 1)-the non-equivalence (asymmetry) of the non-galilean reference frames due to the presence of the background frame $S_V$ that 
    breaks the Lorentz symmetry, i.e., we cannot exchange $v$ for $-v$ by means of inverse transformations, since we cannot achieve
    a rest state ($v_{relative}=0$) for a certain non-galilean reference frame ($S^{\prime}$) with speed $v$ in order to reverse the
    direction of motion of another one ($S_V$) for $-v$. Thus, such non-equivalence (asymmetry) of reference frames in SSR is due to the
    background vector $\vec V$ (see Section 6), having the same modulus $V$ for any direction in space, since $V$ is an invariant 
    minimum speed. 

 2)-the invariance of the speed of light ($c$);

 3)-the covariance of the ultra-referential $S_V$ (background framework) connected to an invariant and unattainable minimum limit
    of speed $V$, i.e., all the non-galilean reference frames with speeds $V<v\leq c$ experience the same background frame $S_V$,
   in the sense that the background energy (vacuum energy) at $S_V$ does not produce a flow $-v$ at any of these referentials.
   Thus, $S_V$ does not work like the newtonian absolute space filled by luminiferous (galilean) ether in the old (classical)
   sense, in spite of $S_V$'s being linked to a background energy that works like a {\it non-galilean (non-luminiferous) 
  aether} (Eq.(44)), leading to the well-known vacuum energy density (cosmological constant), as we will show later (Section 10). 

 The third postulate is directly connected to the second one. Such a connection will be clarified by investigating the new velocity
 transformations to be obtained soon.

 Of course if we consider $V=0$, we recover the well-known two postulates of SR, i.e., we get the equivalence of inertial reference 
frames (Lorentz symmetry), where one can exchange $v$ for $-v$ with appropriate transformations and, consequently, this leads to the
absence of such a background field ($S_V$); however, the constancy of the speed of light is still preserved (2nd. postulate of SR)
without a deeper explanation. 

  Let us assume the reference frame $S^{\prime}$ with a speed $v$ in relation to the ultra-referential $S_V$ according to Fig.2.

\begin{figure}
\includegraphics[scale=0.65]{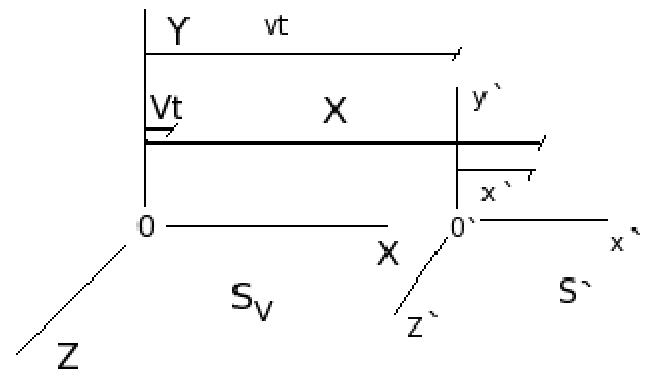}
\caption{$S^{\prime}$ moves in $x$-direction with a speed $v(>V)$ with respect to the background field connected to the ultra-referential
 $S_V$. If $V\rightarrow 0$, $S_V$ is eliminated (empty space) and, thus, the galilean frame $S$ takes place, recovering the Lorentz 
 transformations.}
\end{figure}

So, to simplify, consider the motion only at one spatial dimension, namely $(1+1)D$ space-time with the background field $S_V$.
So we write the following transformations:

  \begin{equation}
 dx^{\prime}=\Psi(dX-\beta_{*}cdt)=\Psi(dX-vdt+Vdt),
  \end{equation}
where $\beta_{*}=\beta\epsilon=\beta(1-\alpha)$, being $\beta=v/c$ and $\alpha=V/v$, so that
$\beta_{*}\rightarrow 0$ for $v\rightarrow V$ or $\alpha\rightarrow 1$.

 \begin{equation}
 dt^{\prime}=\Psi\left(dt-\frac{\beta_{*}dX}{c}\right)=\Psi\left(dt-\frac{vdX}{c^2}+\frac{VdX}{c^2}\right),
  \end{equation}
where $\vec v=v_x{\bf x}$. We have $\Psi=\frac{\sqrt{1-\alpha^2}}{\sqrt{1-\beta^2}}$. If we make
$V\rightarrow 0$ ($\alpha\rightarrow 0$), we recover the Lorentz
transformations in $(1+1)D$, where the ultra-referential $S_V$ is eliminated and simply replaced by the galilean
frame $S$ at rest for a classical observer.

As the transformations in Eq.(45) and Eq.(46) are given in $(1+1)D$ for the simple case of one dimensional motion, they appear in
their scalar form, where the background vector $\vec V$ that breaks the Lorentz symmetry is $\vec V=V{\bf x}$ (Fig.2), which can 
be simply replaced by its scalar $V$ in this case only. However, in the general case $(3+1)D$ or $3$-dimensional motion, the
presence of a background vector $\vec V$ given for any direction of space is needed. Such general case will be investigated 
in the next section.    

 In order to get the transformations in Eq.(45) and Eq.(46), let us consider the following more general transformations:
$x^{\prime}=\theta\gamma(X-\epsilon_1vt)$ and $t^{\prime}=\theta\gamma(t-\frac{\epsilon_2vX}{c^2})$,
 where $\theta$, $\epsilon_1$ and $\epsilon_2$ are factors (functions) to be determined. We hope all these factors
depend on $\alpha$, such that, for $\alpha\rightarrow 0$ ($V\rightarrow 0$), we recover Lorentz transformations as a particular case
 ($\theta=1$, $\epsilon_1=1$ and $\epsilon_2=1$). By using those transformations to perform
$[c^2t^{\prime 2}-x^{\prime 2}]$, we find the identity: $[c^2t^{\prime 2}-x^{\prime 2}]=
\theta^2\gamma^2[c^2t^2-2\epsilon_1vtX+2\epsilon_2vtX-\epsilon_1^2v^2t^2+\frac{\epsilon_2^2v^2X^2}{c^2}-X^2]$.
 Since the metric tensor is diagonal, the crossed terms must vanish and so we assure that
$\epsilon_1=\epsilon_2=\epsilon$. Due to this fact, the crossed terms ($2\epsilon vtX$) are cancelled between
themselves and finally we obtain $[c^2t^{\prime 2}-x^{\prime 2}]=
 \theta^2\gamma^2(1-\frac{\epsilon^2 v^2}{c^2})[c^2t^2-X^2]$. For $\alpha\rightarrow 0$ ($\epsilon=1$ and
$\theta=1$), we reinstate $[c^2t^{\prime 2}-x^{\prime 2}]=[c^2t^2-x^2]$ of SR. Now we write the following
transformations: $x^{\prime}=\theta\gamma(X-\epsilon vt)\equiv\theta\gamma(X-vt+\delta)$ and
$t^{\prime}=\theta\gamma(t-\frac{\epsilon vX}{c^2})\equiv\theta\gamma(t-\frac{vX}{c^2}+\Delta)$, where
we assume $\delta=\delta(V)$ and $\Delta=\Delta(V)$, so that $\delta =\Delta=0$ for $V\rightarrow 0$, which implies $\epsilon=1$.
 So, from such transformations we extract: $-vt+\delta(V)\equiv-\epsilon vt$ and
$-\frac{vX}{c^2}+\Delta(V)\equiv-\frac{\epsilon vX}{c^2}$, from where we obtain
 $\epsilon=(1-\frac{\delta(V)}{vt})=(1-\frac{c^2\Delta(V)}{vX})$. As $\epsilon$ is a dimensionless factor,
we immediately conclude that $\delta(V)=Vt$ and $\Delta(V)=\frac{VX}{c^2}$, so that we find
$\epsilon=(1-\frac{V}{v})=(1-\alpha)$. On the other hand, we can determine $\theta$ as follows: $\theta$ is a
function of $\alpha$ ($\theta(\alpha)$), such that $\theta=1$ for $\alpha=0$, which also leads to $\epsilon=1$ in
order to recover the Lorentz transformations. So, as $\epsilon$ depends on $\alpha$, we conclude that $\theta$ can
also be expressed in terms of $\epsilon$, namely $\theta=\theta(\epsilon)=\theta[(1-\alpha)]$, where
$\epsilon=(1-\alpha)$. Therefore we can write $\theta=\theta[(1-\alpha)]=[f(\alpha)(1-\alpha)]^k$, where the
exponent $k>0$. The function $f(\alpha)$ and $k$ will be estimated by satisfying the following conditions:

i) as $\theta=1$ for $\alpha=0$ ($V=0$), this implies $f(0)=1$.

ii) the function $\theta\gamma =
\frac{[f(\alpha)(1-\alpha)]^k}{(1-\beta^2)^{\frac{1}{2}}}=\frac{[f(\alpha)(1-\alpha)]^k}
{[(1+\beta)(1-\beta)]^{\frac{1}{2}}}$ should have a symmetrical behavior, that is to say it goes to zero closer
to $V$ ($\alpha\rightarrow 1$) in the same way it goes to infinite closer to $c$ ($\beta\rightarrow 1$). In other words, this
means that the numerator of the function $\theta\gamma$, which depends on $\alpha$ should have the same shape of its denumerator,
  which depends on $\beta$. Due to such conditions, we naturally conclude that $k=1/2$ and
$f(\alpha)=(1+\alpha)$, so that $\theta\gamma=
\frac{[(1+\alpha)(1-\alpha)]^{\frac{1}{2}}}{[(1+\beta)(1-\beta)]^{\frac{1}{2}}}=
\frac{(1-\alpha^2)^{\frac{1}{2}}}{(1-\beta^2)^\frac{1}{2}}=\frac{\sqrt{1-V^2/v^2}}{\sqrt{1-v^2/c^2}}=\Psi$, where
$\theta=\sqrt{1-\alpha^2}=\sqrt{1-V^2/v^2}$.

The transformations shown in Eq.(45) and Eq.(46) are the direct transformations
from $S_V$ [$X^{\mu}=(ct,X)$] to $S^{\prime}$ [$x^{\prime\nu}=(ct^{\prime},x^{\prime})$], where
we have $x^{\prime\nu}=\Lambda^{\nu}_{\mu} X^{\mu}$ ($x^{\prime}=\Lambda X$),
so that we write the matrix of transformation for one-dimensional motion in $x$-direction (Fig.(2)), as follows: 

\begin{equation}
\displaystyle\Lambda= 
\begin{pmatrix}
\theta\gamma & -\theta\gamma\beta_* & 0 & 0\\
-\theta\gamma\beta_* & \theta\gamma & 0 &  0\\
  0 & 0 & 1 & 0\\
 0  & 0 & 0 & 1
\end{pmatrix},
\end{equation}
or then simply

\begin{equation}
\displaystyle\Lambda=
\begin{pmatrix}
\Psi & -\Psi\beta_* \\
-\Psi\beta_* & \Psi
\end{pmatrix},
\end{equation}
such that $\Lambda\rightarrow\ L$ (Lorentz matrix of rotation) for $\alpha\rightarrow 0$ ($\Psi\rightarrow\gamma$). We have
$\Psi=\theta\gamma$ and $\beta_{x*}=\beta_*=\beta(1-\alpha)$, as $v=v_x$ for one-dimensional motion in $x$-direction.

We obtain $det\Lambda=\frac{(1-\alpha^2)}{(1-\beta^2)}[1-\beta^2(1-\alpha)^2]$, where $0<det\Lambda<1$. Since
$V$ ($S_V$) is unattainable ($v>V)$, this assures that $\alpha=V/v<1$ and therefore the matrix $\Lambda$
admits inverse ($det\Lambda\neq 0$ $(>0)$). However, $\Lambda$ is a non-orthogonal matrix
($det\Lambda\neq\pm 1$) and so it does not represent a rotation matrix ($det\Lambda\neq 1$) in such a space-time
due to the presence of the privileged frame of background field $S_V$ that breaks strongly the invariance of the norm of the
4-vector (limit $v\rightarrow V$ in Eq.(90) or Eq.(91)). Actually such an effect ($det\Lambda\approx 0$ for $\alpha\approx 1$
or $v\approx V$) emerges from a new relativistic physics of SSR for treating much lower energies at ultra-infrared regime closer to $S_V$ 
(very large wavelengths).

 We notice that $det\Lambda$ is a function of the speed $v$ with respect to $S_V$. In the approximation for
$v>>V$ ($\alpha\approx 0$), we obtain $det\Lambda\approx 1$ and so we practically reinstate the rotation behavior
of Lorentz matrix $L$ as a particular regime for higher energies. If we make $V\rightarrow 0$ ($\alpha\rightarrow 0$),
we recover $det\Lambda\approx det L=1$ (rotation condition). This subject will be explored with more details in Section 7, 
where we will verify whether Eq.(48) forms a group.

The inverse transformations (from $S^{\prime}$ to $S_V$) are

 \begin{equation}
 dX=\Psi^{\prime}(dx^{\prime}+\beta_{*}cdt^{\prime})=\Psi^{\prime}(dx^{\prime}+vdt^{\prime}-Vdt^{\prime}),
  \end{equation}

 \begin{equation}
 dt=\Psi^{\prime}\left(dt^{\prime}+\frac{\beta_{*}
 dx^{\prime}}{c}\right)=\Psi^{\prime}\left(dt^{\prime}+\frac{vdx^{\prime}}{c^2}-
\frac{Vdx^{\prime}}{c^2}\right).
  \end{equation}

In matrix form, we have the inverse transformation $X^{\mu}=\Lambda^{\mu}_{\nu} x^{\prime\nu}$
 ($X=\Lambda^{-1}x^{\prime}$), so that the inverse matrix is

\begin{equation}
\displaystyle\Lambda^{-1}=
\begin{pmatrix}
\Psi^{\prime} & \Psi^{\prime}\beta_* \\
\Psi^{\prime}\beta_* & \Psi^{\prime}
\end{pmatrix},
\end{equation}
where we can show that $\Psi^{\prime}$=$\Psi^{-1}/[1-\beta^2(1-\alpha)^2]=\theta^{-1}\gamma^{-1}/(1-\beta_*^2)$, so that we must satisfy $\Lambda^{-1}\Lambda=I$.

 Indeed we have $\Psi^{\prime}\neq\Psi$ and therefore $\Lambda^{-1}\neq\Lambda(-v)$. This aspect of $\Lambda$ has an important physical
 implication. In order to understand such an implication, let us first consider the rotation aspect of Lorentz matrix in SR. Under SR, we
have $\alpha=0$ ($V=0$), so that $\Psi^{\prime}\rightarrow\gamma^{\prime}=\gamma=(1-\beta^2)^{-1/2}$.
 This symmetry ($\gamma^{\prime}=\gamma$, $L^{-1}=L(-v)$) happens because the galilean reference
frames allow us to exchange the speed $v$ (of $S^{\prime}$) for $-v$ (of $S$) when we are at rest at
$S^{\prime}$. However, under SSR, since there is no rest at $S^{\prime}$, we cannot exchange $v$ (of $S^{\prime}$) for $-v$ (of $S_V$)
due to that asymmetry ($\Psi^{\prime}\neq\Psi$, $\Lambda^{-1}\neq\Lambda(-v)$), leading to the Lorentz symmetry breaking. Due to this fact,
$S_V$ must be covariant, namely $V$ remains invariant for any change of reference frames in such a space-time. Thus we
can notice that the paradox of twins, which appears due to that symmetry by
exchange of $v$ for $-v$ in SR should be naturally eliminated in SSR, where only the
reference frame $S^{\prime}$ can move with respect to $S_V$. So, $S_V$ remains
covariant (invariant for any change of reference frames). Such covariance will be verified soon.

  We have $det\Lambda=\Psi^2[1-\beta^2(1-\alpha)^2]\Rightarrow [(det\Lambda)\Psi^{-2}]=[1-\beta^2(1-\alpha)^2]$. So
we can alternatively write $\Psi^{\prime}$=$\Psi^{-1}/[1-\beta^2(1-\alpha)^2]=\Psi^{-1}/[(det\Lambda)\Psi^{-2}]
=\Psi/det\Lambda$. By inserting this result in Eq.(51) for replacing $\Psi^{\prime}$, we obtain the relationship
between the inverse matrix $\Lambda^{-1}$ and $\Lambda(-v)$, namely $\Lambda^{-1}=\Lambda(-v)/det\Lambda$. 

By dividing Eq.(45) by Eq.(46), we obtain the following speed transformation:

   \begin{equation}
  v_{rel}=\frac{v^{\prime}-v+V}
{1-\frac{v^{\prime}v}{c^2}+\frac{v^{\prime}V}{c^2}},
   \end{equation}
 where we have considered $v_{rel}=v_{relative}\equiv dx^{\prime}/dt^{\prime}$
 and $v^{\prime}\equiv dX/dt$.  $v^{\prime}$ and $v$ are given with
 respect to $S_V$, and $v_{rel}$ is the relative velocity between $v^{\prime}$ and $v$. Let us consider
 $v^{\prime}>v$ (Fig.3). 

\begin{figure}
\includegraphics[scale=0.65]{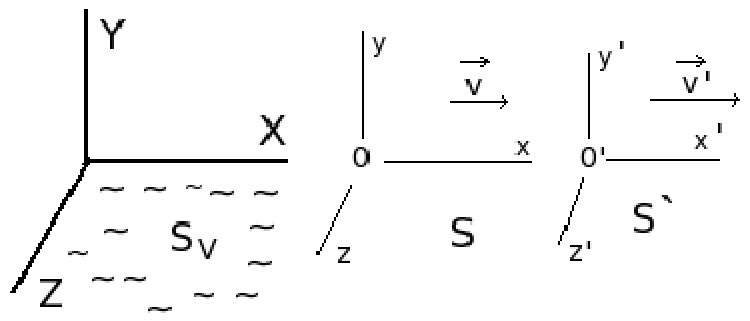}
\caption{$S_V$ is the covariant ultra-referential of background field related to the vacuum energy. $S$ represents the
reference frame for a massive particle with speed $v$ in relation to $S_V$, where $V<v<c$.
  $S^{\prime}$ represents the reference frame for a massive particle with speed $v^{\prime}$
in relation to $S_V$. In this case, we consider $V~(S_V)<v\leq v^{\prime}\leq c$.}
\end{figure}

 If $V\rightarrow 0$, the transformation in Eq.(52) recovers Lorentz velocity transformation where $v^{\prime}$ and $v$ are given in
 relation to a certain galilean frame $S_0$ at rest. Since Eq.(52) implements the ultra-referential $S_V$, the speeds $v^{\prime}$ and
 $v$ are now given with respect to the background frame $S_V$, which is covariant (absolute). Such a covariance is verified if we assume 
 that $v^{\prime}=v=V$ in Eq.(52). Thus, for this case, we obtain $v_{rel}=``V-V"=V$.

  Let us also consider the following cases for Eq.(52):

 {\bf a)} $v^{\prime}=c$ and $v\leq c\Rightarrow v_{rel}=c$. This
 just verifies the well-known invariance of $c$.

 {\bf b)} if $v^{\prime}>v(=V)\Rightarrow v_{rel}=``v^{\prime}-V"=v^{\prime}$. For
 example, if $v^{\prime}=2V$ and $v=V$ $\Rightarrow v_{rel}=``2V-V"=2V$. This
 means that $V$ really has no influence on the speed of the particles. So, $V$ works as if
 it were an ``{\it absolute zero of movement}'', being invariant and having the same value
in all directions of space of the isotropic background field.

 {\bf c)} if $v^{\prime}=v$ $\Rightarrow v_{rel}=``v-v"$($\neq 0)$
$=\frac{V}{1-\frac{v^2}{c^2}(1-\frac{V}{v})}$. From ({\bf c}) let us
consider two specific cases, namely:

  -$c_1$) assuming $v=V\Rightarrow v_{rel}=``V-V"=V$ as verified before.

  -$c_2$) if $v=c\Rightarrow v_{rel}=c$,
 where we have the interval $V\leq v_{rel}\leq c$ for $V\leq v\leq c$.

This last case ({\bf c}) shows us in fact that it is impossible to find the
rest for the particle on its own reference frame $S^{\prime}$, where
$v_{rel}(v)$ ($\equiv\Delta v(v)$) is a function that increases with the increasing of $v$ . However,
 if we make $V\rightarrow 0$, then we would have $v_{rel}\equiv\Delta v=0$ and therefore
it would be possible to find the rest for $S^{\prime}$, which would become simply a galilean
reference frame of SR.

 By dividing Eq.(49) by Eq.(50), we obtain

 \begin{equation}
  v_{rel}=\frac{v^{\prime}+v-V}{1+\frac{v^{\prime}v}{c^2}-\frac{v^{\prime}V}{c^2}}=
 \frac{v^{\prime}+v^*}{1+\frac{v^{\prime}v^*}{c^2}},
   \end{equation}
where we define the notation $v^*=v\epsilon=v(1-\alpha)=v(1-V/v)=v-V$. 

 In Eq.(53), if $v^{\prime}=v=V\Rightarrow ``V+V"=V$. Indeed $V$ is
 invariant, working like an {\it absolute zero state} in SSR. If
 $v^{\prime}=c$ and $v\leq c$, this implies $v_{rel}=c$. For $v^{\prime}>V$
 and considering $v=V$, this leads to $v_{rel}=v^{\prime}$. As a specific example, if $v^{\prime}=2V$
 and assuming $v=V$, we would have $v_{rel} =``2V+V"=2V$. And if
 $v^{\prime}=v\Rightarrow v_{rel}=``v+v"=\frac{2v-V}{1+\frac{v^2}{c^2}(1-\frac{V}{v})}$.
  In newtonian regime ($V<<v<<c$), we recover $v_{rel}=``v+v"=2v$. In relativistic (einsteinian)
regime ($v\rightarrow c$), we reinstate the Lorentz transformation for this case ($v^{\prime}=v$),
 i.e., $v_{rel}=``v+v"=2v/(1+v^2/c^2)$.

 By joining both transformations in Eq.(52) and Eq.(53) into just one, we write the following compact form:

\begin{equation}
  v_{rel}=\frac{v^{\prime}\mp\epsilon v}
{1\mp\frac{v^{\prime}\epsilon v}{c^2}}=\frac{v^{\prime}\mp v(1-\alpha)}
{1\mp\frac{v^{\prime}v(1-\alpha)}{c^2}}=\frac{v^{\prime}\mp v\pm V}
{1\mp \frac{v^{\prime}v}{c^2}\pm \frac{v^{\prime}V}{c^2}},
\end{equation}
where $\alpha=V/v$ and $\epsilon=(1-\alpha)$. For $\alpha=0$ ($V=0$) or $\epsilon=1$, we recover Lorentz speed
transformations.

\begin{figure}
\includegraphics[scale=0.75]{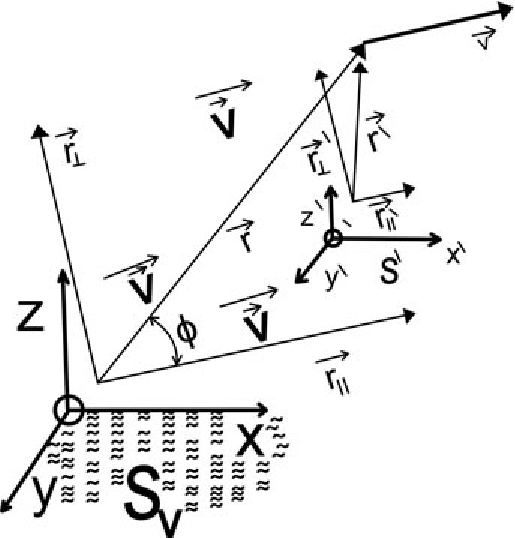}
\caption{$S^{\prime}$ moves with a $3D$-velocity $\vec v=(v_x,v_y,v_z)$ in relation to $S_V$. For the special case of $1D$-velocity
$\vec v=(v_x)$, we recover Fig.(2); however, in this general case of $3D$-velocity $\vec v$, there must be a background vector $\vec V$
(minimum velocity) with the same direction of $\vec v$ as shown in this figure. Such a background vector $\vec V=(V/v)\vec v$ is 
related to the background reference frame (ultra-referential) $S_V$, leading to Lorentz violation. The modulus of $\vec V$ is invariant
at any direction.} 
\end{figure}

\section{General transformations in (3+1)D space-time of SSR-theory}

 In order to obtain the general transformations for the case of $(3+1)D$ space-time, we should replace the one dimensional coordinates
 $X$ and $x^{\prime}$ (Fig.2) by the $3$-vectors $\vec r$ and $\vec r^{\prime}$ (Fig.4), so that we write: 
$\vec r=\vec r_{||}+\vec r_{T}$ ($S_V$), where $\vec r_{||}$ is given in the direction of the motion ($\vec v$) and $\vec r_{T}$ 
is given in the transverse direction of the motion.

 At the frame $S^{\prime}$ with velocity $\vec v$, we have the vector $\vec r^{\prime}$, where we write: $\vec r^{\prime}=\vec
 r^{\prime}_{||}+\vec r^{\prime}_{T}$ ($S^{\prime}$) (Fig.4). 

In classical $(3+1)D$ space-time of SR, of course we should have $\vec r^{\prime}_{T}=\vec r_{T}$, since there is no boost transformation 
for the transverse direction of motion, so that the modulus of $\vec r_{T}$ is always preserved for any reference frame. However, for 
$(3+1)D$ space-time of SSR, there should be a non-classical (without boost) transformation of the transverse vector, such that
 $\vec r^{\prime}_{T}\neq\vec r_{T}$. So let us admit the following
transformation: $\vec r^{\prime}_{T}=\theta\vec r_{T}=\sqrt{1-\alpha^2}\vec r_{T}$, where $\alpha=V/v$. Such a non-classical effect occurs
only due to the existence of a minimum speed $V$ given for any direction in the space, so that the transverse direction could not be
neglected, i.e., in SSR there should be a transformation for the vector at the transverse direction of the motion. Therefore, only if the
speed $v$ is closer to $V$, a drastic dilation of $r_{T}$ occurs, that is to say
$r_{T}\rightarrow\infty$ when $v\rightarrow V$. Such a transverse dilation that occurs only close to the ultra-referential $S_V$ shows us
the isotropic $3$-dimensional aspect of the background frame connected to $S_V$, which could not be simply reduced to $1$-dimensional 
space because the modulus $V$ of the background vector $\vec V$ should remain invariant for any direction in the space, leading to the 
Lorentz symmetry breaking due to the transformation factor $\theta(=\sqrt{1-V^/v^2})$ that acts on the transverse direction as a kind of
scale transformation that will be better justified in the next Section (Eq.(79) and Eq.(80)). 

 So, at the frame $S^{\prime}$, we have

\begin{equation}
\vec r^{\prime}=\vec r^{\prime}_{||}+\theta\vec r_{T},
\end{equation}
where $\vec r^{\prime}_{T}=\theta\vec r_{T}$.

As the direction of motion ($\vec r_{||}$) transforms in a similar way to Eq.(45) for $1$-dimensional case, we write Eq.(55),
as follows:

\begin{equation}
\vec r^{\prime}=\theta[\vec r_{T}+\gamma(\vec r_{||}-\vec v(1-\alpha)t)]=\theta[\vec r_{T}+\gamma(\vec r_{||}-\vec vt+\vec Vt)],
\end{equation}
where we simply have $\vec r^{\prime}_{||}=\theta\gamma(\vec r_{||}-\vec v(1-\alpha)t)=\theta\gamma(\vec r_{||}-\vec vt+\vec Vt)$,
$\vec r_{||}$ being parallel to $\vec v$. We have $\theta\gamma=\Psi$ and $\alpha\vec v=(V/v)\vec v=\vec V$ (background vector for
representing the minimum velocity). 

 From Eq.(56), we can see that $\theta$ appears as a multiplicative factor. Of course, if we make $\alpha=0$ ($V=0$) in Eq.(56), 
this implies $\theta=1$ and, thus, we recover the well-known Lorentz tranformation of the $3$-vector.

 We write $\vec r_T=\vec r-\vec r_{||}$. So, by inserting such relation into Eq.(56) and performing the calculations, we find:

\begin{equation}
\vec r^{\prime}=\theta[\vec r + (\gamma-1)\vec r_{||}-\gamma\vec v(1-\alpha)t]=
\theta[\vec r + (\gamma-1)\vec r_{||}-\gamma\vec vt+\gamma\vec Vt].
\end{equation}

 We have $\vec r_{||}=r\cos\phi\vec e_{||}$, where $\vec e_{||}$ is the unitary vector in the direction of motion, i.e.,
$\vec e_{||}=\frac{\vec v}{v}$. The angle $\phi$ is formed between the two vectors $\vec r$ and $\vec e_{||}$. On the other hand, we can
write: $r\cos\phi=(\vec r.\vec v)/v$. Finally, we get $\vec r_{||}=\frac{(\vec r.\vec v)}{v^2}\vec v$. So we write Eq.(57) in the
following way:

\begin{equation}
\vec r^{\prime}=\theta\left[\vec r + (\gamma-1)\frac{(\vec r.\vec v)}{v^2}\vec v-\gamma\vec v(1-\alpha)t\right]=
\theta\left[\vec r + (\gamma-1)\frac{(\vec r.\vec v)}{v^2}\vec v-\gamma\vec vt+\gamma\vec Vt\right].
\end{equation}

The transformation above in Eq.(58) represents the $3$-vector transformation in $(3+1)D$ space-time with the presence of the
background vector $\vec V$ (minimum velocity) that breaks the Lorentz symmetry. 

From Eq.(58), we can verify that, if we consider $\vec v$ to be in the same direction of $\vec r$, with $r\equiv X$, we obtain
$\frac{(\vec r.\vec v)}{v^2}\vec v=X\frac{\vec v}{v}=X\vec e_x$. So, the transformation in Eq.(58) is reduced to the special case of one
dimensional motion, namely $x^{\prime}=\theta[X + (\gamma-1)X -\gamma v(1-\alpha)t]=\theta\gamma(X - v(1-\alpha)t)=
\Psi(X-vt+Vt)$ (Eq.(45)), where $V$ can be thought of as being a scalar. 

Now, we can realize that the generalization of the transformation in Eq.(46) for the case of $(3+1)D$ space-time leads us to write:

 \begin{equation}
t^{\prime}=\theta\gamma\left[t-\frac{\vec r.\vec v}{c^2}(1-\alpha)\right]=
\theta\gamma\left[t-\frac{\vec r.\vec v}{c^2}+\frac{\vec r.\vec V}{c^2}\right],
\end{equation}
where $\vec V=(\vec v/v)V$. It is easy to verify that, if we have $\vec v||\vec r(\equiv X\vec e_x)$, we recover Eq.(46) for $(1+1)D$ 
spacetime.

 Putting the transformations in Eq.(59) and Eq.(58) into a matricial form, we find the following matrix:

\begin{equation}
\displaystyle\Lambda_{(4X4)}=
\begin{pmatrix}
\theta\gamma  &-\theta\gamma\beta_{x*}  &-\theta\gamma\beta_{y*}  &-\theta\gamma\beta_{z*} \\
 -\theta\gamma\beta_{x*} & \left[\theta+\theta(\gamma-1)\frac{\beta_x^2}{\beta^2}\right]  & \left[\theta(\gamma-1)\frac{\beta_x\beta_y}{\beta^2}\right] & \left[\theta(\gamma-1)\frac{\beta_x\beta_z}{\beta^2}\right]\\
 -\theta\gamma\beta_{y*}                       &\left[\theta(\gamma-1)\frac{\beta_y\beta_x}{\beta^2}\right]  & \left[\theta+\theta(\gamma-1)\frac{\beta_y^2}{\beta^2}\right]  & \left[\theta(\gamma-1)\frac{\beta_y\beta_z}{\beta^2}\right] \\
 -\theta\gamma\beta_{z*}                         & \left[\theta(\gamma-1)\frac{\beta_z\beta_x}{\beta^2}\right]           & \left[\theta(\gamma-1)\frac{\beta_z\beta_y}{\beta^2}\right]    &  \left[\theta+\theta(\gamma-1)\frac{\beta_z^2}{\beta^2}\right]
\end{pmatrix}, 
\end{equation}
where we have defined the compact notations namely $\beta_{x*}=\beta_{x}(1-\alpha)$, $\beta_{y*}=\beta_{y}(1-\alpha)$ and 
$\beta_{z*}=\beta_{z}(1-\alpha)$. 

Writing the general matrix of transformation $\Lambda_{(4X4)}$ (Eq.(60)) in a compact form $(2\times 2)$, we get

\begin{equation}
\displaystyle\Lambda_{(2\times 2)}=
\begin{pmatrix}
\theta\gamma & -\frac{\theta\gamma {\bf v}^T(1-\alpha)}{c} \\
-\frac{\theta\gamma{\bf v}(1-\alpha)}{c} & \left[\theta I+\theta(\gamma-1)\frac{{\bf v}{\bf v^T}}{v^2}\right]
\end{pmatrix},
\end{equation}
where $I=I_{3\times 3}$ is the identity matrix and ${\bf v}^T=(v_x, v_y, v_z)$ is the transposed of ${\bf v}$.

If we make $\alpha=0$ ($V=0$), which implies $\theta=1$, the matrix in Eq.(60) (or Eq.(61)) recovers the general Lorentz matrix.  

Now, in order to obtain the general inverse transformations, we should generalize the inverse transformations in Eq.(49) and Eq.(50)
for the case of $(3+1)D$ space-time. To do that, we firstly consider the following known relations: $\vec r=\vec r_{||}+\vec r_T$ (A)
and $\vec r^{\prime}=\vec r^{\prime}_{||}+\theta\vec r_T$ (B), with $\theta\vec r_T=\vec r^{\prime}_T$ or 
$\vec r_T=\theta^{-1}\vec r^{\prime}_T$ (C).

In the direction of motion, the inverse transformation has the same form of Eq.(49), where we simply replace $X$ by $\vec r_{||}$ and
$x^{\prime}$ by $\vec r^{\prime}_{||}$. So we write:

\begin{equation}
\vec r_{||}=\Psi^{\prime}[\vec r^{\prime}_{||}+\vec v(1-\alpha)t^{\prime}]=
\Psi^{\prime}[\vec r^{\prime}_{||}+\vec vt^{\prime}-\vec Vt^{\prime}],
\end{equation}
where we have shown $\Psi^{\prime}=\Psi^{-1}/[1-\beta^2(1-\alpha)^2]\neq\Psi$.

 Inserting Eq.(62) and the relation (C) into (A), we find

\begin{equation}
\vec r=\theta^{-1}\vec r^{\prime}_T + \Psi^{\prime}[\vec r^{\prime}_{||}+\vec v(1-\alpha)t^{\prime}]=
\theta^{-1}\vec r^{\prime}_T + \Psi^{\prime}[\vec r^{\prime}_{||}+\vec vt^{\prime}-\vec Vt^{\prime}].
\end{equation}

As we have $\vec r^{\prime}_T=\vec r^{\prime}-\vec r^{\prime}_{||}$ (at the frame $S^{\prime}$), thus, by inserting this relation
into Eq.(63) and performing the calculations, we get

\begin{equation}
\vec r=\theta^{-1}\vec r^{\prime} + (\Psi^{\prime}-\theta^{-1})\vec r^{\prime}_{||}+\Psi^{\prime}\vec v(1-\alpha)t^{\prime}=
\theta^{-1}\vec r^{\prime} + (\Psi^{\prime}-\theta^{-1})\vec r^{\prime}_{||}+\Psi^{\prime}\vec vt^{\prime}-\Psi^{\prime}\vec Vt^{\prime},
\end{equation}
where we have $\vec r^{\prime}_{||}=(\frac{\vec r^{\prime}.\vec v}{v^2})\vec v$, and so we write Eq.(64) as follows: 

\begin{equation}
\vec r=\theta^{-1}\vec r^{\prime} + (\Psi^{\prime}-\theta^{-1})\left(\frac{\vec r^{\prime}.\vec v}{v^2}\right)\vec v+\Psi^{\prime}\vec 
vt^{\prime}-\Psi^{\prime}\vec Vt^{\prime}.
\end{equation}

As we already know that $\Psi^{\prime}=\Psi^{-1}/[1-\beta^2(1-\alpha)^2]=\theta^{-1}\gamma^{-1}/[1-\beta^2(1-\alpha)^2]$, 
we can also write Eq.(65) in the following way:

\begin{equation}
\vec r=\theta^{-1}\vec r^{\prime} + \theta^{-1}\left[\left(\frac{\gamma^{-1}}{1-\beta^2_*}-1\right)
\left(\frac{\vec r^{\prime}.\vec v}{v^2}\right)+
\frac{(\gamma^{-1})_*}{1-\beta^2_*}t^{\prime}\right]\vec v,
\end{equation}
where we have used the simplified notation $\beta_*=\beta(1-\alpha)$. We also have $(\gamma^{-1})_*=\gamma^{-1}(1-\alpha)$.

 Now, it is natural to conclude that the time inverse transformation is given as follows:

\begin{equation}
t=\frac{\theta^{-1}\gamma^{-1}}{1-\beta^2(1-\alpha)^2}\left[t^{\prime}+ \frac{\vec r^{\prime}.\vec v}{c^2}(1-\alpha)\right]=
\frac{\theta^{-1}\gamma^{-1}}{1-\beta^2(1-\alpha)^2}\left[t^{\prime}+\frac{\vec r^{\prime}.\vec v}{c^2}-
\frac{\vec r^{\prime}.\vec V}{c^2}\right].
\end{equation}

In Eq.(66) and Eq.(67), if we make $\alpha=0$ (or $V=0$), we recover the $(3+1)D$ Lorentz inverse transformations.

From Eq.(67) and Eq.(66), we obtain the general inverse matrix of transformation as follows:

\begin{equation}
\displaystyle\Lambda^{-1}_{(4\times 4)}=
\begin{pmatrix}
\frac{\theta^{-1}\gamma^{-1}}{1-\beta_*^2}  &\frac{\theta^{-1}\gamma^{-1}\beta_{x*}}{1-\beta_*^2}  &\frac{\theta^{-1}\gamma^{-1}\beta_{y*}}{1-\beta_*^2}  &\frac{\theta^{-1}\gamma^{-1}\beta_{z*}}{1-\beta_*^2} \\
 \frac{\theta^{-1}\gamma^{-1}\beta_{x*}}{1-\beta_*^2} & \left[\theta^{-1}+\theta^{-1}\left(\frac{\gamma^{-1}}{1-\beta_*^2}-1\right)\frac{\beta_x^2}{\beta^2}\right]  & \left[\theta^{-1}\left(\frac{\gamma^{-1}}{1-\beta_*^2}-1\right)\frac{\beta_x\beta_y}{\beta^2}\right] & \left[\theta^{-1}\left(\frac{\gamma^{-1}}{1-\beta_*^2}-1\right)\frac{\beta_x\beta_z}{\beta^2}\right]\\
 \frac{\theta^{-1}\gamma^{-1}\beta_{y*}}{1-\beta_*^2}                       &\left[\theta^{-1}\left(\frac{\gamma^{-1}}{1-\beta_*^2}-1\right)\frac{\beta_y\beta_x}{\beta^2}\right]  & \left[\theta^{-1}+\theta^{-1}\left(\frac{\gamma^{-1}}{1-\beta_*^2}-1\right)\frac{\beta_y^2}{\beta^2}\right]  & \left[\theta^{-1}\left(\frac{\gamma^{-1}}{1-\beta_*^2}-1\right)\frac{\beta_y\beta_z}{\beta^2}\right] \\
 \frac{\theta^{-1}\gamma^{-1}\beta_{z*}}{1-\beta_*^2}                      & \left[\theta^{-1}\left(\frac{\gamma^{-1}}{1-\beta_*^2}-1\right)\frac{\beta_z\beta_x}{\beta^2}\right]           & \left[\theta^{-1}\left(\frac{\gamma^{-1}}{1-\beta_*^2}-1\right)\frac{\beta_z\beta_y}{\beta^2}\right]    &  \left[\theta^{-1}+\theta^{-1}\left(\frac{\gamma^{-1}}{1-\beta_*^2}-1\right)\frac{\beta_z^2}{\beta^2}\right]
\end{pmatrix}, 
\end{equation}
where we have $\beta_{x*}=\beta_{x}(1-\alpha)$, $\beta_{y*}=\beta_{y}(1-\alpha)$, $\beta_{z*}=\beta_{z}(1-\alpha)$
and $\beta_{*}=\beta(1-\alpha)=v(1-\alpha)/c=v_{*}/c$.

Writing the general inverse matrix of transformation $\Lambda^{-1}$ (Eq.(68)) in a compact form $(2\times 2)$, we have

\begin{equation}
\displaystyle\Lambda^{-1}_{(2\times 2)}=
\begin{pmatrix}
\frac{\theta^{-1}\gamma^{-1}}{1-\beta^2_*} & \frac{\theta^{-1}\gamma^{-1}{\bf v}^T_*}{c(1-\beta^2_*)} \\
\frac{\theta^{-1}\gamma^{-1}{\bf v_*}}{c(1-\beta^2_*)} &
\left[\theta^{-1}I+\theta^{-1}(\frac{\gamma^{-1}}{1-\beta^2_*}-1)\frac{{\bf v}{\bf v^T}}{v^2}\right]
\end{pmatrix},
\end{equation}
where ${\bf v}^T_*={\bf v}^T(1-\alpha)$, ${\bf v}_*={\bf v}(1-\alpha)$ and $\beta_*=\beta(1-\alpha)$.

If we make $\alpha=0$ ($V=0$), which implies $\theta=1$, the inverse matrix in Eq.(68) (or Eq.(69)) recovers the general inverse matrix
of Lorentz. 

 Now we can compare the inverse matrix in Eq.(69) with the matrix in Eq.(61) and verify that $\Lambda^{-1}\neq\Lambda^T$,
 in a similar way as made before for the particular case $(1+1)D$ (one dimensional motion). 

 \section{Do the space-time transformations with an invariant minimum speed form a group? What are their deep implications?}

   It is well-known that the Lorentz transformations form a group ($L=L(v)$), since they obey the following conditions, namely:
a) $L_2L_1=L(v_2)L(v_1)=L(v_3)=L_3\in L(v)$ (Closure condition); b) $L_1(L_2L_3)=(L_1L_2)L_3$ (Associativity); 
c) $L_0L=LL_0=L$, such that $L_0=L(0)=I$ (Identity element); d) $L^{-1}L=LL^{-1}=L_0$, where $L^{-1}=L{(-v)}$ (Inverse element). 

    Our goal is to make an analysis of the new transformations in Eq.(45) and Eq.(46) with regard to the conditions above in order to
verify whether they form a group and discuss deeply the results. So, to do that, we first rewrite the matrix $\Lambda$ (Eq.(48)), namely:

\begin{equation}
\displaystyle\Lambda=
\begin{pmatrix}
\Psi & -\Psi\beta^{*} \\
-\Psi\beta^{*} & \Psi
\end{pmatrix},
\end{equation}
where $\Psi=\frac{\sqrt{1-V^2/v^2}}{\sqrt{1-v^2/c^2}}$. We already have defined the notation $\beta^*=\beta\epsilon=\beta(1-\alpha)=
(v/c)[1-V/v]$. If $V\rightarrow 0$ or $\alpha\rightarrow 0$, we recover the Lorentz matrix, i.e., $\Lambda(v)\rightarrow L(v)$,
since $\Psi\rightarrow\gamma$ and $\beta^*\rightarrow\beta$. 

  Now, we have $\Lambda_1=\Lambda(v_1)$ as being 

\begin{equation}
\displaystyle\Lambda_1=
\begin{pmatrix}
\Psi_1 & -\Psi_1\beta_1^{*} \\
-\Psi_1\beta_1^{*} & \Psi_1
\end{pmatrix}= 
\begin{pmatrix}
\Psi_1 & -\Psi_1\frac{v_1^*}{c} \\
-\Psi_1\frac{v_1^*}{c} & \Psi_1
\end{pmatrix}
\end{equation}

  and $\Lambda_2=\Lambda(v_2)$ as being

\begin{equation}
\displaystyle\Lambda_2=
\begin{pmatrix}
\Psi_2 & -\Psi_2\beta_2^* \\
-\Psi_2\beta_2^* & \Psi_2
\end{pmatrix}=
\begin{pmatrix}
\Psi_2 & -\Psi_2\frac{v_2^*}{c} \\
-\Psi_2\frac{v_2^*}{c} & \Psi_2
\end{pmatrix}, 
\end{equation}
so that $\Lambda_2\Lambda_1$ is 

\begin{equation}
\displaystyle\Lambda_2\Lambda_1=[\Psi_2\Psi_1(1+\beta^*_2\beta^*_1)]
\begin{pmatrix}
1 & -\frac{(\beta_1^*+\beta_2^*)}{1+\beta_2^*\beta_1^*} \\
-\frac{(\beta_1^*+\beta_2^*)}{1+\beta_2^*\beta_1^*} & 1
\end{pmatrix},
\end{equation}
where $\beta_1^*=\beta_1\epsilon_1=\beta_1(1-\alpha_1)=(v_1/c)[1-V/v_1]$ and $\beta_2^*=\beta_2\epsilon_2=\beta_2(1-\alpha_2)=
(v_2/c)[1-V/v_2]$.

We obtain that the multiplicative term of the matrix in Eq.(73) is written as $\Psi_2\Psi_1(1+\beta^*_2\beta^*_1)=
\sqrt{(1-V^2/v_2^2)(1-V^2/v_1^2)}\frac{1+(v_1^*v_2^*/c^2)}{\sqrt{1-(v_1^2/c^2+v_2^2/c^2-v_1^2v_2^2/c^4)}}$. Now by inserting 
this term into Eq.(73), we rewrite Eq.(73) in the following way: 

\begin{equation}
\displaystyle\Lambda_2\Lambda_1=\frac{\sqrt{\left(1-\frac{V^2}{v_2^2}\right)\left(1-\frac{V^2}{v_1^2}\right)}
\left(1+\frac{v_1^*v_2^*}{c^2}\right)}{\sqrt{1-\left(\frac{v_1^2}{c^2}+\frac{v_2^2}{c^2}-\frac{v_1^2v_2^2}{c^4}\right)}}
\begin{pmatrix}
1 & -\frac{1}{c}\left(\frac{v_1^*+v_2^*}{1+ \frac{v_1^*v_2^*}{c^2}}\right) \\
-\frac{1}{c}\left(\frac{v_1^*+v_2^*}{1+\frac{v_1^*v_2^*}{c^2}}\right) & 1
\end{pmatrix}
\end{equation}

Now we should note that, if the Eq.(74) satisfies the closure condition, Eq.(74) must be equivalent to 

\begin{equation}
\displaystyle\Lambda_2\Lambda_1=\Lambda_3=\Psi_3
\begin{pmatrix}
1  & -\frac{v_3^*}{c} \\
-\frac{v_3^*}{c} & 1
\end{pmatrix}, 
\end{equation}
 where, by comparing Eq.(74) with Eq.(75), we must verify whether the closure condition is satisfied, i.e., $\Psi_3\equiv\sqrt{(1-V^2/v_2^2)(1-V^2/v_1^2)}\frac{1+(v_1^*v_2^*/c^2)}
{\sqrt{1-(v_1^2/c^2+v_2^2/c^2-v_1^2v_2^2/c^4)}}$ and  $v_3^*\equiv(v_2^*+v_1^*)/[1+(v_2^*v_1^*)/c^2]$. However, we first realize that such
 speed transformation, which should be obeyed in order to satisfy the closure condition, differs from the correct speed transformation 
(Eq.(53)) that has origin from the space-time transformations with a minimum speed given in Eq.(49) and Eq.(50). So, according to Fig.3, 
if we simply redefine $v^{\prime}=v_2$ and $v=v_1$, we rewrite the correct transformation (Eq.(53)) as being
$v_{rel}=v_3=(v_2+v_1^*)/[1+(v_2v_1^*)/c^2]$ with $v_1^*=v_1-V$. Now, we see that the correct transformation for $v_3$ (Eq.(53)) is not the
same transformation given in the matrix above (Eq.(74)), i.e., we have $v_3\neq (v_2^*+v_1^*)/[1+(v_2^*v_1^*)/c^2]$. 

One of the conditions for having the closure relation is that the components outside the diagonal of the matrix in Eq.(74) or Eq.(75) 
must include
$v_3$ given by Eq.(53), which does not occur. Therefore, we are already able to conclude that such condition is not obeyed in a space-time
with a minimum speed (a preferred reference frame) at the subatomic level, i.e., we find $\Lambda_2\Lambda_1\neq\Lambda_3$, which does not 
generate a group. In order to clarify further this question, we just make the approximation $V=0$ or also $v_1\gg V$ and $v_2\gg V$ in
Eq.(74), and thus we recover the closure relation of the Lorentz group, as follows: 

\begin{equation}
\displaystyle(\Lambda_2\Lambda_1)_{V=0}=L_2L_1=\frac{\left(1+\frac{v_1v_2}{c^2}\right)}
{\sqrt{1-\left(\frac{v_1^2}{c^2}+\frac{v_2^2}{c^2}-\frac{v_1^2v_2^2}{c^4}\right)}}
\begin{pmatrix}
1 & -\frac{1}{c}\left(\frac{v_1+v_2}{1+ \frac{v_1v_2}{c^2}}\right) \\
-\frac{1}{c}\left(\frac{v_1+v_2}{1+\frac{v_1v_2}{c^2}}\right) & 1
\end{pmatrix}=L_3, 
\end{equation}
where

\begin{equation}
\displaystyle L_3=\gamma_3
\begin{pmatrix}
 1  &     -\frac{v_3}{c}\\
 -\frac{v_3}{c}  & 1
\end{pmatrix}, 
\end{equation}
which is the closure condition of the Lorentz group, since now it is obvious that the Lorentz transformation of speeds appears outside 
the diagonal of the matrix in Eq.(76), i.e., we find $v_3=(v_1+v_2)/[1+(v_1v_2)/c^2]$. And, in order to complete the verification of the
closure condition above, it is easy to verify that the multiplicative term of the matrix (Eq.(76)) is $\gamma_2\gamma_1(1+\beta_2\beta_1)=
\left(1+\frac{v_1v_2}{c^2}\right)/\sqrt{1-\left(\frac{v_1^2}{c^2}+\frac{v_2^2}{c^2}-\frac{v_1^2v_2^2}{c^4}\right)}=\gamma_3$. To do this,
we have to consider $v_3=(v_1+v_2)/[1+(v_1v_2)/c^2]$, so that we use this transformation to be inserted into $\gamma_3=1/\sqrt{1-v_3^2/c^2}$ and
we finally show that $\gamma_3=1/\sqrt{1-v_3^2/c^2}=\gamma_2\gamma_1(1+\beta_2\beta_1)$. However, now starting from this same procedure
for obtaining $\Psi_3=\frac{\sqrt{1-V^2/v_3^2}}{\sqrt{1-v_3^2/c^2}}$, where we have to use the correct transformation for $v_3$ (Eq.(53)),
we verify that $\Psi_3\neq\Psi_2\Psi_1(1+\beta_2^*\beta_1^*)$ and thus we conclude definitively that the closure condition does not apply to
the new transformations, i.e., indeed we have $\Lambda_2\Lambda_1\neq\Lambda_3$. 

Although we already know that the new transformations do not form a group, it is still important to provide a physical justification 
for such conclusion. To do this with more clarity, we also should investigate whether the identity element and the inverse element 
exist in such a spacetime with an invariant minimum speed, since these two conditions are relevant to give us a clear comprehension
of the conception of motion in this spacetime.\\

 {\bf(I) Identity element} \\

For the case $(1+1)D$, the Lorentz group provides the identity element $L_0=I_{(2X2)}$, since $L_0L=LL_0=L$. As the Lorentz matrix 
is $\displaystyle L_{(2X2)}=\begin{pmatrix}
 \gamma  &     -\beta\gamma \\
-\beta\gamma  & \gamma
\end{pmatrix}$, it is easy to see that, if we make $v=0$ or $\beta=0$ (rest condition), the Lorentz matrix recover the identity matrix $\displaystyle
 I_{(2X2)}=\begin{pmatrix}
 1 &     0 \\
 0  &    1
\end{pmatrix}$, since $\gamma_0=\gamma(v=0)=1$. This trivial condition of rest plus the fact that $det(L)=1$ (rotation matrix $L$)
 shows us the indistinguishability of rest and inertial motion. 

The new transformations are represented by the matrix $\displaystyle\Lambda=
\begin{pmatrix}
\Psi & -\beta(1-\alpha)\Psi \\
-\beta(1-\alpha)\Psi & \Psi
\end{pmatrix}$, where we have $\beta^*=\beta\epsilon=\beta(1-\alpha)$, with $\alpha=V/v$. Now, it is important to notice that there is no any speed $v$
that generates the identity matrix from the new matrix. We would expect that the hypothesis $v=V$ could do that, but,
if we make $v=V$ ($\alpha=1$) inside the new matrix, we find the null matrix, i.e., $\displaystyle\Lambda(V)=\begin{pmatrix}
0 & 0\\
0 & 0
\end{pmatrix}$, since $\Psi(V)=0$. So, we obtain $\Lambda(V)\Lambda=\Lambda_V\Lambda=\Lambda\Lambda_V=\displaystyle\begin{pmatrix}
0 & 0\\
0 & 0
\end{pmatrix}\neq\Lambda$, where we have $\Lambda=\Lambda(v>V)$. Thus, there is no identity element in this spacetime, which means that
there should be a distinction of motion and rest, since there is a preferred reference frame (an invariant minimum speed)  
in respect to which, the motion $v(>V)$ is given, in view of the absence of the rest condition for particles in this spacetime
with a non-null minimum speed $V$.\\

{\bf(II) Inverse element} \\

It is well-known that the inverse element exists in Lorentz transformations that form a group, i.e., we have $L^{-1}(v)=L(-v)$, which means
that we can exchange the observer in the reference frame $S$ at rest by another observer in the reference frame $S^{\prime}$ with speed $v$ in 
respect to $S$, so that the other observer at $S^{\prime}$ simply observes $S$ with a speed $-v$. Such symmetry comes from the galilean
relativity of motion, which it is essentially due to the indistinguishability of rest and inertial motion. Here we must stress that such
indistinguishability is broken down in the new transformations, since the invariant minimum speed related to a background reference
frame introduces a preferential motion $v(>V)$ that cannot be exchanged by $-v$ due to the distinction of motion and rest, since rest
does not exist in this spacetime, where we get $\Lambda^{-1}(v)\neq\Lambda(-v)$, such that we obtain
 $\Lambda(-v)\Lambda(v)=\displaystyle\theta^2
\begin{pmatrix}
\gamma & \beta(1-\alpha)\gamma \\
\beta(1-\alpha)\gamma & \gamma
\end{pmatrix}$ $\times$
$\displaystyle\begin{pmatrix}
\gamma & -\beta(1-\alpha)\gamma \\
-\beta(1-\alpha)\gamma & \gamma
\end{pmatrix}=\displaystyle\Psi^2\begin{pmatrix}
 \left(1-\frac{v^{*2}}{c^2}\right) & 0 \\
 0 & \left(1-\frac{v^{*2}}{c^2}\right)
\end{pmatrix}\neq I_{(2X2)}$. For $V=0$ ($\alpha=0$), we recover the inverse element of the Lorentz group, which is a rotation group.\\

In short, we have verified that the new transformations do not form a group and we have provided a physical explanation for such Lorentz
violation in view of the existence of an invariant minimum speed that breaks down the indistinguishability of rest and motion. 

We have also concluded that the new matrix $\Lambda$ (Eq.(48)) does not represent a rotation matrix (Section 5). In view of this,
we can realize that such transformations are not related with the well-known rotation group $SO(3)$ (Lie group), whose elements
$R(\vec\alpha)$ and $R(\vec\beta)$ should obey a closure condition $R(\vec\alpha)R(\vec\beta)=R(\vec\gamma)$, such that $\vec\gamma=
\gamma(\vec\alpha,\vec\beta)$, with $R(\vec\gamma)$ being a new rotation that belongs to the group, so that
$det(R)=+1$ (rotation condition), while we find $det(\Lambda)=\theta^2\gamma^2\left[1-\frac{v^2(1-\alpha)^2}{c^2}\right]$,
 where $0<det(\Lambda)<1$, violating the rotation condition. 

Although there could be a more complex mathematical structure in order to encompass the new transformations, which should be deeply
investigated, at least, here we will make some interesting mathematical approximations on $\Lambda$ in order to help us to understand
further the nature of the new transformations. 

In a certain approximation, let us show that $\Lambda$ is a combination of rotation and deformation of the space-time interval $ds$, 
reminding the polar decomposition theorem in linear algebra for a ``rigid" body that rotates and deforms. Intuitively, the polar 
decomposition separates a certain matrix $A$ into a component that stretches the space along a set of orthogonal axes, represented by $P$,
and a rotation (with possible reflection) represented by $U$, i.e., $A=UP$. where $U$ is a unitary matrix and $P$ is a Hermitian matrix.

When a rigid body rotates, its length remains invariant. This effect is analogous to the invariance of the space-time interval $ds$ under
the Lorentz transformation $L$ (Lorentz group) due to a rotation.

When a ``rigid'' body deforms, such effect is analogous to a deformation (e.g: stretching) of the spacetime interval $ds$ that occurs close 
to the minimum speed (see Eq.(90) and Eq.(91) for $v\rightarrow V$). Thus, at a first sight, the new transformation $\Lambda$ could be
written simply as a polar decomposition, so that $\Lambda=LD$, where $L$ is a rotation matrix (Lorentz matrix) as a special case of the
unitary matrix $U$ and $D$ is a deformation matrix (symmetric matrix) as a special case of the Hermitian matrix $P$, since Hermitian 
matrices can be understood as complex extensions of real symmetric matrices. However, we can verify that such analogy fails 
quantitatively when one tries to calculate an exact matrix $D_{(2\times 2)}$ that satisfies the linear decomposition $LD=\Lambda$ for any 
speed $v$ (any energy scale), i.e., there is no $D_{(2\times 2)}$ that satisfies such decomposition, because it seems that we have
a kind of non-linear or inseparable combination of rotation and deformation. So, in order to accomplish a stronger analogy with
the polar decomposition, we need to make some mathematical approximations on the matrix $\Lambda$ in such a way that we can be able
to separate both effects of rotation and deformation. To do this, let us first write the matrix $\Lambda$ in the following way:   

\begin{equation}
\displaystyle\Lambda=\frac{\sqrt{1-\alpha^2}}{\sqrt{1-\beta^2}}
\begin{pmatrix}
1  & -\frac{v(1-\alpha)}{c} \\
-\frac{v(1-\alpha)}{c} & 1
\end{pmatrix}
\end{equation}

Instead of making $\alpha=0$ (or $V=0$) in order to recover the Lorentz matrix $L$, here we make an alternative approximation, namely
$v>>V$, which means that, for higher energies, we recover practically the matrix $L$ (rotation). On the other hand, for much lower 
energies, i.e., for $\alpha\approx 1$ (or $v\approx V$), we get 

\begin{equation}
\displaystyle\Lambda_{(v\approx V)}\approx\sqrt{1-\alpha^2}
\begin{pmatrix}
1  & 0 \\
0  & 1
\end{pmatrix}=
\displaystyle\theta
\begin{pmatrix}
1  & 0 \\
0  & 1
\end{pmatrix},
\end{equation}
where we have considered the L'H\^opital's rule, by calculating
 $lim_{\alpha\rightarrow 1}\frac{(1-\alpha)}{\sqrt{1-\alpha^2}}=lim_{\alpha\rightarrow 1}\frac{\sqrt{1-\alpha^2}}{\alpha}=0$. In other
words, this means that $\sqrt{1-\alpha^2}|_{\alpha\approx 1}>>(1-\alpha)|_{\alpha\approx 1}$, so that we can neglect $(1-\alpha)$ with
respect to $\sqrt{1-\alpha^2}$ in such an approximation ($v\approx V$), i.e, we make $\epsilon=(1-\alpha)=0$ into $\Lambda$,
keeping the factor $\theta=\sqrt{1-\alpha^2}$. We obtain $det[\Lambda_{(v\approx V)}]=\theta^2=(1-\alpha^2)=(1-V^2/v^2)$. 

The inverse matrix $\Lambda^{-1}_{(v\approx V)}$ is

\begin{equation}
\displaystyle\Lambda_{(v\approx V)}^{-1}\approx\frac{1}{\sqrt{1-\alpha^2}}
\begin{pmatrix}
1  & 0 \\
0  & 1
\end{pmatrix}=
\displaystyle\theta^{-1}
\begin{pmatrix}
1  & 0 \\
0  & 1
\end{pmatrix},
\end{equation}
where $det[\Lambda_{(v\approx V)}^{-1}]=\theta^{-2}=(1-\alpha^2)^{-1}=(1-V^2/v^2)^{-1}$. 

Both the symmetric matrices in Eq.(79) and Eq.(80) represent deformations, where, for instance, the matrix in Eq.(80) 
leads to a stretching of the space-time interval $ds$ when $v$ is closer to $V$. We realize that such deformations given only 
for much lower energies close to the background frame $S_V$, i.e., the matrices $\Lambda_{v\approx V}=\theta I$ and 
$\Lambda_{v\approx V}^{-1}=\theta^{-1}I$ do not belong to the structure of the Lie group connected to the indentity matrix $I$. 

We conclude that the matrix $\Lambda$ in Eq.(78) already contains effects of deformation ($ds^{\prime}\neq ds$), which become completely evident
for much lower energies ($v\approx V$), where $det(\Lambda)\approx 0$, but, when the speed $v$ increases drastically, i.e., $v>>V$, so, now,
the rotations of Lorentz group are pratically recovered ($det(\Lambda)\approx det(L)=1$) and, thus, we recover the invariance $ds^{\prime}=ds$. With
such approximations, the polar decomposition is practically valid by making $D=\Lambda_{(v\approx V)}$, so that we can verify the
product $LD=L\Lambda_{(v\approx V)}=L(\theta I)\approx\Lambda$, where $\Lambda_{(v\approx V)}$ is a symmetric matrix, which is exactly
the reason for the effects of deformation of $ds$ close to $V$.   

We finally conclude that the set formed by the matrices that appear above does not have a group structure or cannot be considered as a
Poincar\'e's subgroup. This point must be discussed in depth.

Our next step will be to make an investigation of the main effect obtained directly from the violation of the rotation structure at
much lower speeds ($v\approx V$). Such an effect should naturally lead to other deep implications, which will be pointed out, so that 
we will realize that the whole theory contains elements that are connected by a same mathematical and physical structure.

When we make a Lorentz transformation $L(v)$ from the frame $S(v=0)$ to $S^{\prime}$ with speed $v$ with regard to $S$, we have the
well-known ``boost". As the boosts represent rotations, the minimal boost is the identity matrix $L(v=0)=L(0)=I$ connected to the rest
state, such that $L(0)X=X$. However, as such minimal boost does not make sense in this spacetime with a minimum speed that prevents the
rest state, we must stress that the component $\Lambda_{(v\approx V)}(=\theta I)$ in the new transformation 
($\Lambda \approx L\Lambda_{(v\approx V)})$ leads to a non-existence of boosts only in the approximation for much lower energies
($v\rightarrow V$ or  $\alpha\rightarrow 1$), due to the fact that we get $\theta=\sqrt{1-\alpha^2}<<1$. So, only for higher energies 
($v>>V$ or $\alpha\approx 0$), we get $\theta\approx 1$ and, thus, $\Lambda_{(v\approx V)}=\theta I\approx I$, recovering the regime
where the boosts take place (Lorentz group). 

In short, the effects of ``boosts" are generally weakened in this spacetime, mainly in the regime when $v\approx V$, i.e., much closer to the background 
frame $S_V$. So, it is important to stress that, in such special regime, there are no boosts and, therefore, the transformation 
$\theta I$ has another meaning because it does not lead to the change of reference frames in the usual context.
 We will go deeper into this issue.

Actually, the symmetric matrix $\theta I$ (Eq.(79)) is the reason of breaking the structure of rotation group (boosts) and it should be interpreted
just as a scale transformation ($\theta$) that provides a variation of the usual space-time interval ($ds$) in function of speed $v$, 
especially when $v$ is close to $V$. The deep physical implication directly related to the effect of variation of $ds$ will be
investigated below. 

Since the matrix $\theta I$ just deforms the interval $ds$, this transformation does not act for changing reference frames (boosts). 
So, in view of this, we use the following notation to represent such a scale transformation, namely: 

\begin{equation}
 x^{*\mu}=\theta I x^{\mu}, 
\end{equation}
where $x^{*\mu}$ is the deformed vector, $\theta$ being a scale factor, since $\Lambda_{(v\approx V)}x^{\mu}=\theta x^{\mu}=x^{*\mu}$, 
so that we get

 \begin{equation}
ds^{*2}=ds^2(v)=dx^{*\mu}dx^*_{\mu}=\theta^{2}ds^{2}=det[\Lambda_{(v\approx V)}]ds^2, 
 \end{equation}
where $ds^{2}=dx^{\mu}dx_{\mu}$ is the usual squared space-time interval of SR and $ds^{*2}$ is the deformed squared space-time interval
due to new relativistic effects closer to $V$ (no boosts). 

As the usual interval $ds$ does not remain invariant in this spacetime, specially when $v\approx V$; so according to Eq.(82), we realize
that the deformed interval $ds^{*}$ should be the new invariant interval under the change of reference frames in this flat space-time
with the presence of the background frame $S_V$, such that $ds^{*\prime}=ds^{*}$. In doing this, we introduce a new invariance of deformed
intervals in SSR, namely: 

\begin{equation}
ds^{*\prime 2}=ds^{*2}=g_{\mu\nu}dx^{\mu}dx^{\nu}, 
\end{equation}
where $ds^{*2}=(1-\alpha^2)dx^{\mu}dx_{\mu}=\theta^2 ds^2$ and $ds^{*\prime 2}=(1-\alpha^2)dx^{\prime\mu}dx^{\prime}_{\mu}=\theta^2
ds^{\prime 2}$. Of course if we make $V\rightarrow 0$ or $\alpha\rightarrow 0$ in Eq.(83), we recover the invariance of the usual 
(non-deformed) $ds$ of SR, i.e., $ds^{\prime 2}=ds^{2}=g_{\mu\nu}dx^{\mu}dx^{\nu}$. 

Indeed we realize that the deformed interval $ds^{*}=\theta ds=\sqrt{1-V^2/v^2}ds$ remains finite (Eq.83), since, in the limit 
of $\theta\rightarrow 0$ ($v\rightarrow V$), the usual interval $ds$ undergoes a very large stretching, i.e., $ds\rightarrow\infty$.

From Eq.(83), we obtain 

\begin{equation}
ds^{*\prime}=ds^{*}=\sqrt{g_{\mu\nu}dx^{\mu}dx^{\nu}}=\sqrt{c^2dt^2-dx^2}, 
\end{equation}
where, in this case, we have $dy=dz=0$ and $ds^{*\prime}=\sqrt{(1-\alpha^2)dx^{\prime\mu}dx^{\prime}_{\mu}}=\sqrt{(1-\alpha^2)(c^2dt^{\prime2}-dx^{\prime2})}$. Now, if
we make $dx^{\prime}=0$ (or $x^{\prime}=0$, i.e., at the origin of the reference frame $S^{\prime}$), from Eq.(84) we obtain 

\begin{equation}
ds^{*\prime}=cd\tau^{*}=\sqrt{1-\alpha^2}cd\tau=\sqrt{1-\alpha^2}cdt^{\prime}=\sqrt{c^2dt^2-dx^2}, 
\end{equation}
where we have considered $dt^{\prime}=d\tau$ and so $dt^{\prime *}=d\tau^{*}$, with $d\tau^{*}(=ds^{*\prime}/c=\sqrt{g_{\mu\nu}dx^{\mu}dx^{\nu}}/c)$ being the deformed 
proper time interval, where we have $d\tau^{*}=\theta d\tau=\sqrt{1-\alpha^2}d\tau$. This result has a deep physical implication that has origin in the
breakdown of the structure of Lorentz group.

 Now we are ready to investigate the physical implication from Eq.(85). So, by simply making $dx=vdt$ in Eq.(85) and performing the
calculations, we finally obtain

\begin{equation}
 d\tau\sqrt{1-\alpha^2}=dt\sqrt{1-\beta^2}
\end{equation}

and, then

\begin{equation}
 \Delta\tau\sqrt{1-\frac{V^2}{v^2}}=\Delta t\sqrt{1-\frac{v^2}{c^2}},
\end{equation}
where $\Delta\tau$ is the proper time interval and $\Delta t$ is the improper one. Eq.(87) is the immediate physical implication of 
the violation of Lorentz group by means of the symmetric matrix $\theta I$ (Eq.(79)) that deforms the proper time, so that 
we can also write Eq.(87) as being $\sqrt{det(\theta I)}\Delta\tau=\theta\Delta\tau=\Delta t\sqrt{1-v^2/c^2}$, where $\theta=\sqrt{1-V^2/v^2}$. 

It is important to call attention to the fact that Eq.(87) shows us that the proper time interval $\Delta\tau$ depends on speed $v$ and, thus, now
it can also be deformed (dilated) like the improper time interval. So, we realize that Eq.(87) reveals a perfect symmetry in the sense 
that both intervals of time $\Delta t$ and $\Delta\tau$ can dilate, namely $\Delta t$ dilates for $v\rightarrow c$ and, on the other hand,
$\Delta\tau$ dilates for $v\rightarrow V$. But, if we make $V\rightarrow 0$, we break down such new symmetry of SSR and so we recover the 
well-known time equation of SR, where only $\Delta t$ dilates whereas $\Delta\tau$ remains invariant.

From Eq.(87) we notice that, if we make $v=v_0=\sqrt{cV}$ (a geometric average between $c$ and $V$), we find exactly the equality 
$\Delta\tau$ (at $S^{\prime}$)=$\Delta t$ (at $S$), namely this is a newtonian result where the time intervals are the same. Thus we conclude 
that $v_0$ represents a special intermediary speed in SSR ($V<<v_0<<c$) such that, if: 

a) $v>>v_0$ (or $v\rightarrow c$), we get $\Delta\tau<<\Delta t$. This is the well-known {\it improper time dilation}.

b) $v<<v_0$ (or $v\rightarrow V$), we get $\Delta\tau>>\Delta t$. Let us call such a new effect as {\it improper time contraction} or the
{\it dilation of the proper time interval $\Delta\tau$ with respect to the improper time interval $\Delta t$}.

 This new effect (b) becomes
more evident only for $v~(S^{\prime})\rightarrow V~(S_V)$, so that, in this limit, we have $\Delta\tau\rightarrow\infty$ for a certain $\Delta t$ fixed 
as being finite. In other words, this means that the proper time ($S^{\prime}$) can now elapse much faster than the improper one.

 It is interesting to notice that we restore the newtonian regime when $V<<v<<c$, which represents a regime of intermediary speeds, 
 so that we get the newtonian approximation from Eq.(87), i.e., $\Delta\tau\approx\Delta t$.

 Squaring both members of Eq.(87) ($\Delta t=\Psi\Delta\tau=\theta\gamma\Delta\tau$) and manipulating the result, we write Eq.(87) as
follows: 
 
   \begin{equation}
    c^2\Delta\tau^2=\frac{1}{(1-\frac{V^2}{v^2})}[c^2\Delta t^2-v^2\Delta t^2]
   \end{equation}

 By placing Eq.(88) in a differential form and manipulating it, we obtain

   \begin{equation}
   c^2\left(1-\frac{V^2}{v^2}\right)\frac{d\tau^2}{dt^2} + v^2=c^2
   \end{equation}

 Eq.(89) shows us that both speeds related to the marching of time  (``temporal-speed''$v_t=c\sqrt{1-V^2/v^2}d\tau/dt$)
 and the spatial speed $v$ form the vertical and horizontal legs of a rectangular triangle respectively (Fig.5). The hypotenuse of the
 triangle is $c=(v_t^2+v^2)^{1/2}$, which represents the spatio-temporal speed of any particle. If $V\rightarrow 0$ in Eq.(89), we recover
 the well-known time equation of SR written alternatively as $c^2(d\tau^2/dt^2)+v^2=c^2$. 

Looking at Fig.5, now we see clearly three important cases, namely:

 a) If $v\approx c$, $v_t\approx 0$ (the marching of proper time in $S^{\prime}$ is much slower than in $S$), such that 
$\Delta t>>\Delta\tau$, with $\Psi\approx\gamma>>1$, leading to the well-known dilation of the improper time. 

 b) If $v=v_0=\sqrt{cV}$, $v_t=\sqrt{c^2-v_0^2}$, i.e., the marching of time in $S^{\prime}$ is faster, but it is still 
 in an intermediary regime, such that $\Delta t=\Delta\tau$, with $\Psi=\Psi_0=\Psi(v_0)=1$ (newtonian regime). 

 c) If $v\approx V(<<v_0)$, $v_t\approx\sqrt{c^2-V^2}=c\sqrt{1-V^2/c^2}$ (the marching of proper time is even faster), such that 
$\Delta t<<\Delta\tau$, with $\Psi\approx\theta<<1$ (dilation of the proper time). To illustrate this new effect of proper time dilation,
 let us consider a box that contains an ideal gas with $N$ particles in the frame $S$ of a laboratory. Since the minimum speed 
 $V$ has a microscopic origin, then by considering an average speed per particle of the gas (atom or molecule), we should have such average
 speed $v_{rms}$ ($S^{\prime}$) close to $V$ ($S_V$) only when the temperature of the gas is $T\rightarrow 0$K 
($v_{rms}=\sqrt{\left<v\right>^{2}_N}\rightarrow V$). Thus, an imaginary clock in thermal equilibrium with such ultra-cold system 
($S^{\prime}$) should measure a dilated time interval ($\Delta\tau$) with respect to the (shorter) time interval ($\Delta t$) measured 
in the observer's clock (in laboratory $S$), or in other words, we could say that ultra-cold systems ``grow old" more rapidly, contrary 
to higher energies when one grows old more slowly. A great experimental effort should be made in order
to detect the effect of proper time dilation close to $V$ or when $T\rightarrow 0$K. To do that, we can propose an experimental
route to test the hypothesis that the atomic decay rate of a radiative sample with $N_0$ atoms (e.g: $^{137}C$) at $t=0$, 
thermalized at a very low temperature could be increased due to the improper time contraction (observer's clock in laboratory), i.e., 
$N\approx N_0 exp(-\mu\theta^{-1}t)$, where $\theta^{-1}t\approx\tau$, with $\theta=\sqrt{1-V^2/\left<v^2\right>}$, $\left<v^2\right>$ being 
the average squared speed per atom (or molecule) and $\mu$ the atomic decay constant. This topic will be investigated elsewhere. 

\begin{figure}
\includegraphics[scale=0.70]{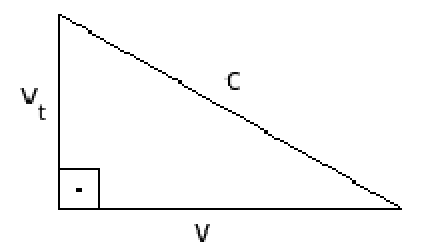}
\caption{We see that the horizontal leg represents the spatial-speed $v$, while the vertical leg represents the temporal-speed $v_t$
(marching of time), where $v_t=\sqrt{c^2-v^2}=c\sqrt{1-v^2/c^2}=c\sqrt{1-V^2/v^2}d\tau/dt$ (see Eq.(89)), so that we always have 
$v^2+v_t^2=c^2$. In SR, when $v=0$, the horizontal leg vanishes (no spatial speed) and so the vertical leg becomes maximum
($v_t=v_{tmax}=c$). However, now according to SSR, due to the existence of a minimum limit of spatial speed ($V$), we can never nullify
the horizontal leg, so that the maximum temporal speed (maximum vertical leg) is $v_{tmax}=\sqrt{c^2-V^2}=c\sqrt{1-V^2/c^2}<c$. On the 
other hand $v_t$ (the vertical leg) cannot be zero since $v=c$ is forbidden for massive particles. So we conclude that the rectangular
triangle is always preserved since both temporal and spatial speeds cannot vanish and, thus, they always coexist. In this sense, we
realize that there is a strong symmetry in SSR.}
\end{figure}

 \subsection{Flat space-time metric with the background frame $S_V$}

From Eq.(83), we obtain 

\begin{equation}
ds^2=\frac{1}{\left(1-\frac{V^2}{v^2}\right)}g_{\mu\nu}dx^{\mu}dx^{\nu},
\end{equation}
where we have $ds^{*2}=\theta^2 ds^2=(1-\alpha^2)ds^2=(1-V^2/v^2)ds^2=g_{\mu\nu}dx^{\mu}dx^{\nu}$ (Eq.(83)). 

Eq.(90) is written as

\begin{equation}
ds^2=\Theta g_{\mu\nu}dx^{\mu}dx^{\nu},
\end{equation}
where $\Theta=\theta^{-2}=\left(1-\frac{V^2}{v^2}\right)^{-1}$. 

The presence of the ultra-referential $S_V$ deforms the Minkowsky metric (Eq.(91)) and works like a uniform background field 
(a non-luminiferous aether) that fills the whole flat space-time as a perfect fluid, playing the role of a kind of de-Sitter (dS)
space-time ($\Lambda>0$)\cite{52}. 

The function $\Theta$ can be understood as being a scale factor that increases for very large wavelengths (cosmological scales)
governed by vacuum (dS), that is to say for much lower energies ($v\rightarrow V$) where we have $\Theta\rightarrow\infty$. In a recent
paper\cite{11}, it was shown that such factor $\Theta(=\theta^{-2})$ breaks strongly the invariance of $ds$ only for very large distances 
governed by vacuum of the ultra-referential $S_V$, leading to the cosmological anti-gravity governed by the tiny positive value of the 
cosmological constant. In this regime of vacuum-$S_V$ ($v\rightarrow V$ or $\Theta\rightarrow\infty$), the interval $ds$ diverges. 

On the other hand, we have $\Theta\rightarrow 1$ for smaller scales of length, namely for higher energies ($v>>V$), 
where dS space-time approximates to the Minkowski metric as a special case, restoring the Lorentz symmetry and the invariance of $ds$. 

We realize that the presence of the background frame $S_V$ deforms the metric $g_{\mu\nu}$ by means of the scale factor $\Theta$, so that
we define a deformed flat metric $G_{\mu\nu}=\Theta g_{\mu\nu}$ that remains a diagonal matrix, but now having $\Theta$ in its diagonal,
namely: $\displaystyle G_{\mu\nu}=\Theta g_{\mu\nu}=
\begin{pmatrix}
\frac{1}{\left(1-\frac{V^2}{v^2}\right)}  & 0  & 0  & 0 \\
  0                           & -\frac{1}{\left(1-\frac{V^2}{v^2}\right)}  & 0   & 0 \\
  0                           & 0  & -\frac{1}{\left(1-\frac{V^2}{v^2}\right)}   & 0 \\
  0                           & 0           & 0     & -\frac{1}{\left(1-\frac{V^2}{v^2}\right)}
\end{pmatrix}$. Therefore, we simply write Eq.(91) as being $ds^2=G_{\mu\nu}dx^{\mu}dx^{\nu}$. If we make $v>>V$, this implies
$\Theta\rightarrow 1$ and, thus, we recover the Minkowsky metric $g_{\mu\nu}$. 

Now we are already able to conclude that there should be the same universal factor $\theta=\sqrt{det(\theta I)}(<1)$ that deforms all
the invariant scalars of SR as, for instance, the space-time interval, i.e., $\theta ds=ds^{*}$, and the proper time interval, i.e., 
$\theta d\tau=d\tau^{*}$. 

In Section 8, we will see that the mass, energy and momentum are also deformed by the same factor $\theta$, i.e., 
$m_{(0,\alpha)}=\theta m_0$, $E=\theta mc^2=\theta\gamma m_0c^2=\Psi m_0c^2$ and $p=\theta\gamma m_0v=\Psi m_0v$. Thus, we already can conclude that all those
invariant quantities of SR and others like the rest mass (Section 8) are abandoned in SSR, since they are modified by the factor $\theta$
due to the presence of the ultra-referential $S_V$ connected to the own invariance of the minimum speed $V$. 

In sum, we should understand that, as the invariance of $c$ leads us to break down the newtonian invariance
of the improper time interval ($\Delta t=\Delta\tau$), by introducing the dilation of the improper time interval 
($\Delta t=\gamma\Delta\tau$), which still preserves the invariance of the proper time and the space-time interval, now with a further step 
towards a new invariance of a minimum speed $V$, we are led to break down such invariant quantities of SR, since the proper time interval
can also dilate by means of the new factor $\theta$, i.e. $\Delta t=\gamma\theta\Delta\tau$. 

\subsection{Light cone and dark cone in SSR-theory}

If we want to represent the front of a light beam sent in any direction of the $xy$ plane simply by rotating the line $x=ct$
($(1+1)D$) around the $t$-axis we get the well-known light cone surface in $(2+1)D$, which is the place of all rays of light, starting from
the origin and located in the $xy$ plane ($2D$). By other side, by rotating an internal line $x=Vt$ ($(1+1)D$) around $t$ axis we get an 
internal (dark) cone surface in $(2+1)D$ as shown in Fig.6. The hypercone $(3+1)D$ is more complicated to show here, whereas the case 
$(2+1)D$ (Fig.6) facilitates our visualization of both cones.  

If we put an observer at the origin of the coordinate system in Fig.6, the origin represents the event $1$, i.e., the event 
{\it here-now} ($x=0$, $y=0$ and $ct=0$). Any event $2$ on the light cone surface is connected to the event $1$ at the origin by
means of a null interval ($\Delta s_{12}^2=0$), because they are connected by a signal of light (light-like vector). For another event $3$
outside the surface of the light cone, we have a space-like vector, where $\Delta s_{13}^2<0$. So we note that a light signal leaving 
the event $1$ cannot reach $3$ before the occurrence of the event $3$, so that there is no causal relationship between them. Such region
outside the light cone is well-known {\it ``elsewhere''}. 

For the case of an event $4$ inside the light cone, by taking into account that we must have the condition $V<v_4<c$, so we have 
a time-like vector, where $\Delta s_{14}^2>0$, which means that $(c\Delta t)^2>(\Delta x^2+\Delta y^2)$, i.e., the part of the time interval
is greater than the spatial part and we say that is the standard time interval. However, we must stress that SSR-theory predicts that,  
even inside the light cone, there should be a forbidden region where a given event $5$ would lead to a dilated time-like interval which
diverges to $+\infty$. This would happen exactly on the surface of the internal cone represented by the 
minimum speed $V$ (Fig.6), so that such hypothetical event $5$ on the surface of the dark cone ($v=V$) would be distantly but still 
temporally connected to event $1$, i.e., we would have $\Delta s_{15}^2\rightarrow +\infty$. In order to show such divergence, we define
the scalar product of quadri-vectors $\vec x$ in SSR, leading to the space-time interval $ds^2$ according to Eq.(90) and Eq.(91), i.e., we
have the squared modulus of the quadri-vector $\vec x$ given by $\vec x.\vec x=s^2=\Theta g_{\mu\nu}x^{\mu}x^{\nu}=\Theta(c^2t^2-x^2-y^2)$, 
where we have $\vec x=(ct,\vec r)=(ct,x,y)$ in this case and $G_{\mu\nu}=\Theta g_{\mu\nu}=(1-V^2/v^2)^{-1}g_{\mu\nu}$, $G_{\mu\nu}$ 
being the $(3\times 3)$-metric in $(2+1)D$ space-time of SSR-theory (Fig.6). 

\begin{figure}
\begin{center}
\includegraphics[scale=0.70]{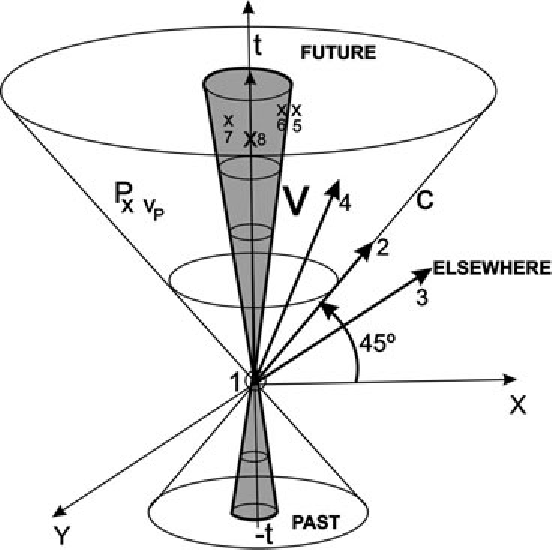}
\end{center}
\caption{The external and internal conical surfaces represent respectively the speed of light $c$ and the unattainable minimum 
speed $V$, which is a definitely prohibited boundary for any particle. For a point $P$ in the world line
 of a particle, in the interior of the two conical surfaces, we obtain a corresponding internal conical surface, 
 such that we must have $V<v_p\leq c$.}
\end{figure}

Basing on $(2+1)D$ space-time interval $\Delta s^2=(1-V^2/v^2)^{-1}[(c\Delta t)^2-\Delta x^2-\Delta y^2]=
\Theta[(c\Delta t)^2-\Delta x^2-\Delta y^2]$ (Fig.6), let us now investigate three important cases that happen with $\Delta s^2$ 
when $v$ is close to $V$ and $v<V$, as follow: 

$a)$ close to $V$ by right ($v>V$, where $v\rightarrow V^{+}$ or slightly outside the dark cone). In this case, we have the event $5$ so that 
$\Delta s^2=\Delta s_{15}^2\rightarrow +\infty$ (Fig.6), since $lim_{v\rightarrow V^{+}}\Theta(v)\rightarrow +\infty$. Let
us call such interval as a {\it very long time-like interval}, which does not violate causality, but it is practically forbidden 
because $V$ is unattainable. 

$b)$ close to $V$ by left ($v<V$, where $v\rightarrow V^{-}$ or slightly inside the dark cone). In this case, we have an event $6$ so that
$\Delta s^2=\Delta s_{16}^2\rightarrow -\infty$ (Fig.6), since $lim_{v\rightarrow V^{-}}\Theta(v)\rightarrow -\infty$. It is interesting to note
that, when we are slightly inside the dark cone, being close to its surface, only the sign of the interval $\Delta s^2$ changes,
but its modulus remains infinite. Let us call such interval as a {\it limitless space-like interval}, which violates strongly causality. Therefore, this
region inside the dark cone is definitly forbbiden, since it plays essentially the role of the space-like intervals, which is similar to 
the extreme case $v\rightarrow\infty(>>c)$, i.e, $\Delta s^2\rightarrow -\infty$. 

Thus we can see that there is an enormous discontinuity between time-like effects slightly outside the dark cone
($\Delta s^2\rightarrow +\infty$) and space-like effects slightly inside the dark cone ($\Delta s^2\rightarrow -\infty$). In other
words, this means that, if we imagine slightly below the fundamental level of vacuum energy density ($S_V$) for the present time of
the universe, causality would be strongly broken. It is also interesting to realize that, for $v<V$ or away from the internal surface
of the dark cone, causality is still broken, but weaker than the result given in (b), namely:

$c)$ For $v<V$ (inside the dark cone), we have an event $7$ (Fig.6), so that $\Delta s^2=\Delta s_{17}^2<0$ (negative, but finite). 
This is an usual space-like interval, also breaking causality in a similar way to $v>c$ ($\Delta s^2<0$ or $(ct)^2<dr^2$).

Although the cases $(b)$ and $(c)$ $(v<V)$ are physically inaccessible due to the causality breaking, we should consider an important 
subcase of $(c)$, namely:

$c_1)$ consider an event $8$ almost over the $t$-axis inside the dark cone (Fig.6), where $v\rightarrow 0$, such that we have
 $lim_{v\rightarrow 0}\Theta(v)\rightarrow 0^-$, leading to $\Delta s^2=\Delta s_{18}^2\rightarrow 0^-$. Although this result reminds us
 a light-like interval ($\Delta s^2=0$), it breaks causality. As we have the interval $\Delta s_{18}\rightarrow\sqrt{0^-}$, which is always an
 imaginary pure number with a very small modulus tending to zero, then let us call it as an {\it imaginary light-like interval} for
representing the absolute rest in the newtonian conception of absolute space, which is prevented by the invariant (absolute) minimum 
speed.  

In Fig.6, it should be noted that, given an event $1$ at the origin $0$ ($t=0$), the cone of light and the dark cone classify all events 
in space-time of SSR-theory in $8$ distinct categories, as follow: 

$(1)$ Events on the surface of the future light cone of $1$ ($t>0$), i.e., like-light intervals; 

$(2)$ Events so close to the surface of the dark cone of a distant future of $1$ ($t>>0$), since $v\rightarrow V^+$; 

$(3)$ Events on the surface of the past light cone of $1$ ($t<0$), i.e., like-light intervals; 

$(4)$ Events so close to the surface of the dark cone of a distant past of $1$ ($t<<0$), since $v\rightarrow V^+$;  

$(5)$ Events inside the future light cone of $1$ ($t>0$), which are those that are affected by material particles emitted in $1$, i.e.,
like-time intervals; 

$(6)$ Events inside the past light cone of $1$ ($t<0$), which are those who may have emitted material particles and so affect what
 happens in $1$, obeying causality, i.e., like-time intervals; 

$(7)$ All other events that are elsewhere of $1$, i.e, all events that never affect or may be affected by $1$ (outside the light cone or $v>c$), 
i.e., like-space intervals; 

$(8)$ Still other events inside the dark cone of $1$, i.e., all events that never affect or may be affected by $1$ ($v<V$). 

According to Fig.6, we conclude that all events that obey causality in SSR must be placed between the dark and light cone, i.e.,
$V<v\leq c$. Here we must stress that there is no particle exactly on the surface of the dark cone, since $v>V$, whereas, on the other hand, massless particles like photons
are on the surface of the light cone, i.e., $v=c$. Such asymmetry shows that the ultra-referential $S_V$ is inaccessible for any 
massive particle with finite mass. 

We also conclude that Fig.6 reduces to the usual (light) cone of SR in $(2+1)D$ if we make $V=0$ ($\Theta=1$), such that the dark cone
would vanish, i.e., we recover $\Delta s^2=(c\Delta t)^2-\Delta x^2-\Delta y^2$ and thus the well-known $5$ distinct categories that
classify all events in the space-time of SR-theory.  

\section{Energy and momentum with the presence of the minimum speed}

Let us firstly define the $4$-velocity in the presence of the background frame $S_V$ connected to the invariant minimum speed $V$, 
as follows:

\begin{equation}
 U^{\mu}=\left[U^{\alpha},U^{4}\right]=\left[\frac{v_{\alpha}\sqrt{1-\frac{V^2}{v^2}}}{c\sqrt{1-\frac{v^2}{c^2}}}~ , 
~\frac{\sqrt{1-\frac{V^2}{v^2}}}{\sqrt{1-\frac{v^2}{c^2}}}\right], 
\end{equation}
where $\mu=1,2,3,4$ and $\alpha=1,2,3$. If $V\rightarrow 0$, we recover the well-known 4-velocity of SR. From Eq.(92), it is interesting 
to observe that the 4-velocity of SSR vanishes in the limit of $v\rightarrow V$ ($S_V$), i.e., $U^{\mu}=(0,0,0,0)$, whereas in SR, 
for $v=0$ we find $U^{\mu}=(1,0,0,0)$.

The $4$-momentum is
\begin{equation}
 p^{\mu}=m_0cU^{\mu},
   \end{equation}
where $U^{\mu}$ is given in Eq.(92). So we find

\begin{equation}
 p^{\mu}=\left[p^{\alpha},p^{4}\right]=\left[\frac{m_0v_{\alpha}\sqrt{1-\frac{V^2}{v^2}}}{\sqrt{1-\frac{v^2}{c^2}}}~ , 
~ \frac{m_0c\sqrt{1-\frac{V^2}{v^2}}}{\sqrt{1-\frac{v^2}{c^2}}}\right],
\end{equation}
where $p^4=E/c$, such that

\begin{equation}
E=cp^4=mc^2=m_0c^2\frac{\sqrt{1-\frac{V^2}{v^2}}}{\sqrt{1-\frac{v^2}{c^2}}},
\end{equation}
where $E$ is the total energy of the particle with speed $v$ in relation to the background reference frame (ultra-referential $S_V$). 
From Eq.(95), we observe that, if $v\rightarrow c\Rightarrow E\rightarrow\infty$. If $v\rightarrow V\Rightarrow E\rightarrow 0$ and, if
$v=v_0=\sqrt{cV}\Rightarrow E=E_0=m_0c^2$ (proper energy in SSR), where we should stress that $m_0c^2$ requires a non-zero motion
 $v(=v_0)$ in relation to $S_V$. Figure 7 shows us the graph for the energy $E$. 

\begin{figure}
\begin{center}
\includegraphics[scale=0.65]{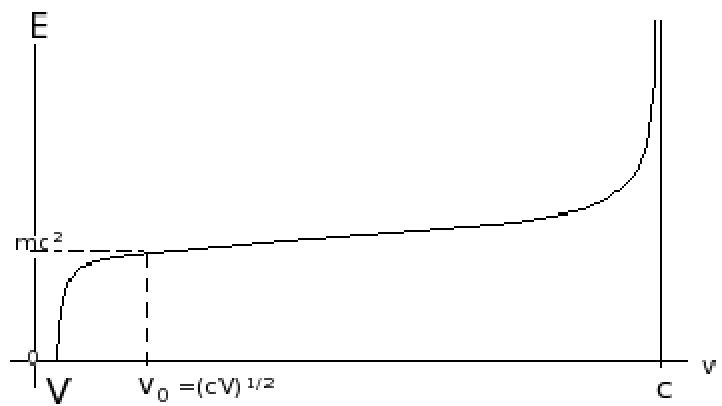}
\end{center}
\caption{$v_0=\sqrt{cV}$ is a speed such that we get the proper energy of the particle ($E_0=m_0c^2$) in SSR, where $\Psi_0=\Psi(v_0)=1$. 
For $v<<v_0$ or closer to $S_V$ ($v\rightarrow V$), a new relativistic correction on energy arises, so that $E\rightarrow 0$. On the other 
hand, for $v>>v_0$, i.e., $v\rightarrow c$, we find $E\rightarrow\infty$.}
\end{figure}

From Eq.(94) we also obtain the $3$(spatial)-momentum, namely:

\begin{equation}
\vec p = m_0\vec v\frac{\sqrt{1-\frac{V^2}{v^2}}}{\sqrt{1-\frac{v^2}{c^2}}},
\end{equation}
where $\vec v=(v_1,v_2,v_3)$.

From Eq.(94), performing the quantity $p^{\mu}p_{\mu}$, we obtain the energy-momentum relation of SSR, as follows:

\begin{equation}
p^{\mu}p_{\mu}=\frac{E^2}{c^2}-\vec p^{2}=m_0^2c^2\left(1-\frac{V^2}{v^2}\right)=(\theta m_0)^2c^2,
\end{equation}
where $\vec p^2=p_1^2+p_2^2+p_3^2$. 

In SR theory represented by the Lorentz group, some elements are preserved under rotations, as for instance, the 
$4$-interval $ds^2(=g_{\mu\nu}dx^{\mu}dx^{\nu})$ and also the rest mass by means of the inner product $p^{\mu}p_{\mu}=m_0^2c^2$,
which is the dispersion relation, where the rest mass is conserved. This means that the rest condition and the rest mass are fundamental
in SR, since they are independent of the state of motion, i.e., we have the well-known indistinguishability of motion and rest. However,
the new dispersion relation given in Eq.(97) shows us that the rest condition does not exist, since now the mass depends on
its preferred state of motion $v$ with respect to the background frame $S_V$ connected to the invariant minimum speed $V$. This is the 
reason why we find the massive term as a function of $\alpha$, i.e., we get $m_0^2c^2(1-\alpha^2)$ in Eq.(97). 

In Eq.(97), when $\alpha\rightarrow 1$ ($v\rightarrow V$), we find $p^{\mu}p_{\mu}\rightarrow 0$, however, we can never nullify 
$p^{\mu}p_{\mu}$, since the minimum speed $V$ is unattainable. Therefore, we can conclude that a certain massive term in the space-time
of SSR has connection with its state of motion with respect to the preferred frame-$S_V$, where, according to Eq.(97), we can write the
effective mass as $m_{(0,\alpha)}=m_0\theta=m_0\sqrt{1-V^2/v^2}$, which does not represent simply the rest mass $m_0$, since we always have
$v>V$. In view of this, we can also write the total energy, as follows: 

\begin{equation}
E=m_{(0,\alpha)}c^2 + K=\gamma m_{(0,\alpha)}c^2=\frac{\sqrt{1-\frac{V^2}{v^2}}}{\sqrt{1-\frac{v^2}{c^2}}}m_0c^2,
\end{equation}
where $K$ is the knetic energy and $E=\gamma m_{(0,\alpha)}c^2=\gamma\theta m_0c^2=\Psi m_0c^2$. So, from Eq.(98) we obtain $K$, namely: 

\begin{equation}
K= m_{(0,\alpha)}c^2(\gamma-1)=m_0c^2\sqrt{1-\frac{V^2}{v^2}}\left(\frac{1}{\sqrt{1-\frac{v^2}{c^2}}}- 1\right), 
\end{equation}

where $K\rightarrow 0$ if $v\rightarrow V$. If $V\rightarrow 0$ in Eq.(99), we recover the relativistic knetic energy, i.e.,
$K=m_0c^2(\gamma-1)$. 

Making an expansion in Eq.(99) and consider the approximation $v<<c$, we find

\begin{equation}
K=m_0c^2\sqrt{1-\frac{V^2}{v^2}}\left(1+\frac{v^2}{2c^2}+...-1\right)\approx\frac{1}{2}\left(m_0\sqrt{1-\frac{V^2}{v^2}}\right)v^2
=\frac{1}{2}m_{(0,\alpha)}v^2, 
\end{equation}
where $m_{(0,\alpha)}=m_0\theta(v)=m_0\sqrt{1-\frac{V^2}{v^2}}$.

Now, also making the approximation $v>>V$ ($\alpha\approx 0$) in Eq.(100), i.e., $m_{(0,\alpha)}\approx m_{(0,0)}=m_0$, we finally obtain
the approximation $V<<v<<c$, so that we find 

\begin{equation}
K\approx\frac{1}{2}m_0v^2,  
\end{equation}
which is the newtonian knetic energy, being recovered only in the approximation for intermediary speeds. 

\subsection{The Lagrangian of a single particle in SSR-theory}

 It is already known that the relativistic Lagrangian for a free particle is

\begin{equation}
 L=-m_0c^2\sqrt{1-\frac{v^2}{c^2}}.
\end{equation}

If we consider a single particle acted on by conservative forces independent of velocity, we have $L=-m_0c^2\sqrt{1-\beta^2}-U$, where
$U$ is the potential depending only upon position. 

It is also known that, if $L$ does not contain the time explicitly, there exists a constant of motion for such a single particle, namely:

\begin{equation}
h=\dot q_j p_j-L=\frac{mv_jv_j}{\sqrt{1-\frac{v^2}{c^2}}}+m_0c^2\sqrt{1-\frac{v^2}{c^2}}+U, 
\end{equation}
which reduces to

\begin{equation}
 h=\frac{m_0c^2}{\sqrt{1-\frac{v^2}{c^2}}}+U=E,
\end{equation}
where the quantity $h$ is thus seen to be the total energy $E=\gamma m_0c^2$, which is therefore a constant of motion under these
conditions. We simply write $v_jv_j=v^2$ for a single particle. 

Now, following the path above in order to obtain firstly the Lagrangian ${\it L}$ for a free particle in SSR, let us start from 
$h=E=\Psi m_0c^2$ to be the constant of motion (total energy) in SSR. In doing this and knowing that $p=\Psi m_0v$ is the momentum
in SSR, we write: 

\begin{equation}
 E=m_0c^2\frac{\sqrt{1-\frac{V^2}{v^2}}}{\sqrt{1-\frac{v^2}{c^2}}}=h=m_0v^2\frac{\sqrt{1-\frac{V^2}{v^2}}}{\sqrt{1-\frac{v^2}{c^2}}}-
{\it L},
\end{equation}
where $v^2=v_jv_j$. So, from Eq.(105), we extract ${\it L}$, namely:

\begin{equation}
 {\it L}=-m_0c^2\theta\sqrt{1-\frac{v^2}{c^2}}=-m_0c^2\sqrt{1-\frac{V^2}{v^2}}\sqrt{1-\frac{v^2}{c^2}}, 
\end{equation}
which recovers the relativistic Lagrangian for a free particle (Eq.(102)) when $V\rightarrow 0$ ($\theta=1$) or $v>>V$. 

Finally, we write the Lagrangian of SSR for a single particle in the presence of a potential $U$, as follows: 

\begin{equation}
 {\it L}=-m_0c^2\sqrt{1-\frac{V^2}{v^2}}\sqrt{1-\frac{v^2}{c^2}}-U.
\end{equation}

Our next step is to consider the case of velocity-dependent potentials that produces no particular difficulty here and can be performed
in exactly the same manner as given for relativistic mechanics. Thus the Lagrangian of SSR for a single particle in an electromagnectic
field is 

\begin{equation}
 {\it L}=-m_0c^2\sqrt{1-\frac{V^2}{v^2}}\sqrt{1-\frac{v^2}{c^2}}-q\phi+\frac{q}{c}\vec A.\vec v. 
\end{equation}

If $V\rightarrow 0$ above, we recover the relativistic Lagrangian of a particle in an electromagnetic field. 

Many specific cases of conservative and non-conservative forces and their equations of motion can be treated within the Lagrangian
formulation of SSR. This extensive topic deserves a detailed investigation elsewhere.

\subsection{Transformations of momentum-energy in the presence of the ultra-referential $S_V$}
 
 By considering the quadri-vector of momentum-energy given in Eq.(94), we have $p^{\mu}=[p^{\alpha}, E/c]$. Since we already have
considered the motion in only one dimension (e.g: $x$), we obtain the vector $[p^{1}, E/c]$, where $p^{1}=p_x$.

Now, as we want to investigate how $p^{\mu}$ transforms in such a spacetime with the presence of the ultra-referential $S_V$, we have
to make those two transformations by using the matrix $\Lambda$ (Eq.(48)) and its inverse $\Lambda^{-1}$ (Eq.(51)). So, by first
considering $\Lambda$, we rewrite 

\begin{equation}
\displaystyle\Lambda=
\begin{pmatrix}
\Psi & -\beta (1-\alpha)\Psi \\
-\beta (1-\alpha)\Psi & \Psi
\end{pmatrix},
\end{equation}
such that the direct matricial transformation $p^{\prime\nu}=\Lambda^{\nu}_{\mu} p^{\mu}$ ($S_V\rightarrow S^{\prime}$) leads to the 
new momentum-energy transformations, as follow: 

\begin{equation} 
 p^{\prime}_x=\Psi\left[p_x-\frac{v(1-\alpha)E}{c^2}\right]=\Psi\left(p_x-\frac{vE}{c^2}+\frac{VE}{c^2}\right),  
\end{equation}
where $p^{\prime}_y=p_y$ and  $p^{\prime}_z=p_z$.

\begin{equation} 
 E^{\prime}=\Psi\left[E-v(1-\alpha)p_x\right]=\Psi\left(E-vp_x+Vp_x\right)
\end{equation}

We know that the inverse matrix (Eq.(51)) that transforms $S^{\prime}\rightarrow S_V$ is 

\begin{equation}
\displaystyle\Lambda^{-1}=
\begin{pmatrix}
\Psi^{\prime} & \beta (1-\alpha)\Psi^{\prime} \\
\beta (1-\alpha)\Psi^{\prime} & \Psi^{\prime}
\end{pmatrix},
\end{equation}
where we find $\Psi^{\prime}=\Psi^{-1}/[1-\beta^2(1-\alpha)^2]$. Thus, the inverse matricial transformation
$p^{\nu}=\Lambda^{-1\nu}_{\mu} p^{\prime\mu}$ ($S^{\prime}\rightarrow S_V$) leads to the following momentum-energy transformations, 
namely: 

\begin{equation} 
 p_x=\Psi^{\prime}\left[p^{\prime}_x+\frac{v(1-\alpha)E^{\prime}}{c^2}\right]=
\Psi^{\prime}\left(p^{\prime}_x+\frac{vE^{\prime}}{c^2}-\frac{VE^{\prime}}{c^2}\right),  
\end{equation}
where $p_y=p^{\prime}_y$ and  $p_z=p^{\prime}_z$.

\begin{equation} 
 E=\Psi^{\prime}\left[E^{\prime}+v(1-\alpha)p^{\prime}_x\right]=\Psi^{\prime}\left(E^{\prime}+vp^{\prime}_x-Vp^{\prime}_x\right)
\end{equation}

The Lorentz transformations of the energy-momentum $p^{\mu}$ are simply recovered if we make $V=0$.  

\section{Transformations of the electromagnetic field in SSR-theory}

\subsection{The case $(1+1)D$}

   According to the relativistic electrodynamics in SR, it is well-known that the electromagnetic fields represented by the tensor 
$F_{\mu\nu}$ are transformed by the change of reference frames, namely it is a similarity transformation, i.e., we write
$F^{\prime}_{\mu\nu}=L_{\mu\sigma}F_{\sigma\rho}L^{T}_{\rho\nu}$, where $L^{T}_{\rho\nu}=L_{\nu\rho}$, or, in the matricial form, we write
$F^{\prime}=LFL^{T}$, $L$ being the Lorentz matrix. $F^{\prime}$ is given at the reference frame $S^{\prime}$ that
moves with speed $v$ in relation to the frame $S$ at rest, where we have $F$. So, here we have a direct field transformation, i.e.,
a transformation from $S$ to $S^{\prime}$. 

 The electromagnetic field tensor $F_{\mu\nu}$ is given as follows:

\begin{equation}
\displaystyle F_{\mu\nu}=
\begin{pmatrix}
0 & E_x/c & E_y/c & E_z/c\\
-E_x/c  & 0 & -B_z &  B_y\\
-E_y/c   & B_z & 0 & -B_x\\
-E_z/c  & -B_y & B_x & 0 
\end{pmatrix}
\end{equation}
    
Now, as we want to know how the fields $F_{\mu\nu}$ are transformed in $(1+1)D$ space-time of SSR, where we consider the motion only
in a certain axis (e.g: $x$-axis), we replace the Lorentz matrix by the non-orthogonal matrix $\Lambda$ (Eq.(47)) in order to obtain a new 
similarity transformation, where the galilean reference frame $S$ at rest must be replaced by the background reference frame, i.e.,
the ultra-referential $S_V$ connected to the unattainable minimum speed $V$. So we write the following transformation:

\begin{equation}
F^{\prime}=\Lambda F\Lambda^{T}, 
\end{equation}
where $F$ is given at $S_V$ and $F^{\prime}$ is given to a particle at $S^{\prime}$ moving in the $x$-axis with speed $v$ in relation
to $S_V$. 

From Eq.(116), after performing the products of matrices, we extract six field transformations in the presence of the background
frame $S_V$, namely: 

\begin{equation}
 E_y^{\prime}=\Psi[E_y-\beta c(1-\alpha)B_z]=\Psi(E_y-\beta c B_z+\xi c B_z), 
\end{equation}

\begin{equation}
 B_y^{\prime}=\Psi\left[B_y+\frac{\beta}{c}(1-\alpha)E_z\right]=\Psi\left(B_y+\frac{\beta}{c}E_z-\frac{\xi}{c} E_z\right);  
\end{equation}

\begin{equation}
 E_z^{\prime}=\Psi[E_z+\beta c(1-\alpha)B_y]=\Psi(E_z+\beta c B_y-\xi c B_y), 
\end{equation}

\begin{equation}
 B_z^{\prime}=\Psi\left[B_z-\frac{\beta}{c}(1-\alpha)E_y\right]=\Psi\left(B_z-\frac{\beta}{c}E_y+\frac{\xi}{c} E_y\right); 
\end{equation}

\begin{equation}
 E_x^{\prime}=\Psi^{2}[1-\beta^2(1-\alpha)^2]E_x
\end{equation}

and

\begin{equation}
 B_x^{\prime}=B_x, 
\end{equation}

where the velocity is $\vec v=v{\bf x}$, $v=\beta c$ being the speed with respect to $S_V$. We have $V=\xi c$. 

If we make $V\rightarrow 0$ in the transformations above, this implies $\alpha\rightarrow 0$ or $\xi\rightarrow 0$ and the ultra-referential $S_V$ is simply replaced by the galilean frame
$S$, thus leading to $\Psi\rightarrow\gamma$, with $v$ given now with respect to $S$, so that we recover the Lorentz transformations for 
the electromagnetic fields, namely: $E_y^{\prime}=\gamma(E_y-vB_z)$, $B_y^{\prime}=\gamma[B_y+(v/c^2)E_z]$, 
$E_z^{\prime}=\gamma(E_z+vB_y)$, $B_z^{\prime}=\gamma[B_z-(v/c^2)E_y]$, $E_x^{\prime}=E_x$ and $B_x^{\prime}=B_x$.

  As the motion is given in the direction of the $x$-axis, it is easy to understand that the fields $E_y$ (or $E_y^{\prime}$) and 
$E_z$ (or $E_z^{\prime}$) are in the directions perpendicular to the direction of motion ($x$), so that we can write
$\vec E_{\perp}=\vec E_y+\vec E_z$ (or $\vec E^{\prime}_{\perp}=\vec E^{\prime}_y+\vec E^{\prime}_z$). The same reasoning is used for
the magnetic fields, i.e., $\vec B_{\perp}=\vec B_y+\vec B_z$ (or $\vec B^{\prime}_{\perp}=\vec B^{\prime}_y+\vec B^{\prime}_z$).

Thus, if we sum Eq.(117) with Eq.(119) in their vectorial forms, we can put them into a compact form, as follows: 

\begin{equation}
\vec E_{\perp}^{\prime}=\Psi(\vec E_{\perp}+\vec v\times\vec B_{\perp}-\vec V\times\vec B_{\perp}),
\end{equation}
where $c\vec\xi=\vec V=V{\bf x}$. 

Now, if we sum Eq.(118) with Eq.(120) in their vectorial forms, we find 

\begin{equation}
\vec B_{\perp}^{\prime}=\Psi\left(\vec B_{\perp}-\frac{\vec v}{c^2}\times\vec E_{\perp}+\frac{\vec V}{c^2}\times\vec E_{\perp}\right). 
\end{equation}

We also obtain

\begin{equation}
 \vec E_{\parallel}^{\prime}=\Psi^{2}\left[1-\frac{v^2}{c^2}\left(1-\frac{V}{v}\right)^2\right]\vec E_{\parallel}
\end{equation}

and

\begin{equation}
 \vec B_{\parallel}^{\prime}=\vec B_{\parallel}. 
\end{equation}

If we make $V=0$ ($\alpha=0$ and $\xi=0$), we recover the Lorentz direct transformations
for the fields, i.e., $\vec E_{\perp}^{\prime}=\gamma(\vec E_{\perp}+\vec v\times\vec B_{\perp})$,
$\vec B_{\perp}^{\prime}=\gamma\left(\vec B_{\perp}-\frac{\vec v}{c^2}\times\vec E_{\perp}\right)$, 
$\vec E_{\parallel}^{\prime}=\vec E_{\parallel}$ and $\vec B_{\parallel}^{\prime}=\vec B_{\parallel}$. 

It is important to note that the transformations in Eq.(123) and Eq.(124) pose the electric and magnetic fields in a perfect equal-footing,
in the sense that one cannot find any reference frame in which the magnetic field of the particle cancels, since the minimum speed 
prevents the rest state. This impossibility to find only an electric field is revealed under the additional term
``$\vec V\times\vec B_{\perp}$'' in Eq.(123), which can never be cancelled, since we always find $v>V$. Thus, we also conclude that the
existence of such a weak background field ($\rho^{(2)}_{em}$) of gravitational origin leads to an unbreakable coexistence of the electric
and magnetic fields, in such a way that the invariant minimum speed emerges from this new aspect of the gravitation at low energy scales
and prevents to cancel any magnetic field of the moving charge by any change of reference frames in SSR.

In summary, we say that the breakdown of the Lorentz symmetry due to the presence of a background field at the ultra-referential $S_V$ further 
enhances the symmetry between electric and magnetic fields of the charged particles moving in gravitational fields.

 As it is already known that the matrix $\Lambda$ is a non-orthogonal matrix, i.e., $\Lambda^{-1}\neq\Lambda^{T}$, we have obtained
$\Lambda^{-1}$ [Eq.(51)]. Thus, as we want to obtain the inverse transformation of $F$, i.e., from $F^{\prime}$ ($S^{\prime}$) 
to $F$ ($S_V$), we write

 \begin{equation}
 F=\Lambda^{-1}F^{\prime}(\Lambda^{-1})^{T},
 \end{equation}
where, for the special case of the Lorentz matrix $L$, such that $L^{-1}=L^{T}$, then we write the inverse transformation 
above as being simply $F=L^{-1}F^{\prime}(L^{-1})^{T}= L^{T}F^{\prime}L$, since $L$ is orthogonal and represents a rotation of $4$-vector
in space-time of SR; however, the matrix $\Lambda$ is not a rotation matrix due to the presence of the background field ($S_V$) 
that breaks the Lorentz symmetry, i.e., $\Lambda^T\neq\Lambda^{-1}$. 

 After performing the products of matrices in Eq.(127), we obtain six inverse transformations, namely: 

\begin{equation}
 E_y=\Psi^{\prime}[E_y^{\prime}+\beta c(1-\alpha)B_z^{\prime}]=\Psi^{\prime}(E_y^{\prime}+\beta c B_z^{\prime}-\xi c B_z^{\prime}), 
\end{equation}

\begin{equation}
 B_y=\Psi^{\prime}\left[B_y^{\prime}-\frac{\beta}{c}\left(1-\alpha\right)E_z^{\prime}\right]=
\Psi^{\prime}\left(B_y^{\prime}-\frac{\beta}{c} E_z^{\prime}+\frac{\xi}{c} E_z^{\prime}\right);  
\end{equation}

\begin{equation}
 E_z=\Psi^{\prime}[E_z^{\prime}-\beta c(1-\alpha)B_y^{\prime}]=\Psi^{\prime}(E_z^{\prime}-\beta c B_y^{\prime}+\xi c B_y^{\prime}), 
\end{equation}

\begin{equation}
 B_z=\Psi^{\prime}\left[B_z^{\prime}+\frac{\beta}{c}\left(1-\alpha\right)E_y^{\prime}\right]=
\Psi^{\prime}\left(B_z^{\prime}+\frac{\beta}{c}E_y^{\prime}-\frac{\xi}{c} E_y^{\prime}\right); 
\end{equation}

\begin{equation}
 E_x=\frac{\Psi^{-2}}{1-\beta^2(1-\alpha)^2}E_x^{\prime}
\end{equation}

and

\begin{equation}
 B_x=B_x^{\prime}, 
\end{equation}
where we already have obtained $\Psi^{\prime}=\frac{\Psi^{-1}}{1-\beta^2(1-\alpha)^2}$. We find $\Psi^{\prime}\neq\Psi$ due to the
breakdown of the Lorentz symmetry (rotation) in the presence of the ultra-referential $S_V$, since the matrix $\Lambda$ is non-orthogonal. 

Finally, by adding Eq.(128) with Eq.(130) in their vectorial forms and also adding Eq.(129) with Eq.(131), then the six inverse 
transformations above are reduced to four transformations, namely: 

\begin{equation}
\vec E_{\perp}=\Psi^{\prime}(\vec E_{\perp}^{\prime}-\vec v\times\vec B_{\perp}^{\prime}
+\vec V\times\vec B_{\perp}^{\prime}),
\end{equation}

\begin{equation}
\vec B_{\perp}=\Psi^{\prime}\left(\vec B_{\perp}^{\prime}+\frac{\vec v}{c^2}\times\vec E_{\perp}^{\prime}
-\frac{\vec V}{c^2}\times\vec E_{\perp}^{\prime}\right); 
\end{equation}

\begin{equation}
\vec E_{\parallel}=\frac{\Psi^{-2}}{1-\frac{v^2}{c^2}\left(1-\frac{V}{v}\right)^2}\vec E_{\parallel}^{\prime}
\end{equation}

and

\begin{equation}
\vec B_{\parallel}=\vec B_{\parallel}^{\prime}. 
\end{equation}

If we make $V=0$ ($\alpha=0$ and $\xi=0$), we recover the Lorentz inverse transformations
for the fields, i.e., $\vec E_{\perp}=\gamma(\vec E_{\perp}^{\prime}-\vec v\times\vec B_{\perp}^{\prime})$,
$\vec B_{\perp}=\gamma\left(\vec B_{\perp}^{\prime}+\frac{\vec v}{c^2}\times\vec E_{\perp}^{\prime}\right)$, 
$\vec E_{\parallel}=\vec E_{\parallel}^{\prime}$ and $\vec B_{\parallel}=\vec B_{\parallel}^{\prime}$, with $\gamma^{\prime}=\gamma$,
since the Lorentz matrix $L$ is orthogonal, so that $\gamma$ is preserved under the inverse transformations. 

\subsection{The extended case $(1+1+1)D$, i.e., the case $(x,x_{\perp},t)$}

It is important to note that, when we try to reduce the general matrix of transformation (Eq.(60)) to the case of one dimensional 
motion in $x$-direction (Eq.(47)) by simply making $\beta_y=0$ and $\beta_z=0$ in Eq.(60), we find the following matrix: 

\begin{equation}
\displaystyle\Lambda= 
\begin{pmatrix}
\theta\gamma & -\theta\gamma\beta_* & 0 & 0\\
-\theta\gamma\beta_* & \theta\gamma & 0 &  0\\
  0 & 0 & \theta & 0\\
 0  & 0 & 0 & \theta
\end{pmatrix},
\end{equation}
where $\beta{_x*}=\beta_*$, since $v_x=v$ ($\beta{_x}=\beta$). However, we see that the matrix in Eq.(138) is not the same matrix 
in Eq.(47), because we get the new components $\Lambda_{(3\times 3)}=\Lambda_{(4\times 4)}=\theta$ (Eq.(138)) instead of simply
$\Lambda_{(3\times 3)}=\Lambda_{(4\times 4)}=1$ (Eq.(47)). This means that any space-time transformation in SSR-theory must
preserve the existence of the minimum speed $V$ at any direction of space, as $V$ is invariant, thus leading to the presence of a 
transversal dimension $x_{\perp}$ that cannot be neglected, although the case $(1+1)D$ has been considered before. 

 The presence of $\theta$ as being the components $\Lambda_{(3\times 3)}$ and $\Lambda_{(4\times 4)}$ in the diagonal of the 
matrix in Eq.(138) shows that the transverse directions are transformed as follow: $y^{\prime}=\theta y$, $z^{\prime}=\theta z$, 
or then $x^{\prime}_{\perp}=\theta x_{\perp}$ (Section 6). Such transformations do not represent boosts (translational
motion), since we already know that $\theta$ works like a kind of scale factor that appears due to the presence of the background field 
(vacuum) connected to the ultra-referential $S_V$, breaking the Lorentz symmetry (Section 7). 

In fact, Lorentz symmetry is recovered only if $\theta\rightarrow 1$ ($V\rightarrow 0$) or even when $v>>V$, which means that particles with higher 
speeds do not ``feel'' the effects of the background field, so that their motions are practically reduced to one dimension ($x$), i.e., 
we have $x^{\prime}_{\perp}\approx x_{\perp}$. However, if $v\rightarrow V$ ($\theta^{-1}\rightarrow\infty$), a particle would ``feel'' strongly
the quantum effects of the background field. 

Actually, since there is no boost in the transverse direction of motion, it is natural to think that a particle (e.g: electron) has a
rotational (circulatory) motion with a radius $x_{0\perp}$ around the $x$-axis of the translational motion ($v_x=v$), so that these 
two combined motions lead to a helical motion or a kind of {\it ``Zitterbewegung''} (trembling motion)\cite{53}\cite{54}. Therefore,
we can conclude that such a trembling motion is the result of fluctuations on the position in the transverse directions (plane $yz$), thus being essentially 
quantum effects due to the presence of the background field connected to the minimum speed $V$ having gravitational origin 
($V\propto G^{1/2}$). So, it is interesting to realize that SSR must be built on a quasi-flat space-time such that a weak 
gravity naturally arises from the minimum speed $V$ in order to provide a fundamental explanation for the trembling motion. In 
other words, we say that the {\it ``Zitterbewegung''} ({\it zbw}) investigated by Schroedinger {\it et al} \cite{55}\cite{56} 
could have origin in a new scenario of quantum-gravity based on SSR-theory already used to provide the foundation for the uncertainty
principle\cite{12}. This open question initially investigated in ref.\cite{12} should be deeper explored elsewhere, where we will search 
for new connections between SSR and QM. 

In order to obtain the transformations of the electromagnetic fields $F_{\mu\nu}$ (Eq.(115)) based on the extended space-time $(1+1+1)D$
of SSR represented by the matrix of transformation $\Lambda$ in Eq.(138), let us first consider the direct transformations  
$F^{\prime}=\Lambda F\Lambda^T$, from where, by taking into account Eq.(115) and Eq.(138) and after performing the calculations, we 
find 

\begin{equation}
 E_y^{\prime}=\theta\Psi[E_y-\beta c(1-\alpha)B_z]=\theta\Psi(E_y-\beta c B_z+\xi c B_z), 
\end{equation}

\begin{equation}
 B_y^{\prime}=\theta\Psi\left[B_y+\frac{\beta}{c}(1-\alpha)E_z\right]=\theta\Psi\left(B_y+\frac{\beta}{c}E_z-\frac{\xi}{c} E_z\right);  
\end{equation}

\begin{equation}
 E_z^{\prime}=\theta\Psi[E_z+\beta c(1-\alpha)B_y]=\theta\Psi(E_z+\beta c B_y-\xi c B_y), 
\end{equation}

\begin{equation}
 B_z^{\prime}=\theta\Psi\left[B_z-\frac{\beta}{c}(1-\alpha)E_y\right]=\theta\Psi\left(B_z-\frac{\beta}{c}E_y+\frac{\xi}{c} E_y\right); 
\end{equation}

\begin{equation}
 E_x^{\prime}=\Psi^{2}[1-\beta^2(1-\alpha)^2]E_x
\end{equation}

and

\begin{equation}
 B_x^{\prime}=\theta^2 B_x 
\end{equation}

If we simply make $\theta=1$, we recover the direct field transformations in the simple case $(1+1)D$ of SSR.

Here, it is interesting to note that the transformation in Eq.(144) given to the magnetic field in the direction of motion shows exactly
the presence of the transverse scale factor ($\theta^2$) for representing the trembling motion that affects directly such magnetic field, 
so that this is consistent with the idea that such a transverse motion could be thought of as being a circulatory motion of the charged 
particle around the $x$-axis with a radius $x_{0\perp}$, creating a magnetic moment in the direction of motion ($x$), which seems 
to have a connection with the electron spin, 
reminding us the interpretation of Schroedinger about the free-particle wave packet solutions of the Dirac equation, where the {\it zbw} 
is attributed to ``interference'' between positive and negative energy states oscillating with a circular frequency
$w_0=2m_0c^2/\hbar=1.6\times 10^{21}s^{-1}$\cite{56}. But then how could the {\it zbw} be the origin of spin, which is surely significant in 
the non-relativistic domain? This paradox was resolved by Hestenes\cite{56}, however SSR aims to go depper into this question, since it 
goes beyond the non-relativistic domain, i.e., close to the vacuum domain of gravitational origin with very low energies 
($v\approx V$ or $\theta\approx 0$). A mathematical and physical description for the connection between SSR, {\it zbw} and perhaps the
electron spin at all scales of energy ($v>V$), not only the relativistic domain, should be deeply investigated in a future research 
denominated as a new interpretation of {\it zbw} due to the presence of the vacuum energy $S_V$ of gravitational origin.

 Adding Eq.(139) with Eq.(141) in their vectorial forms and also adding Eq.(140) with Eq.(142), then the six direct transformations
(Eq.(139) to Eq.(144)) are reduced to four transformations, namely: 

\begin{equation}
\vec E_{\perp}^{\prime}=\theta\Psi(\vec E_{\perp}+\vec v\times\vec B_{\perp}-\vec V\times\vec B_{\perp}), 
\end{equation}

\begin{equation}
\vec B_{\perp}^{\prime}=\theta\Psi\left(\vec B_{\perp}-\frac{\vec v}{c^2}\times\vec E_{\perp}+\frac{\vec V}{c^2}\times\vec E_{\perp}\right); 
\end{equation}

\begin{equation}
\vec E_{\parallel}^{\prime}=\Psi^{2}\left[1-\frac{v^2}{c^2}\left(1-\frac{V}{v}\right)^2\right]\vec E_{\parallel}
\end{equation}

and

\begin{equation}
\vec B_{\parallel}^{\prime}=\theta^2\vec B_{\parallel}. 
\end{equation}

Now, in order to find the inverse transformations for the electromagnetic field $F_{\mu\nu}$ in this case, we first need to calculate 
the inverse of the matrix given in Eq.(138), or simply making $\beta_y=0$ and $\beta_z=0$ with $\beta_x=\beta$ in Eq.(68). In doing this,
we obtain 

\begin{equation}
\displaystyle\Lambda^{-1}_{(4\times 4)}= 
\begin{pmatrix}
\frac{\theta^{-1}\gamma^{-1}}{1-\beta_*^2} & -\left(\frac{\theta^{-1}\gamma^{-1}}{1-\beta_*^2}\right)\beta_* & 0 & 0\\
-\left(\frac{\theta^{-1}\gamma^{-1}}{1-\beta_*^2}\right)\beta_* & \frac{\theta^{-1}\gamma^{-1}}{1-\beta_*^2} & 0 &  0\\
  0 & 0 & \theta^{-1} & 0\\
 0  & 0 & 0 & \theta^{-1}
\end{pmatrix},
\end{equation}
where $\frac{\theta^{-1}\gamma^{-1}}{1-\beta_*^2}=\frac{\Psi^{-1}}{1-\beta^2(1-\alpha^2)}=\Psi^{\prime}$, knowing that 
$\Lambda^{-1}_{(4\times 4)}\neq\Lambda_{(4\times 4)}^T$. Thus, performing the inverse similarity transformation 
 $F=\Lambda^{-1}F^{\prime}(\Lambda^{-1})^{T}$ (Eq.(127)), by taking into account Eq.(149), we find

\begin{equation}
 E_y=\theta^{-1}\Psi^{\prime}[E_y^{\prime}+\beta c(1-\alpha)B_z^{\prime}]=\theta^{-1}\Psi^{\prime}(E_y^{\prime}+\beta c B_z^{\prime}-\xi c B_z^{\prime}), 
\end{equation}

\begin{equation}
 B_y=\theta^{-1}\Psi^{\prime}\left[B_y^{\prime}-\frac{\beta}{c}(1-\alpha)E_z^{\prime}\right]=\theta^{-1}\Psi^{\prime}\left(B_y^{\prime}-\frac{\beta}{c}E_z^{\prime}+\frac{\xi}{c} E_z^{\prime}\right);  
\end{equation}

\begin{equation}
 E_z=\theta^{-1}\Psi^{\prime}[E_z^{\prime}-\beta c(1-\alpha)B_y^{\prime}]=\theta^{-1}\Psi^{\prime}(E_z^{\prime}-\beta c B_y^{\prime}+\xi c B_y^{\prime}), 
\end{equation}

\begin{equation}
 B_z=\theta^{-1}\Psi^{\prime}\left[B_z^{\prime}+\frac{\beta}{c}(1-\alpha)E_y^{\prime}\right]=\theta^{-1}\Psi^{\prime}\left(B_z^{\prime}+\frac{\beta}{c}E_y^{\prime}-\frac{\xi}{c} E_y^{\prime}\right); 
\end{equation}

\begin{equation}
 E_x=\frac{\Psi^{-2}}{1-\beta^2(1-\alpha)^2}E_x^{\prime}
\end{equation}

and

\begin{equation}
 B_x=\theta^{-2} B_x^{\prime} 
\end{equation}

Finally, by adding Eq.(150) with Eq.(152) in their vectorial forms and also adding Eq.(151) with Eq.(153), then the six inverse
transformations (Eq.(150) to Eq.(155)) are reduced to four transformations as follow: 

\begin{equation}
\vec E_{\perp}=\theta^{-1}\Psi^{\prime}(\vec E_{\perp}^{\prime}-\vec v\times\vec B_{\perp}^{\prime}+\vec V\times\vec B_{\perp}^{\prime}), 
\end{equation}

\begin{equation}
\vec B_{\perp}=\theta^{-1}\Psi^{\prime}\left(\vec B_{\perp}^{\prime}+\frac{\vec v}{c^2}\times\vec E_{\perp}^{\prime}-\frac{\vec V}{c^2}\times\vec E_{\perp}^{\prime}\right); 
\end{equation}

\begin{equation}
\vec E_{\parallel}=\frac{\Psi^{-2}}{1-\frac{v^2}{c^2}\left(1-\frac{V}{v}\right)^2}\vec E_{\parallel}^{\prime}
\end{equation}

and

\begin{equation}
\vec B_{\parallel}=\theta^{-2}\vec B_{\parallel}^{\prime}=\Theta\vec B_{\parallel}^{\prime},
\end{equation}
where we already know that $\Theta=\theta^{-2}=\left(1-\frac{V^2}{v^2}\right)^{-1}$, which is exactly the component that appears in the
diagonal of the metric $G_{\mu\nu}=\Theta g_{\mu\nu}$ (Section 7). 

Here it is important to call attention to the fact that the transformations of momentum-energy made for the case $(1+1)D$ (subsection B
of Section 8) should be also extended to this case $(1+1+1)D$. To do that, we
use the matrix $\Lambda$ in Eq.(138) and its inverse $\Lambda^{-1}$ in Eq.(149) to calculate the direct transformation 
$p^{\prime\nu}=\Lambda^{\nu}_{\mu} p^{\mu}$ and its inverse $p^{\nu}=\Lambda^{-1\nu}_{\mu} p^{\prime\mu}$, so that
we find the same transformations already obtained in Eq.(110), Eq.(111) and Eq.(113), Eq.(114) respectively, except the transformations
of momentum in the transverse directions like the $y$-axis and $z$-axis, where we obtain $p_y^{\prime}=\theta p_y$, 
$p_z^{\prime}=\theta p_z$ and $p_y=\theta^{-1} p_y^{\prime}$, $p_z=\theta^{-1} p_z^{\prime}$. 
 
As the transformations of the electromagnetic field in $(1+1+1)D$ already contain all the basic ingredients of SSR due to the presence
of the {\it zbw}, which is missed in the simple case $(1+1)D$, the transformations in $(3+1)D$ that include boosts in $y$-axis ($\beta_y$)
and in $z$-axis ($\beta_z$) do not reveal relevant physical effects beyond the {\it zbw}. In view of this, the case $(3+1)D$ will be 
considered elsewhere. 

\subsection{Covariance of Maxwell wave equation in the presence of the ultra-referential $S_V$}

Let us assume a light ray emitted from the frame $S^{\prime}$. Its equation of electrical wave at this reference frame is

\begin{equation}
\frac{\partial^2\vec E(x^{\prime},t^{\prime})}{\partial x^{\prime 2}}-
\frac{1}{c^2}\frac{\partial^2\vec E(x^{\prime},t^{\prime})}
{\partial t^{\prime 2}}=0
\end{equation}

 As it is already known, when we make the exchange by conjugation on the
spatial and temporal coordinates, we obtain respectively the following
operators: $X\rightarrow\partial/\partial t$ and $t\rightarrow\partial/\partial X$;
also $x^{\prime}\rightarrow\partial/\partial t^{\prime}$ and $t^{\prime}\rightarrow\partial/
\partial x^{\prime}$. Thus the transformations in Eq.(45) and Eq.(46) for such differential operators are

\begin{equation}
\frac{\partial}{\partial t^{\prime}}=
\Psi\left[\frac{\partial}{\partial t}-\beta
 c(1-\alpha)\frac{\partial}{\partial X}\right],
\end{equation}

\begin{equation}
\frac{\partial}{\partial x^{\prime}}=
\Psi\left[\frac{\partial}{\partial X}-\frac{\beta}{c}(1-\alpha)\frac{\partial}{\partial t}\right],
\end{equation}
where $\beta =v/c$ and $\alpha=V/v$ (Fig.2).

By squaring Eq.(161) and Eq.(162), inserting into Eq.(160) and after performing the calculations, we will finally obtain

\begin{equation}
det\Lambda\left(\frac{\partial^2\vec E}{\partial X^2}
-\frac{1}{c^2}\frac{\partial^2\vec E}{\partial t^2}\right)=0,
\end{equation}
where $det\Lambda=\Psi^2[1-\beta^2(1-\alpha)^2]$. 

  As the ultra-referential $S_V$ is definitely inaccessible for any particle, we always have
$\alpha<1$ (or $v>V$), which always implies $det\Lambda=\Psi^2[1-\beta^2(1-\alpha)^2]>0$. And as we
already have shown in Section 5, this result is in agreement with the fact that we must
have $det\Lambda>0$. Therefore this will always assure

  \begin{equation}
 \frac{\partial^2\vec E}{\partial X^2}
-\frac{1}{c^2}\frac{\partial^2\vec E}{\partial t^2}=0
 \end{equation}

By comparing Eq.(164) with Eq.(160), we verify the covariance of the equation
of the electromagnetic wave propagating in the background field of the ultra-referential $S_V$. So, indeed we can conclude that the
space-time transformations in SSR also preserve the covariance of the Maxwell equations in vacuum as well
as the Lorentz transformations. This leads us to think that $S_V$ works like a relativistic background field (a non-galilean ``ether''),
which is compatible with electromagnetism, other than it was the Galilean (luminiferous) ether of pre-einsteinian physics, breaking the
covariance of the Maxwell equations under the change of reference frames.

\section{Cosmological implications of the invariant mimimum speed $V$: the cosmological constant and vacuum energy density}

\subsection{Energy-momentum tensor in the presence of the ultra-referential-$S_V$}

  Let us rewrite the 4-velocity (Eq.(92)) in the following alternative way: 
   \begin{equation}
 U^{\mu}=\left[U^0, U^{\alpha}\right]=\left[\frac{\sqrt{1-\frac{V^2}{v^2}}}{\sqrt{1-\frac{v^2}{c^2}}}~ , ~
\frac{v_{\alpha}\sqrt{1-\frac{V^2}{v^2}}}{c\sqrt{1-\frac{v^2}{c^2}}}\right], 
   \end{equation}
where now we have $\mu=0,1,2,3$ and $\alpha=1,2,3$. If $V\rightarrow 0$, we recover the $4$-velocity of SR.

The well-known energy-momentum tensor to deal with perfect fluid is of the form
   \begin{equation}
  T^{\mu\nu}=(p+\epsilon)U^{\mu}U^{\nu} - pg^{\mu\nu},
   \end{equation}
where $U^{\mu}$ is given in Eq.(165). Here it represents the quadri-velocity of the fluid, $p$ represents its pressure 
and $\epsilon$ is its energy density.

As spacetime has been treated as a superfluid (``aether'')\cite{57} in order to go beyond GR-theory by including quantum gravity effects, let
us use the energy-momentum tensor for a perfect fluid (Eq.(166)) combined with the quadri-velocity of SSR (Eq.(165)) in order
to obtain the cosmological constant $\Lambda_0$ as the effect of the vacuum (``aether'') $S_V$ in the limit of lower energies 
($v\rightarrow V$). 

From Eq.(165) and Eq.(166), by performing the new component $T^{00}$, we obtain
   \begin{equation}
  T^{00}=\frac{\epsilon(1-\frac{V^2}{v^2})+p(\frac{v^2}{c^2}-\frac{V^2}{v^2})}{(1-\frac{v^2}{c^2})}
   \end{equation}

 If $V\rightarrow 0$, we recover the old component $T^{00}$.

Now, as we are interested only in obtaining $T^{00}$ in absence of matter, i.e.,
 the vacuum limit (lower energies) connected to the ultra-referential $S_V$, we perform the limit of Eq.(167) as follows:
 \begin{equation}
 lim_{v\rightarrow V} T^{00}= T^{00}_{vacuum}=\frac{p(\frac{V^2}{c^2}-1)}{(1-\frac{V^2}{c^2})}= -p.
  \end{equation}

 From Eq.(167), we notice that the term $\epsilon\gamma^2(1-V^2/v^2)$ for representing matter naturally vanishes
 in the limit of vacuum-$S_V$ ($v\rightarrow V$), and therefore just the contribution of vacuum prevails. As we always
 must have $T^{00}>0$, we get $p<0$ in Eq.(168). This implies a negative pressure for vacuum energy density of the
ultra-referential $S_V$. So we verify that a negative pressure emerges naturally from such a new component $T^{00}$ (Eq.(167)) 
in the limit of vacuum $S_V$ ($v\rightarrow V$), without being an {\it ad-hoc} condition used in classical spacetime. 

  We can obtain $T^{\mu\nu}_{vacuum}$ by calculating the limit of vacuum-$S_V$ for Eq.(166), by considering Eq.(165), as follows:

 \begin{equation}
 T^{\mu\nu}_{vacuum}= lim_{v\rightarrow V}T^{\mu\nu}= -pg^{\mu\nu},
 \end{equation}
 where we conclude that $\epsilon=-p$. In Eq.(165), we see that the new $4$-velocity vanishes in the limit of the vacuum-$S_V$ ($v\rightarrow V$),
 namely $U^{\mu}_{vac.}=(0,0)$. So, $T^{\mu\nu}_{vac.}$ is in fact a diagonalized tensor as we hope to be. The vacuum-$S_V$ that is
 inherent to such a space-time with an invariant minimum speed works like a {\it sui generis} fluid in equilibrium with negative pressure, 
leading to a cosmological anti-gravity, i.e., the invariant minimum speed connected to a universal background field in the preferred
 frame $S_V$ leads naturally to the well-known equation of state of the cosmological constant $p=w\epsilon$, with $w=-1$\cite{58}. 

\subsection{The cosmological constant $\Lambda$ and the vacuum energy density $\rho$}

 The well-known relation\cite{58} between the cosmological constant $\Lambda$ and the vacuum energy density $\rho_{(\Lambda)}$ is

\begin{equation}
\rho_{(\Lambda)}=\frac{\Lambda c^2}{8\pi G}
\end{equation}

Let us consider a spherical universe with Hubble radius filled by a uniform vacuum energy density. On the surface of such a
sphere (frontier of the observable universe), the bodies (galaxies) experience an accelerated expansion (anti-gravity) due to
the whole ``dark mass (energy)" of vacuum inside the sphere. So we could think that each galaxy is a proof body interacting with that big sphere
like in the simple case of two bodies interaction. However, we need to show that there is an anti-gravitational interaction between the 
ordinary proof mass $m_0$ and the big sphere with a ``dark mass" of vacuum ($M_{\Lambda}$). To do that, let us first start from the well-known 
simple model of a massive proof particle $m_0$ that escapes from a newtonian gravitational potential $\phi$
on the surface of a big sphere of matter, namely $E=m_0c^2(1-v^2/c^2)^{-1/2}\equiv m_0c^2(1+\phi/c^2)$, where $E$ is its relativistic
energy. Here the interval of escape velocity $0\leq v<c$ is associated with the interval of potential $0\leq\phi<\infty$, where we 
stipulate $\phi>0$ to be the attractive (classical) gravitational potential. 

Now we can show that the influence of the background field (vacuum energy inside the sphere) connected to the ultra-referential
$S_V$ (see Eq.(168)) leads to a strong repulsive (negative) gravitational potential ($\phi<<0$) for very low energies ($E\rightarrow 0$). 
In order to see this non-classical aspect of gravitation\cite{59}, we use Eq.(95) just taking into account the new approximation given
for very low energies $(v(\approx V)<<c)$, as follows: 

\begin{equation}
E\approx\theta m_0c^2=m_0c^2\sqrt{1-\frac{V^2}{v^2}}\approx m_0c^2\left(1+\frac{\phi}{c^2}\right),
\end{equation}
where $\phi<0$ (repulsive). For $v\rightarrow V$, this implies $E\rightarrow 0$, which leads to $\phi\rightarrow -c^2$.
So, the non-classical (most repulsive) minimum potential $\phi(V)(=-c^2)$
connected to the vacuum-$S_V$ is responsible for the cosmological anti-gravity (see also Eq.(168) and Eq.(169)). We interpret this result
assuming that only an exotic ``particle" of the vacuum energy at $S_V$ could escape from the anti-gravity ($\phi=-c^2$) generated by the 
vacuum energy inside the sphere (consider $v=V$ in Eq.(171)). Therefore, ordinary bodies like galaxies and any matter on the surface 
of such a sphere cannot escape from its anti-gravity, being accelerated far away.

 According to Eq.(171), it is interesting to note that such an exotic ``particle" of vacuum (at $S_V$) has an infinite mass 
($m_0=\infty$), since we consider $v=V$ ($\theta=0$) in order to get a finite value of $E$, other than the photon ($v=c$), that is a
 massless particle (see Eq.(95)). So we conclude that an exotic ``particle" of the vacuum $S_V$ works like a counterparty of the photon,
 namely an infinitely massive particle, i.e., it can be an extremely massive boson. 

  We consider that the most negative (repulsive) potential ($\phi=-c^2$ for $v=V$, in Eq.(171)) is related to the cosmological
constant (vacuum energy density), since we have shown in Eq.(168) and Eq.(169) that the background reference frame $S_V$ plays the role of 
the vacuum energy density with a negative pressure, working like the cosmological constant $\Lambda$ ($p=-\epsilon=-\rho_{(\Lambda)}$).
So we write

\begin{equation}
\phi_{\Lambda}=\phi(\Lambda)=\phi(V)=-c^2
\end{equation}

Let us consider the simple model of spherical universe with a radius $R_u$, being filled by a uniform vacuum energy
density $\rho_{(\Lambda)}$, so that the total vacuum energy inside the sphere $E_{\Lambda}=\rho_{(\Lambda)}V_u=-pV_u=M_{\Lambda}c^2$.
$V_u$ is its volume and $M_{\Lambda}$ is the total dark mass associated with the dark energy for $\Lambda$ 
(vacuum energy: $w=-1$\cite{58}). Therefore, we find a newtonian and repulsive gravitational potential $\phi_{\Lambda}$ on the surface of 
such a sphere (universe), since it is filled by a low density of dark ``mass'' (dark energy) that is close to its total
density of matter, so that we can use the newtonian approximation for such potential, i.e.: 

\begin{equation}
\phi_{\Lambda}=-\frac{GM_{\Lambda}}{R_u}=-\frac{G\rho_{(\Lambda)}V_u}{R_uc^2}=\frac{4\pi GpR_u^2}{3c^2},
\end{equation}
where $p=-\rho_{(\Lambda)}$, with $w=-1$ (cosmological anti-gravity)\cite{58}. 

 By introducing Eq.(170) into Eq.(173), we find

\begin{equation}
\phi_{\Lambda}=\phi(\Lambda)=-\frac{\Lambda R_u^2}{6}
\end{equation}

Finally, by comparing Eq.(174) with Eq.(172), we extract

\begin{equation}
\Lambda=\frac{6c^2}{R_u^2}=\frac{6V^2}{\xi^2 R_u^2},
\end{equation}
so that we find $\Lambda S_u=24\pi c^2$, where $S_u=4\pi R_u^2$. We have used $V=\xi c$. 

And, by comparing Eq.(173) with Eq.(172), we have

\begin{equation}
\rho_{(\Lambda)}=-p=\frac{3c^4}{4\pi G R_u^2},
\end{equation}
where $\rho_{(\Lambda)} S_u=3c^4/G$. We can verify that Eq.(176) and Eq.(175) satisfy Eq.(170).

 In Eq.(175), $\Lambda$ is a kind of {\it cosmological scalar field (parameter)}, extending the old concept of
Einstein about the cosmological constant for stationary universe. From Eq.(175), by considering the Hubble radius, with
$R_{u}=R_{H_0}\sim 10^{26}$m, we obtain $\Lambda=\Lambda_0\sim (10^{17}m^2s^{-2}/10^{52}m^2)\sim 10^{-35}s^{-2}$.
To be more accurate, we know the age of the universe $T_0=13.7$ Gyr, such that $R_{H_0}=cT_0\approx
1.3\times 10^{26}$m, which leads to $\Lambda_0\approx 3\times 10^{-35}s^{-2}$. It is interesting to notice that this tiny positive value
of $\Lambda_0$ is in agreement with the observational data\cite{60}\cite{61}\cite{62}\cite{63}\cite{64}. The vacuum energy
density\cite{65}\cite{66} given in Eq.(176) for $R_{H_0}$ is $\rho_{(\Lambda_0)}\approx
2\times 10^{-29}g/cm^{3}$, which is also in agreement with observations. 

Here it must be stressed that our assumption for obtaining the tiny value of $\Lambda_0$ starts from first principles related to
a new symmetry in space-time, since we have introduced the idea of a background reference frame $S_V$ for representing the vacuum energy 
connected to an invariant minimum speed in order to solve the question about the cosmological constant problem, where the quantum 
field theory fails\cite{58}. 

\section{Prospects}

The present research has various implications which shall be investigated in coming articles. We should thoroughly explore many 
interesting consequences of SSR and its new dispersion relation in quantum field theories (QFT), since the existence of a mimimum speed
for lower energies with the same status of the speed of light for higher energies leads to a new metric for describing such deformed 
(symmetric) space-time, allowing us to build a modified QFT, where the Lorentz symmetry is broken down. This kind of metric
$\Theta(v)g_{\mu\nu}$ in Eq.(91) is a special case of metric that has already been explored in the literature and it seems to lead to
the Finsler's geometry, namely a Finslerian space with a metric that depends on the position and also the velocity, 
i.e., $G_{\mu\nu}(x,\dot{x})$\cite{67}\cite{68}\cite{69}\cite{70}.

Beyond the experiment proposed in Section $7$ for detecting $V$ by investigating the change in the decay rate of a radiative sample
thermalized close to $T=0$K, another experiment, which could be used to test for the existence of $V$, would be done by measuring the
 speed of light of a {\it laser} in an ultra-cold gas at temperatures closer and closer to the absolute zero ($0$K). As previously obtained in laboratory, the speed of light
was drastically reduced inside the condensate. So, we expect the speed of light in such a medium comes close to a non-zero minimum value
$V$ when its temperature approaches to $T=0$K, i.e., $c(T\approx 0K)=c^{\prime}\approx V(\sim 10^{-14}$m/s), thus leading to the emergence
of a very high refraction index ($n$) of the gas close to absolute zero, namely $n(T\rightarrow 0)\rightarrow\sigma=\xi^{-1}=c/V$. Since
$V$ is the minimum speed that light could reach inside a condensate close to $T=0$K, there should be a maximum refraction index,
being finite and very large, i.e., $n_{max}=\sigma=c/V\sim 10^{22}$. Although there may be many technical difficulties for making such
experiment with very high accuracy, a thorough investigation of this result would be quite worthwhile.

As the last term of Eq.(33) ($\rho^{(2)}_{em}$) provides an electromagnetic effect of purely gravitational origin, such an effect being 
capable of increasing significantly the magnitude of the electric force between two charges immersed only in a strong gravitational field, 
the idea of reciprocity by replacing a strong gravitational field by a strong static electric field, in which two masses are immersed,
can lead to a very small change in the gravitational force acting on the test mass. To achieve such a slight deviation on gravity,  
it would be feasible to search for such an effect by using a torsional balance with a discharged test mass immersed in a very
intense electrostatic field. This could be a possible experiment to test any crosstalk between gravitation and electromagnetism still
motivated by an interesting recent article\cite{15}.\\

{\noindent\bf  Acknowledgements}

I am specially grateful to Em\'ilio C. M. Guerra and Wladimir Guglinski for interesting discussions. I thank J. Alves Correia Jr., 
Alisson Xavier, C\'assio Guilherme Reis, Carlos Magno Leiras, G. Vicentini, P. R. Silva, Paulo R. Souza Coelho,
A. C. Amaro de Faria Jr., Jonas Durval Cremasco, Augusto C. Lobo and Fl\'avio Cassino for their special comprehension of this
work's significance, where one proposes the solution of the so-called {\it``The Cosmological Constant Problem''}\cite{58}, i.e., 
the puzzle of the very low vacuum energy density and the tiny value of the cosmological constant. 

I dedicate this work to the memory of the great scientist Albert Einstein who dedicated his life to understand the foundations of nature 
in a relentless search for a fundamental theory to describe all natural processes through first principles by symmetries imposed by nature.

\end{document}